# Semiclassical asymptotics of orthogonal polynomials, Riemann-Hilbert problem, and universality in the matrix model

By Pavel Bleher and Alexander Its

## Abstract

We derive semiclassical asymptotics for the orthogonal polynomials $P_n(z)$ on the line with respect to the exponential weight $\exp(-NV(z))$, where $V(z)$ is a double-well quartic polynomial, in the limit when $n, N \to \infty$. We assume that $\varepsilon \le (n/N) \le \lambda_{\mathrm{cr}} - \varepsilon$ for some $\varepsilon > 0$, where $\lambda_{\mathrm{cr}}$ is the critical value which separates orthogonal polynomials with two cuts from the ones with one cut. Simultaneously we derive semiclassical asymptotics for the recursive coefficients of the orthogonal polynomials, and we show that these coefficients form a cycle of period two which drifts slowly with the change of the ratio $n/N$. The proof of the semiclassical asymptotics is based on the methods of the theory of integrable systems and on the analysis of the appropriate matrix Riemann-Hilbert problem. As an application of the semiclassical asymptotics of the orthogonal polynomials, we prove the universality of the local distribution of eigenvalues in the matrix model with the double-well quartic interaction in the presence of two cuts.

## Contents

## 1. Introduction and formulation of the main theorem

About thirty years ago Freeman Dyson found an exact solution for the scaling limit of correlations between eigenvalues in the Gaussian unitary ensemble of random matrices. He conjectured that this scaling limit should appear in a much broader class of non-Gaussian unitary ensembles of random matrices. This constitutes the famous conjecture of universality in the theory of random matrices. Dyson found also a remarkable formula which expresses the eigenvalue correlations for finite $N$ (before the scaling limit) in terms of orthogonal polynomials on the line with respect to an exponential weight (see [Dys]; see also [Meh], [BIPZ], and others). This reduces the universality conjecture to semiclassical asymptotics of orthogonal polynomials.

The theory of orthogonal polynomials is a classical branch of mathematical analysis and it finds applications in different areas of pure and applied mathematics. Among the most exciting recent applications of orthogonal polynomials related to the random matrices are those to quantum gravity, string theory, and integrable PDEs (see, e.g., [ASM], [Dem], [Wit] and references therein). In the present paper we show how these recent developments can be beneficial for the theory of orthogonal polynomials itself and, in particular, for the aspects of the theory related to the matrix models.

The Dyson formula gives the eigenvalue correlations for finite $N$. To get the universal correlation functions in the limit when $N \to \infty$ one needs to obtain semiclassical asymptotics for orthogonal polynomials. For the Gaussian matrix model the problem reduces to the semiclassical asymptotics for the Hermite polynomials which were obtained in the classical work of Plancherel and Rotach [PR]. The Hermite polynomials are the eigenfunctions of the quantum linear oscillator (up to a Gaussian function) and their asymptotics follow from the semiclassical formulae for the Schrödinger operator with quadratic potential. Using the Plancherel-Rotach asymptotics in the inner region, between the turning points, Dyson evaluated the scaling limit of correlations in the Gaussian matrix model inside the support of the limiting spectral measure. The scaling limit is expressed in terms of the sine-kernel. Dyson's conjecture of universality states that the same scaling limit in the bulk of the spectrum should appear in the general case of non-Gaussian unitary ensembles of random matrices. The conjecture of universality was then extended to the scaling limit of correlations at the edges of the support of the limiting spectral measure. For the Gaussian matrix model the scaling limit of correlations at the edge is expressed in terms of the Airy kernel (see [BB], [For], [TW2]), and this follows from the Plancherel-Rotach asymptotics near the turning points. The conjecture of universality states that the same scaling limit is to appear at all regular edges of the support of the limiting spectral measure for a gen-eral unitary matrix model. Here regular means that the spectral density has a square-root singularity at the edge. The conjecture of universality has, in fact, an even more general formulation. One should expect universal scaling limits to exist at regular critical points, tricritical points, etc., each being related to its own class of universality. The critical points do not occur in the Gaussian matrix model but they do in non-Gaussian matrix models, and the critical points correspond usually to bifurcations of the support of the limiting spectral measure. In this paper we prove the universality conjecture inside the spectrum and at the edges for the matrix model with quartic interaction. We will discuss the universality at the critical points in another publication. A very different approach using estimates for resolvents, to the universality conjecture for a large class of potentials has been given in the paper [PS] by Pastur and Shcherbina.

We derive the scaling limit of correlations from the semiclassical asymptotics of orthogonal polynomials. For the general matrix model the problem of semiclassical asymptotics is much more difficult than for the Gaussian model. The corresponding Schrödinger operator is very complicated and, what is even more important, it depends on unknown recursive coefficients for orthogonal polynomials. So one has to solve the two problems simultaneously: to find semiclassical asymptotics for the recursive coefficients and for the orthogonal polynomials. The recursive coefficients for orthogonal polynomials satisfy a nontrivial nonlinear equation which was discovered independently in mathematics and in physics. Mathematicians call it the Freud equation (and its generalizations), in the name of Géza Freud who first derived it for orthogonal polynomials with some exponential weights on the line. For physicists this equation is known as a discrete string equation, and it was derived and studied in the papers by Bessis, Brézin, Itzykson, Parisi, and Zuber, and others (see, e.g., [BPIZ], [BIZ], [IZ]), devoted to the problem of enumeration of Feynman graphs in string theory. The Freud equation does not admit a direct solution, and this is the main reason why despite many interesting and deep achievements in the theory of orthogonal polynomials (see, e.g., [LS], [Mag4], [Nev3] and references therein), the semiclassical asymptotics for general exponential weights remains one of the central unsolved problems in the theory. Our new technique is based on the important observation made in [FIK1] that the Freud equation is integrable in the sense that it admits a Lax pair. This motivates use of the powerful tools of the theory of integrable systems, the method developed in the present paper. The central role in our scheme is played by the Riemann-Hilbert setting of orthogonal polynomials suggested in [FIK4].

To formulate the problem let us consider a polynomial $V(z)$ with real coefficients,

$$V(z) = a_0 + a_1 z + \cdots + a_{2p} z^{2p}, \qquad a_{2p} > 0,$$

and orthogonal polynomials on the real line,

$$P_n(z) = z^n + \dots, \qquad n = 0, 1, 2, \dots,$$

with respect to the exponential weight $e^{-NV(z)}dz$, i.e.,

$$(1.1) \qquad \int_{-\infty}^{\infty} P_n(z)P_m(z)e^{-NV(z)}dz = h_n\delta_{nm}, \qquad h_n > 0.$$

We normalize the orthogonal polynomials by taking the leading coefficient equal to 1. The number $N$ in the weight is a large parameter and we are interested in the asymptotics of the polynomials $P_n(z)$ in the limit when $n, N \to \infty$ in such a way that

$$\varepsilon < \frac{n}{N} < \varepsilon^{-1},$$

for some fixed $\varepsilon > 0$. In this approach the large parameter $N$ plays the same role as the inverse Planck constant plays in the semiclassical asymptotics of quantum mechanics.

The orthogonal polynomials $P_n(z)$ satisfy the basic recursive equation

$$(1.2) \qquad zP_n(z) = P_{n+1}(z) + Q_nP_n(z) + R_nP_{n-1}(z),$$

where $Q_n$ and $R_n$ are some coefficients which depend on $N$. The semiclassical asymptotics of the coefficients $Q_n$ and $R_n$ are tied to one of the polynomials $P_n(z)$ and we solve these two problems together.

Although our approach is quite general and can be applied to a general polynomial $V(z)$, in this paper we will work out all the details for the quartic polynomial

$$(1.3) \qquad V(z) = \frac{tz^2}{2} + \frac{gz^4}{4}, \qquad g > 0,$$

and we will consider the most interesting case when $t < 0$ (a double-well potential). Since $V(z)$ is even, (1.2) is simplified to

$$(1.4) \qquad zP_n(z) = P_{n+1}(z) + R_nP_{n-1}(z),$$

where

$$(1.5) \qquad\qquad R_n = \frac{h_n}{h_{n-1}}.$$

In addition, integration by parts gives

$$(1.6) \qquad P_n'(z) = NR_n[t + g(R_{n-1} + R_n + R_{n+1})]P_{n-1}(z)$$
$$+ g(NR_{n-2}R_{n-1}R_n)P_{n-3}(z), \qquad (') \equiv \frac{d}{dz}.$$

Since $P_n'(z) = nz^{n-1} + \dots$, this implies the Freud equation

$$(1.7) \qquad n = NR_n[t + g(R_{n-1} + R_n + R_{n+1})], \quad n \geq 1,$$

(cf. [Fre]). The difference equation (1.7) is supplemented by the following initial conditions:

$$R_0 = 0, \quad R_1 = \frac{\int_0^\infty z^2 e^{-NV(z)} dz}{\int_0^\infty e^{-NV(z)} dz} \equiv \frac{h_1}{h_0}.$$

From (1.5) and (1.7) it follows that

$$(1.8) \qquad 0 < R_n < \frac{-t + \sqrt{t^2 + 4\lambda g}}{2g}, \qquad \lambda = \frac{n}{N}.$$

Let

$$(1.9) \qquad \psi_n(z) = \frac{1}{\sqrt{h_n}} P_n(z) e^{-NV(z)/2}.$$

Then

$$(1.10) \qquad \int_{-\infty}^\infty \psi_n(z) \psi_m(z) \, dz = \delta_{nm}.$$

Our main goal is to prove semiclassical asymptotics for the functions $\psi_n(z)$ and for the coefficients $R_n$ in the limit when $N, n \to \infty$ in such a way that there exists $\varepsilon > 0$ such that the ratio $\lambda = n/N$ satisfies the inequalities

$$(1.11) \qquad \varepsilon < \lambda < \lambda_{\mathrm{cr}} - \varepsilon, \qquad \lambda = \frac{n}{N},$$

where

$$(1.12) \qquad \lambda_{\mathrm{cr}} = \frac{t^2}{4g}.$$

In what follows the potential function

$$U(z) = z^2 \left[ \frac{(gz^2 + t)^2}{4} - \lambda g \right],$$

is important. We introduce the turning points $z_1$ and $z_2$ as zeros of $U(z)$,

$$(1.13) \qquad z_{1,2} = \left( \frac{-t \mp 2\sqrt{\lambda g}}{g} \right)^{1/2}.$$

The condition (1.11) implies that $z_1$ and $z_2$ are real, and $z_2 > z_1 > C\sqrt{\varepsilon}$. We prove the following main theorem.

THEOREM 1.1. *Assume that $N, n \to \infty$ in such a way that (1.11) holds. Then there exists $C = C(\varepsilon) > 0$ such that*

$$(1.14) \qquad \left| R_n - \frac{-t - (-1)^n \sqrt{t^2 - 4\lambda g}}{2g} \right| \le C N^{-1}, \qquad \lambda = \frac{n}{N}.$$

*In addition, for every $\delta > 0$, in the interval $z_1 + \delta < z < z_2 - \delta$,*

$$(1.15) \quad \psi_n(z) = \frac{2C_n\sqrt{z}}{\sqrt{\sin\phi}} \left\{ \cos\left[ \frac{\left(n + \frac{1}{2}\right)}{2}\left(\frac{\sin 2\phi}{2} - \phi\right) - \frac{(-1)^n\chi}{4} + \frac{\pi}{4} \right] + O\left(N^{-1}\right) \right\},$$

*where*

$$\phi = \arccos q, \qquad \chi = \arccos r,$$

*and*

$$(1.16) \quad q = \frac{gz^2 + t}{2\sqrt{\lambda' g}}, \qquad r = \frac{2\sqrt{\lambda' g} - tq}{2\sqrt{\lambda' g}\, q - t}, \qquad \lambda' = \frac{n + \frac{1}{2}}{N}.$$

*If $z > z_2 + \delta$ or $0 \leq z < z_1 - \delta$, then*

$$(1.17) \quad \psi_n(z) = (-1)^\sigma \frac{C_n\sqrt{z}}{\sqrt{\sinh\phi}} \exp\left\{ -\frac{\left(n + \frac{1}{2}\right)}{2}\left[ \frac{\sinh(2\phi)}{2} - \phi \right] \right.$$
$$\left. + \frac{(-1)^n\chi}{4} + O\left(\frac{1}{N(1 + |z|)}\right) \right\},$$

*where*

$$\sigma = \begin{cases} 0 & \text{if} \quad z > z_2 + \delta, \\ \left[\dfrac{n}{2}\right] = k & \text{if} \quad 0 \leq z < z_1 - \delta \quad \text{and} \quad n = 2k \quad \text{or} \quad 2k+1, \end{cases}$$

*and*

$$\phi = \cosh^{-1}|q|, \qquad \chi = \cosh^{-1}|r|$$

*where $q, r$ are given by (1.16).*

*If $z_j - \delta \leq z \leq z_j + \delta, \quad j = 1, 2,$ then*

$$(1.18) \quad \psi_n(z) = \frac{D_n\, z}{\sqrt{|w'(z)|}} \left[ (1 + r_1(z))\, \mathrm{Ai}\left(N^{2/3}w(z)\right) + r_2(z)\, \mathrm{Ai}'\left(N^{2/3}w(z)\right) \right],$$

*where $r_1(z) = O\left(N^{-1}\right)$, $r_2(z) = O\left(N^{-4/3}\right)$, $\mathrm{Ai}(z)$ is the Airy function, and $w(z)$ is an analytic function on $[z_j - \delta, z_j + \delta]$ such that for $(z - z_j^{(N)})(-1)^j \geq 0$,*

$$(1.19) \quad w(z) = \left[ \frac{3}{2}\left| \int_{z_j^N}^z \sqrt{U_N(v)}\, dv \right| \right]^{2/3}, \qquad j = 1, 2,$$

*where $z_j^N = z_j + O(N^{-1})$ is the closest to $z_j$ zero of the polynomial*

$$(1.20) \quad U_N(z) = U(z) + N^{-1}\left( -\frac{gz^2}{2} + \frac{t}{2} + gR_n^0 \right)$$
$$= z^2\left[ \frac{(gz^2 + t)^2}{4} - \lambda' g \right] + N^{-1}\left( \frac{t}{2} + gR_n^0 \right),$$

*and*

$$R_n^0 = \frac{-t - (-1)^n \sqrt{t^2 - 4\lambda g}}{2g}.$$

*The constant factor $C_n$ in (1.15) and (1.17) satisfies the asymptotic equation*

$$(1.21) \qquad C_n = \frac{1}{2\sqrt{\pi}} \left(\frac{g}{\lambda}\right)^{1/4} (1 + O(N^{-1})),$$

*and $D_n$ in (1.18) is*

$$(1.22) \qquad D_n = N^{1/6} \sqrt{g} \, (-1)^{\sigma_0}, \qquad \sigma_0 = (2 - j) \left[\frac{n}{2}\right].$$

*Finally, the asymptotics of $h_n$ in (1.1), (1.3) is*

$$(1.23) \qquad h_n = 2\pi \sqrt{R_n^0} \exp\left[\frac{Nt^2}{4g} - \frac{N\lambda}{2}\left(1 + \ln\frac{g}{\lambda}\right) + O(N^{-1})\right].$$

The asymptotic formulae (1.15), (1.17), and (1.18) are an extension of the classical Plancherel–Rotach asymptotics for the Hermite polynomials (see [PR] and [Sze]), to the orthogonal polynomials with respect to the weight $e^{-NV(z)}$ where $V(z)$ is the quartic polynomial (1.3). These formulae are extended into the complex plane in $z$ as well (see Theorems 7.5, 7.7 in Section 7 below). The formula (1.15) can be rewritten as the semiclassical formula

$$(1.24) \qquad \psi_n(z) = \frac{z\sqrt{g/\pi}}{|U(z)|^{1/4}}\left[\cos\left(N\int_{z_2^N}^z |U_N(v)|^{1/2}\, dv + \frac{\pi}{4}\right) + O(N^{-1})\right],$$

where $U_N(z)$ is as defined in (1.20). Asymptotics (1.14) of the coefficients $R_n$ are Freud-type asymptotics. For the homogeneous function $V(z) = |z|^\alpha$ and some of its generalizations, the asymptotics of $R_n$ are obtained in the papers of Freud [Fre], Nevai [Nev1], Magnus [Mag1,2], Lew and Quarles [LQ], Máté, Nevai, and Zaslavsky [MNZ], Bauldry, Máté, and Nevai [BMN]. Semiclassical asymptotics of the functions $\psi_n(z)$ are proved then for $V(z) = z^4$ by Nevai [Nev1] and for $V(z) = z^6$ by Sheen [She]. For general homogeneous $V(z)$, somewhat weaker asymptotics are obtained in the works of Lubinsky and Saff [LS], Lubinsky, Mhaskar, and Saff [LMS], Lubinsky [Lub], Levin and Lubinsky [LL], Rahmanov [Rah], and others. Application of these asymptotics to random matrices is discussed in the work of Pastur [Pas]. The distribution of zeros and related problems for orthogonal polynomials corresponding to general homogeneous $V(z)$ are studied in the recent work [DKM] by Deift, Kriecherbauer, and McLaughlin. Many results and references on the asymptotics of orthogonal polynomials are given in the comprehensive review article [Nev3] of Nevai. The problem of finding asymptotics of $R_n$ for a quartic nonconvex polynomial is discussed in [Nev2, 3], and is known as "Nevai's problem."

Equation (1.14) shows that if $0 < \lambda < \lambda_{\mathrm{cr}}$ then

(1.25)
$$\lim_{N\to\infty;\ (2m)/N\to\lambda} R_{2m} = L(\lambda) = \frac{-t - \sqrt{t^2 - 4\lambda g}}{2g},$$

$$\lim_{N\to\infty;\ (2m+1)/N\to\lambda} R_{2m+1} = R(\lambda) = \frac{-t + \sqrt{t^2 - 4\lambda g}}{2g}.$$

Both $L(\lambda)$ and $R(\lambda)$ satisfy the quadratic equation

$$gu^2 + tu + \lambda = 0,$$

so that, when $n$ grows, $R_n$ jumps back and forth from one sheet of the parabola to another (see Fig.1). In other words, the sequence $\{\, R_n,\ n = 0, 1, 2, \ldots \,\}$ forms a period two cycle which is slowly drifting with the change of the ratio $n/N$. At $\lambda = \lambda_{\mathrm{cr}}$ the two sheets of the parabola merge, i.e., $L(\lambda_{\mathrm{cr}}) = R(\lambda_{\mathrm{cr}})$. For $\lambda > \lambda_{\mathrm{cr}}$,

(1.26)
$$\lim_{N\to\infty;\ n/N\to\lambda} R_n = Q(\lambda),$$

where $u = Q(\lambda)$ satisfies the quadratic equation

$$3gu^2 + tu - \lambda = 0$$

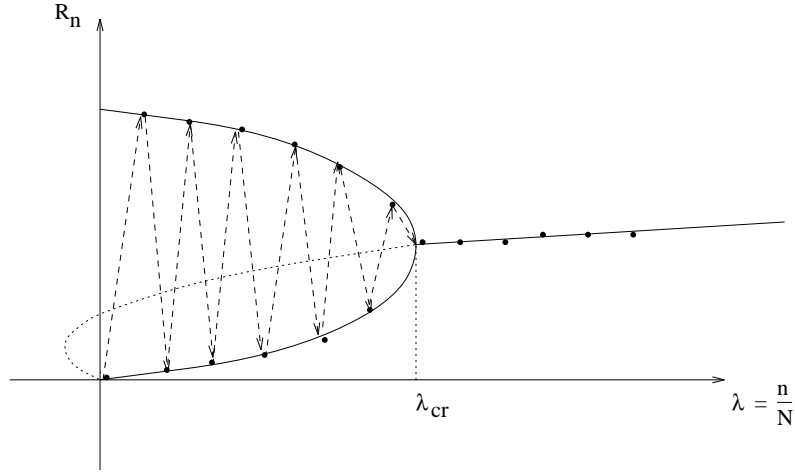

Figure 1. The qualitative behavior of the recurrence coefficients

(which follows from the Freud equation (1.7) if we put $u = R_{n-1} = R_n = R_{n+1}$). We consider semiclassical asymptotics for $\lambda > \lambda_{\mathrm{cr}}$ and in the vicinity of $\lambda_{\mathrm{cr}}$ (double scaling limit) in a separate work. The difference in the asymptotics between the cases $\lambda < \lambda_{\mathrm{cr}}$ and $\lambda > \lambda_{\mathrm{cr}}$ is that for $\lambda < \lambda_{\mathrm{cr}}$ the function $\psi_n(z)$ is concentrated on two intervals, or two cuts, $[-z_2, -z_1]$ and $[z_1, z_2]$, and it is exponentially small outside of these intervals, while for $\lambda > \lambda_{\mathrm{cr}}$, $z_1$ becomes pure imaginary, and $\psi_n(z)$ is concentrated on one cut $[-z_2, z_2]$. The transition from a two-cut to a one-cut regime is discussed in physical papers by Cicuta, Molinari, and Montaldi [CMM], Crnković and Moore [CM], Douglas, Seiberg, Shenker [DSS], Periwal and Shevitz [PeS], and others.

Another proof of the limits (1.25) was recently given by Albeverio, Pastur, and Shcherbina (see [APS]) in a completely different approach based on some estimates of the Stieltjes transform of the spectral measure.

A general ansatz on the structure of the semiclassical asymptotics of the functions $\psi_n(z)$ for a "generic" polynomial $V(z)$ was proposed in the work [BZ] of Brézin and Zee. They considered $n$ close to $N$, $n = N + O(1)$, and suggested that for these $n$'s,

$$(1.27) \qquad \psi_n(z) = \frac{1}{\sqrt{f(z)}} \cos\big(N\zeta(z) - (N - n)\varphi(z) + \chi(z)\big),$$

for some functions $f(z)$, $\zeta(z)$, $\varphi(z)$, and $\chi(z)$. This fits in well the asymptotics (1.15), except for the factor $(-1)^n$ at $\chi$ in (1.15), which is related to the two-cut structure of $\psi_n(z)$.

Equation (1.7) also appears in the planar Feynman diagram expansions of Hermitian matrix models, which were introduced and studied in the classical papers [BIPZ], [BIZ], [IZ] by Brézin, Bessis, Itzykson, Parisi, and Zuber and in the well-known recent works by Brézin, Kazakov [BK], Douglas, Shenker [DS], and Gross, Migdal [GM] devoted to the matrix models for 2D quantum gravity (see also [Dem] and [Wit]). In fact, it is the latter context that broadened the interest to the Freud equation (1.7) and brought to the area new powerful analytic methods from the theory of integrable systems. It turns out [FIK1,2] that equation (1.7) admits a $2 \times 2$ matrix Lax pair representation (see equation (3.15) below), which allows one to identify the Freud equation (1.7) as a discrete Painlevé I equation and imbeds it in the framework of the isomonodromy deformation method suggested in 1980 by Flaschka and Newell [FN] and by Jimbo, Miwa, and Ueno [JMU] (about analytical aspects of the method see, e.g., [IN] and [FI]). The relevant Riemann-Hilbert formalism for (1.7) was developed in [FIK1,2] as well. It was used in [FIK1-3] together with the isomonodromy method for the asymptotic analysis of the solution of (1.7), which is related to the double-scaling limit in the 2D quantum gravity studied in [BK], [DS], [GM].

As a matter of fact, the solution of (1.7) which is analysed in [FIK] corresponds to the system of orthogonal polynomials on certain rays in the complex plane. Nevertheless, the basic elements of the Riemann-Hilbert isomonodromy scheme suggested in [FIK] can be easily extended to an arbitrary system of orthogonal polynomials corresponding to a rational potential $V(z)$ (cf. [FIK4]; see also Appendix E below). The proof of Theorem 1.1 is based on the approach of [FIK] combined with the nonlinear steepest descent method proposed recently by Deift and Zhou [DZ1] for analyzing the asymptotics of oscillatory matrix Riemann-Hilbert problems. We appeal to the Deift-Zhou method in Section 7 where we construct explicitly and then justify rigorously the asymptotic solution of the master Riemann-Hilbert problem associated to the orthogonal polynomials (see conditions (i)–(iii) and (5.16)–(5.18) below).

The paper is organized as follows:

In the next section we use the results of Theorem 1.1 for proving the universality of the local distribution of eigenvalues in the matrix model with quartic potential. We prove both the sine-kernel universality at internal points and the Airy-kernel universality at the end-points of the limiting spectral measure. It is important to note that for internal points the universality was recently proved by a different technique in the paper by Pastur and Shcherbina [PS], for a general class of matrix models.

In Sections 3–5 we reproduce in a slightly different way the results of [FIK] concerning the Lax pair representation of equation (1.7) and the matrix Riemann-Hilbert reformulation of the orthogonal polynomials. In particular, we show that there is an exact and simple relation between orthogonal polynomials $P_n(z)$ and the $2 \times 2$ matrix-valued function $\Psi_n(z)$ which solves the following matrix Riemann-Hilbert problem on a line (the problem (5.16)–(5.18) below):

(i)  $\Psi_n(z)$ is analytic in $\mathbb{C} \setminus \mathbb{R}$, and it has a jump at the real line.

(ii)  $\Psi_n(z) \sim \left( \sum_{k=0}^{\infty} \dfrac{\Gamma_k}{z^k} \right) e^{-\left( \frac{NV(z)}{2} - n \ln z + \lambda_n \right) \sigma_3}, \quad z \to \infty, \quad$ where

$$\lambda_n = \frac{1}{2} \ln h_n, \quad \sigma_3 = \begin{pmatrix} 1 & 0 \\ 0 & -1 \end{pmatrix},$$

$$\Gamma_0 = \begin{pmatrix} 1 & 0 \\ 0 & R_n^{-1/2} \end{pmatrix}, \qquad \Gamma_1 = \begin{pmatrix} 0 & 1 \\ R_n^{1/2} & 0 \end{pmatrix}.$$

(iii)  $\Psi_{n+}(z) = \Psi_{n-}(z)S, \qquad \text{Im } z = 0, \qquad S = \begin{pmatrix} 1 & -2\pi i \\ 0 & 1 \end{pmatrix}.$

As explained at the end of Section 5, in the setting of the Riemann-Hilbert problem (i)–(iii), the real quantities $R_n$ and $\lambda_n$ *are not the given data*. They are evaluated via the solution $\Psi_n(z)$, which is determined by conditions (i)–(iii) uniquely without any prior specification of $R_n$ and $\lambda_n$. Simultaneously, the function $\Psi_n(z)$ satisfies the Lax pair (see equation (3.15) below) whose second equation is the linear differential equation:

$$(1.28) \qquad \frac{d\Psi_n(z)}{dz} = N A_n(z)\Psi_n(z),$$

where

$$A_n(z) = \begin{pmatrix} -(\frac{tz}{2} + \frac{gz^3}{2} + gzR_n) & R_n^{1/2}[t + gz^2 + g(R_n + R_{n+1})] \\ -R_n^{1/2}[t + gz^2 + g(R_{n-1} + R_n)] & \frac{tz}{2} + \frac{gz^3}{2} + gzR_n \end{pmatrix}.$$

The jump matrix $S$ in (iii) constitutes the only nontrivial Stokes matrix (for more details see Sections 4, 5) corresponding to the system (1.28) with $R_n$ generated by (1.4) and (1.5). This reduces the problem of the asymptotic analysis of the quantities $P_n(z)$ and $R_n$ to the asymptotic solution of the matrix Riemann-Hilbert problem (i)–(iii), i.e. to the asymptotic solution of the corresponding inverse monodromy problem for differential equation (1.28).

The short but important Section 6 provides a formal asymptotic ansatz indicated in (1.14) for the recurrence coefficients $R_n$.

The proof of Theorem 1.1 is given in Section 7. The central point of the proof is a construction of an approximate solution $\Psi^0(z)$ to the Riemann–Hilbert problem (i)–(iii). To that end we consider an approximate differential equation (1.28) in which the coefficients $R_n$ are replaced by the numbers

$$R_n^0 = \frac{-t - (-1)^n \sqrt{t^2 - 4(n/N)g}}{2g},$$

that are taken from our formal analysis in Section 6. The function $\Psi^0(z)$ is constructed then as a semiclassical solution to this approximate differential equation with the large parameter $N$. Our semiclassical construction is based on the version of the complex WKB method which was recently suggested in [Kap] for asymptotic solution of the direct monodromy problems for the $2 \times 2$ systems with rational coefficients. We partition the whole complex plane into several regions which separate different turning points of the approximate differential equation, and we construct WKB- and turning point semiclassical solutions in each of these regions (see details in Section 7). This provides us with an *explicit* matrix-valued function $\Psi^0(z)$ which solves asymptotically, as $N \to \infty$, the basic Riemann-Hilbert problem (i)–(iii). Then we prove that the quotient $\Psi_n(z)\left[\Psi^0(z)\right]^{-1}$ is equal to $I + O(N^{-1}(1+|z|)^{-1})$ and this completes the proof of Theorem 1.1. We emphasize that we only use equation (1.28) to motivate our choice of the function $\Psi^0(z)$. The uniform estimate for the difference, $\Psi_n(z)\left[\Psi^0(z)\right]^{-1} - I$, is proved by means independent of the WKB theory of differential equations. The main ideas and technique used in Section 7 are based on the Deift-Zhou nonlinear steepest descent method [DZ].

The Riemann-Hilbert reformulation of the orthogonal polynomials (1.1), (1.3) suggested in [FIK] plays the central role in our approach and in its extention to the general rational potentials $V(z)$. For that reason we decided to present with more detail the scheme of [FIK] in Appendix E.

As mentioned above, the Freud equation (1.7) has a meaning similar to the discrete Painlevé I equation. We refer the reader to the papers [FIZ], [NPCQ], [GRP], [Mag3,4], [Meh2] for more on the subject. As first noticed by Kitaev, equation (1.7) can also be interpreted as the Backlund-Schlezinger transform of the classical Painlevé IV equation so that the coefficients $R_n$ coincide, in fact, with the special PIV function (see [FIK1,3] for more details). This PIV function, in turn, can be expressed in terms of certain $n \times n$ determinants involving the parabolic cylinder functions (see [Mag4]). In this work however we do not use these algebraic connections to the modern Painlevé theory. Instead, we use its analytical methods.

The present paper is a revised and shortened version of our earlier preprint [BI].

## 2. Universality of the local distribution of eigenvalues in the matrix model

Theorem 1.1 can be applied to proving the universality of the local distribution of eigenvalues in the matrix model with quartic potential. The matrix model is defined as follows. Let $M = (M_{jk})_{j,k=1,\dots N}$ be a Hermitian random matrix, with the probability distribution

$$(2.1) \qquad \mu_N(dM) = Z_N^{-1} e^{-N\operatorname{Tr} V(M)} dM,$$

where

$$V(M) = a_0 + a_1 M + \dots + a_{2p} M^{2p}, \qquad a_{2p} > 0,$$

is a polynomial,

$$dM = \prod_{j<k} (d\operatorname{Re} M_{jk}\, d\operatorname{Im} M_{jk}) \prod_j dM_{jj},$$

is the Lebesgue measure on the space of Hermitian matrices, and

$$Z_N = \int e^{-N\operatorname{Tr} V(M)} dM$$

is the grand partition function. Let $\lambda_1 \leq \dots \leq \lambda_N$ be eigenvalues of $M$. Consider the distribution function of the eigenvalues,

$$F_N(z) = N^{-1} \operatorname{E} \# \{j \colon \lambda_j \leq z\}.$$

and the density function

$$p_N(z) = F_N'(z).$$

In the matrix model we are interested in the following problems:

(1) To calculate the limit density $p(z) = \lim_{N \to \infty} p_N(z)$.

(2) To calculate the limit local distribution (scaling limit) of eigenvalues at regular points, where $p(z)$ is positive, and at end-points, where $p(z)$ vanishes.

(3) To calculate the free energy

$$f(a_0, \ldots, a_{2p}) = -\lim_{N \to \infty} \frac{\log Z_N(a_0, \ldots, a_{2p})}{N^2}$$

and to find the points of nonanalyticity of $f$ (critical points) in the space of the parameters $a_0, \ldots, a_{2p}$. A further problem is to calculate the critical asymptotics of the recursive coefficients $R_n$ and of the local distribution of eigenvalues (double scaling limit).

Dyson [Dys] (see also [Meh1] and [TW1]) proved a formula which expresses the correlations between the eigenvalues of $M$ in terms of orthogonal polynomials. Namely, the $m$-point correlation function is written as

(2.2) $$K_{Nm}(z_1, \ldots, z_m) = \det \big( Q_N(z_j, z_k) \big)_{j,k=1,\ldots,m}$$

where

(2.3) $$Q_N(z, w) = \sum_{j=1}^{N} \psi_j(z) \psi_j(w),$$

and $\psi_j(z)$ is as defined in (1.9). When $m = 1$ the correlation function reduces to the function $N p_N(z)$; hence

$$p_N(z) = N^{-1} \sum_{j=1}^{N} \psi_j^2(z).$$

By the Christoffel-Darboux formula (see, e.g., [Sze]), the kernel $Q_N(z, w)$ can be written as

(2.4) $$Q_N(z, w) = \frac{\sqrt{R_{N+1}} \left[ \psi_{N+1}(z) \psi_N(w) - \psi_N(z) \psi_{N+1}(w) \right]}{z - w},$$

and

(2.5) $$p_N(z) = \frac{\sqrt{R_{N+1}} \left[ \psi'_{N+1}(z) \psi_N(z) - \psi'_N(z) \psi_{N+1}(z) \right]}{N}.$$

The formula (1.24) is valid in a complex neighborhood of the interval $[z_1 + \delta, z_2 - \delta]$ and this allows us to differentiate it. We will assume that

$$t < t_{\mathrm{cr}} = -2\sqrt{g}$$

(two-cut case); hence we can use $n = N$ in the asymptotic formulae (1.15)–(1.18). For the sake of brevity we rewrite (1.24), (1.15) as

$$(2.6) \qquad \psi_n = \frac{Cz}{\sqrt{\zeta_z}} \cos(N\zeta + \eta),$$

where

(2.7)

$$C = \sqrt{g/\pi}\,; \qquad \zeta = \zeta(z; \lambda') = \int_{z_2}^{z} |U_0(v; \lambda')|^{1/2} dv + \frac{\pi}{4N}\,;$$

$$\zeta_z = \frac{\partial \zeta(z; \lambda')}{\partial z} = |U_0(z; \lambda')|^{1/2}; \quad U_0(z; \lambda') = z^2 \left[\frac{(gz^2 + t)^2}{4} - \lambda' g\right];$$

$$\eta = -\frac{(-1)^n}{4} \chi(z; \lambda') = -\frac{(-1)^n}{4} \arccos r, \quad r = \frac{2\sqrt{\lambda' g} - tq}{2\sqrt{\lambda' g}\, q - t}, \quad q = \frac{gz^2 + t}{2\sqrt{\lambda' g}},$$

and we drop terms of the order of $N^{-1}$. In addition, (1.24) gives that modulo terms of the order of $N^{-1}$,

$$(2.8) \qquad \psi_{n\pm 1} = \frac{Cz}{\sqrt{\zeta_z}} \cos(N\zeta \pm \xi - \eta),$$

where

$$(2.9) \qquad \xi = \frac{\partial \zeta(z; \lambda')}{\partial \lambda'} = -\frac{1}{2} \arccos q, \qquad q = \frac{gz^2 + t}{2\sqrt{\lambda' g}}\,.$$

The functions $\psi_n$ satisfy the recursive equation

$$z\psi_n = \sqrt{R_{n+1}}\, \psi_{n+1} + \sqrt{R_n}\, \psi_{n-1}$$

(see (1.4)), hence from (2.6) and (2.8) we obtain that

$$z\cos(N\zeta - \eta)\cos(2\eta) - z\sin(N\zeta - \eta)\sin(2\eta)$$
$$= \sqrt{R_{n+1}}\cos(N\zeta - \eta)\cos\xi - \sqrt{R_{n+1}}\sin(N\zeta - \eta)\sin\xi$$
$$+ \sqrt{R_n}\cos(N\zeta - \eta)\cos\xi + \sqrt{R_n}\sin(N\zeta - \eta)\sin\xi.$$

Equating the coefficients at $\cos(N\zeta - \eta)$ and $\sin(N\zeta - \eta)$, we obtain that

$$(2.10) \qquad z\cos 2\eta = (\sqrt{R_{n+1}} + \sqrt{R_n})\cos\xi,$$
$$z\sin 2\eta = (\sqrt{R_{n+1}} - \sqrt{R_n})\sin\xi.$$

These formulae can be checked directly from (1.14), (2.7) and (2.9). Differentiating (2.6) and (2.8) in $z$, we get that

$$\psi_n' = -Cz\sin(N\zeta + \eta)\, N\sqrt{\zeta_z} + O(1),$$
$$\psi_{n+1}' = -Cz\sin(N\zeta + \xi - \eta)\, N\sqrt{\zeta_z} + O(1);$$

hence by (2.5), modulo terms of the order of $N^{-1}$,

(2.11)
$$p_N = \sqrt{R_{N+1}}\, C^2 z^2 \left[-\sin(N\zeta + \xi - \eta)\cos(N\zeta + \eta)\right.$$
$$\left. + \cos(N\zeta + \xi - \eta)\sin(N\zeta + \eta)\right]$$
$$= \sqrt{R_{N+1}}\, C^2 z^2 \sin(2\eta - \xi) = \sqrt{R_{N+1}}\, C^2 z^2 (\sin 2\eta \cos \xi - \cos 2\eta \sin \xi),$$

and by (2.10),

$$p_N = \sqrt{R_{N+1}}\, C^2 z \left[(\sqrt{R_{N+1}} - \sqrt{R_N})\sin \xi \cos \xi\right.$$
$$\left. - (\sqrt{R_{N+1}} + \sqrt{R_N})\sin \xi \cos \xi\right]$$
$$= -\sqrt{R_{N+1}R_N}\, C^2 z \sin 2\xi.$$

Since modulo terms of the order of $N^{-1}$,

$$R_{N+1}R_N = \frac{1}{g}; \qquad C^2 = \frac{g}{\pi}; \qquad \sin 2\xi = \sin(-\arccos q) = -\sqrt{1 - q^2},$$

we obtain that

$$p_N = \frac{\sqrt{g}}{\pi} z \sqrt{1 - q^2} + O(N^{-1}).$$

Substitution of the value of $q$ gives that

$$p_N(z) = p(z) + O(N^{-1}),$$

where

(2.12)
$$p(z) = \frac{1}{\pi}|U_0(z;1)|^{1/2} = \frac{|z|}{\pi}\left[g - \left(\frac{gz^2 + t}{2}\right)^2\right]^{1/2} = \frac{g|z|}{2\pi}\sqrt{(z^2 - z_1^2)(z_2^2 - z^2)}$$

and

(2.13)
$$z_{1,2} = \left(\frac{-t \mp 2\sqrt{g}}{g}\right)^{1/2}.$$

This gives an explicit formula for the limiting density $p = p(z)$ of eigenvalues (integrated density of states). In a completely different approach, based on the Coulomb gas representation of the matrix model, this formula is derived in the work [BPS] of Boutet de Monvel, Pastur, and Shcherbina, as an application of the proven (in [BPS]) variational principle for the integrated density of states.

The scaling limit of the correlation function $K_{Nm}(z_1, \ldots, z_m)$ at a regular point $z$, where $p(z) > 0$, is defined as

$$K_m(u_1, \ldots, u_m) = \lim_{N \to \infty}[Np(z)]^{-m} K_{Nm}\left(z + \frac{u_1}{Np(z)}, \ldots, z + \frac{u_m}{Np(z)}\right).$$

Observe that $K_m(u_1, \ldots, u_m)$ is the limiting $m$-point correlation function of the rescaled eigenvalues

$$\mu_j = Np(z)(\lambda_j - z).$$

The rescaling reduces the mean value of the spacing $\mu_{j+1} - \mu_j$ to 1. From Dyson's formula (2.2),

$$(2.14) \qquad K_m(u_1, \ldots, u_m) = \det\big(Q(u_j, u_k)\big)_{j,k=1,\ldots,m},$$

where

$$(2.15) \qquad Q(u, v) = \lim_{N \to \infty} \big[Np(z)\big]^{-1} Q_N\left(z + \frac{u}{Np(z)}, z + \frac{v}{Np(z)}\right).$$

By (2.4),

$$(2.16) \quad \big[Np(z)\big]^{-1} Q_N\left(z + \frac{u}{Np(z)}, z + \frac{v}{Np(z)}\right)$$
$$= \frac{\sqrt{R_{N+1}}}{u - v} T_N\left(z + \frac{u}{Np(z)}, z + \frac{v}{Np(z)}\right),$$

where

$$T_N(z, w) = \psi_{N+1}(z)\psi_N(w) - \psi_N(z)\psi_{N+1}(w).$$

By (2.6) and (2.8), modulo terms of the order of $N^{-1}$,

$$(2.17)$$
$$\psi_N\left(z + \frac{u}{Np(z)}\right) = \frac{Cz}{\sqrt{\zeta_z}} \cos(N\zeta + \alpha + \eta), \qquad \alpha = \frac{\zeta_z u}{Np(z)},$$
$$\psi_{N+1}\left(z + \frac{u}{Np(z)}\right) = \frac{Cz}{\sqrt{\zeta_z}} \cos(N\zeta + \alpha + \xi - \eta);$$

hence

$$(2.18)$$
$$T_N\left(z + \frac{u}{Np(z)}, z + \frac{v}{Np(z)}\right)$$
$$= \frac{C^2 z^2}{\zeta_z} \left[\cos(N\zeta + \alpha + \xi - \eta)\cos(N\zeta + \beta + \eta)\right.$$
$$\left. - \cos(N\zeta + \alpha + \eta)\cos(N\zeta + \beta + \xi - \eta)\right]$$
$$= \frac{C^2 z^2}{2\zeta_z} \left[\cos(\alpha + \xi - \beta - 2\eta) - \cos(\alpha - \xi - \beta + 2\eta)\right]$$
$$= \frac{C^2 z^2}{\zeta_z} \sin(2\eta - \xi)\sin(\alpha - \beta),$$

where

$$(2.19) \qquad \alpha = \frac{\zeta_z u}{p(z)} = \frac{|U_0(z)|^{1/2} u}{|U_0(z)|^{1/2}\pi^{-1}} = \pi u, \qquad \beta = \pi v.$$

By (2.11) and (2.12),

$$\sqrt{R_{N+1}}\, C^2 z^2 \sin(2\eta - \xi) = p(z) = \frac{1}{\pi}\sqrt{|U_0(z;1)|} = \zeta_z(z;1);$$

hence (2.18) implies that

$$\sqrt{R_{N+1}}\, T_N\left(z + \frac{u}{Np(z)}, z + \frac{v}{Np(z)}\right) = \sqrt{R_{N+1}}\,\frac{C^2 z^2}{\zeta_z}\,\sin(2\eta - \xi)\sin(\alpha - \beta)$$

$$= \frac{\sin(\alpha - \beta)}{\pi} = \frac{\sin\pi(u-v)}{\pi},$$

and, by (2.15), (2.16),

$$Q(u,v) = \frac{\sin\pi(u-v)}{\pi(u-v)}\,.$$

This proves the Dyson sine-kernel for the local distribution of eigenvalues at a regular point $z$. In a completely different approach, the sine-kernel at regular points is proved in [PS].

*Remark.* It follows from the Dyson sine-kernel, due to the Gaudin formula (see, e.g., [Meh1]), that the spacing distribution of eigenvalues is determined by the Fredholm determinant $\det(1 - Q(x,y))_{x,y\in J}$. The asymptotics of this determinant as $|J| \to \infty$ has been studied intensively since the classical works by des Cloizeaux, Dyson, Gaudin, Mehta, and Widom (see [Meh1] for the history of the subject). The Riemann-Hilbert approach to this asymptotics has been developed in the paper [DIZ].

At the endpoints of the spectrum we use the semiclassical asymptotics (1.18), and it leads to the Airy kernel (cf. the papers of Bowick and Brézin [BB], Forrester [For], Moore [Mo], and Tracy and Widom [TW2], where the Airy kernel is discussed for the Gaussian matrix model and some other related models, and, in addition, some nonrigorous arguments are given for general matrix models). Consider for the sake of definiteness $z = z_2$.

By (1.18),

$$(2.20) \qquad \psi_n = \frac{DN^{1/6}z}{\sqrt{w'}}\left[\mathrm{Ai}\left(N^{2/3}w\right) + O(N^{-1})\right], \qquad D = \sqrt{g},$$

where $w$ is defined as in (1.19). From (1.19),

$$\sqrt{w}\,\frac{\partial w}{\partial \lambda'} = \frac{\partial}{\partial \lambda'}\int_{z_2^{(N)}}^{z}\sqrt{U_N(v)}\,dv.$$

This allows us to derive from (2.20) that

(2.21)
$$\psi_n = \frac{DN^{1/6}z}{\sqrt{\varphi_0'}} \left[ \operatorname{Ai}\left(N^{2/3}\varphi_0 + N^{-1/3}\omega\right) + O(N^{-1})\right],$$

$$\psi_{n\pm 1} = \frac{DN^{1/6}z}{\sqrt{\varphi_0'}} \left[ \operatorname{Ai}\left(N^{2/3}\varphi_0 \pm N^{-1/3}\rho - N^{-1/3}\omega\right) + O(N^{-1})\right],$$

where

(2.22)
$$\varphi_0 = \varphi_0(z;\lambda') = \left(\frac{3}{2}\int_{z_2}^z \sqrt{U_0(v;\lambda')}\,dv\right)^{2/3},$$

$$\rho = \rho(z;\lambda') = \frac{\xi(z;\lambda')}{\sqrt{\varphi_0(z;\lambda')}},\qquad \omega = \omega(z;\lambda') = \frac{\eta(z;\lambda')}{\sqrt{\varphi_0(z;\lambda')}},$$

and

(2.23)
$$\xi(z;\lambda') = -\frac{\cosh^{-1}q}{2},\qquad q = \frac{gz^2+t}{2\sqrt{\lambda'g}};$$

$$\eta(z;\lambda') = -\frac{(-1)^n}{4}\cosh^{-1}r,\qquad r = \frac{2\sqrt{\lambda'g}-tq}{2\sqrt{\lambda'g}q-t}.$$

The formulae (2.22), (2.23) define the functions $\varphi_0(z;\lambda')$, $\rho(z;\lambda')$ and $\omega(z;\lambda')$ for $z \geq z_2$. It is easy to check that these functions are analytic in $z$ at $z = z_2$, and they can be continued analytically to the interval $z > z_1$. In addition,

(2.24)
$$U_0(z_2;\lambda') = 0,\qquad \frac{\partial U_0}{\partial z}(z_2;\lambda') = \varkappa = 2(\lambda')^{1/2}g^{3/2}z_2^3;$$

$$\varphi_0(z_2;\lambda') = 0,\qquad \frac{\partial \varphi_0}{\partial z}(z_2;\lambda') = \varkappa^{1/3} = 2^{1/3}(\lambda')^{1/6}g^{1/2}z_2;$$

$$\rho(z_2) = -2^{-2/3}(\lambda')^{-1/3};$$

$$\omega(z_2) = -\frac{(-1)^n}{4}2^{1/3}(\lambda')^{-1/3}z_1 z_2^{-1}.$$

We will consider

(2.26)
$$z = z_2 + N^{-2/3}\alpha,\qquad w = z_2 + N^{-2/3}\beta,$$

where $\alpha$ and $\beta$ are fixed.

Substitution of (2.21) into the recursive equation

$$z\psi_n = \sqrt{R_{n+1}}\psi_{n+1} + \sqrt{R_n}\psi_{n-1}$$

gives the equations

(2.27)
$$z_2 = \sqrt{R_{n+1}} + \sqrt{R_n},$$

$$z_2\omega = \sqrt{R_{n+1}}(\rho - \omega) + \sqrt{R_n}(-\rho - \omega),$$

from whence

(2.28) $$(\sqrt{R_{n+1}} + \sqrt{R_n})\, 2\omega = (\sqrt{R_{n+1}} - \sqrt{R_n})\, \rho.$$

Similarly,

$$z_1 = (-1)^n (\sqrt{R_{n+1}} - \sqrt{R_n});$$

hence

(2.29) $$2\omega = \frac{(-1)^n z_1 \rho}{z_2},$$

which agrees with (2.24).

Substituting the formulae (2.21) into (2.4) and throwing away terms of the lower order, we obtain that

(2.30)
$$Q_N(z,w) = \frac{\sqrt{R_{n+1}}\, D^2 N^{1/3} z_2^2}{(z-w)\varphi_0'}$$
$$\times \Big[\, \mathrm{Ai}\left(N^{2/3}\varphi_0(z) + N^{-1/3}\rho - N^{-1/3}\omega\right) \mathrm{Ai}\left(N^{2/3}\varphi_0(w) + N^{-1/3}\omega\right)$$
$$-\; \mathrm{Ai}\left(N^{2/3}\varphi_0(z) + N^{-1/3}\omega\right) \mathrm{Ai}\left(N^{2/3}\varphi_0(w) + N^{-1/3}\rho - N^{-1/3}\omega\right) \Big],$$

where $\varphi_0'$, $\rho$ and $\omega$ are taken at $z_2$. Taking the linear part of Ai we obtain that

$$Q_N(z,w) = \frac{\sqrt{R_{n+1}}\, D^2 N^{1/3} z_2^2}{(z-w)\varphi_0'} \left[\, \mathrm{Ai}\,(u)\, \mathrm{Ai}\,'(v) - \mathrm{Ai}\,'(u)\, \mathrm{Ai}\,(v) \right] (2\omega - \rho) N^{-1/3},$$

where

$$u = \varphi_0'\,\alpha, \qquad v = \varphi_0'\,\beta.$$

By (2.28) and (2.24), modulo terms of the order of $N^{-1/3}$,

$$\sqrt{R_{n+1}}\,(2\omega - \rho) = \sqrt{R_{n+1}}\,\frac{(-2\sqrt{R_n})}{\sqrt{R_{n+1}} + \sqrt{R_n}}\,(-2^{-2/3}) = 2^{1/3} g^{-1/2} z_2^{-1};$$

hence

$$Q_N(z,w) = N^{2/3} 2^{1/3} g^{1/2} z_2 \,\frac{\mathrm{Ai}\,(u)\, \mathrm{Ai}\,'(v) - \mathrm{Ai}\,'(u)\, \mathrm{Ai}\,(v)}{u - v} + O(N^{1/3}).$$

Thus,

$$\lim_{N\to\infty} \frac{1}{cN^{2/3}} Q_N\left(z_2 + \frac{u}{cN^{2/3}}, z_2 + \frac{v}{cN^{2/3}}\right) = \frac{\mathrm{Ai}\,(u)\, \mathrm{Ai}\,'(v) - \mathrm{Ai}\,'(u)\, \mathrm{Ai}\,(v)}{u - v},$$

where

$$c = \varphi_0'(z_2; 1) = 2^{1/3} g^{1/2} z_2.$$

This proves the Airy kernel at the endpoint $z_2$. The endpoint $z_1$ is treated similarly.

### 3. The Lax pair for the Freud equation

Let

$$(3.1) \qquad \psi_n(z) = \frac{1}{\sqrt{h_n}} \, P_n(z) e^{-NV(z)/2} \, .$$

Then

$$(3.2) \qquad \int_{-\infty}^{\infty} \psi_n(z) \psi_m(z) \, dz = \delta_{nm}.$$

A recursive equation for $\psi_n(z)$ follows from (1.4):

$$(3.3) \qquad z\psi_n(z) = R_{n+1}^{1/2} \psi_{n+1}(z) + R_n^{1/2} \psi_{n-1}(z).$$

In addition,

$$(3.4)$$

$$\begin{aligned}
\psi_n'(z) = &- \left( N \frac{g}{2} \, R_{n+1}^{1/2} R_{n+2}^{1/2} R_{n+3}^{1/2} \right) \psi_{n+3}(z) \\
&- \left[ N \frac{t}{2} \, R_{n+1}^{1/2} + N \frac{g}{2} \, R_{n+1}^{1/2} (R_n + R_{n+1} + R_{n+2}) \right] \psi_{n+1}(z) \\
&+ \left[ N \frac{t}{2} \, R_n^{1/2} + N \frac{g}{2} \, R_n^{1/2} (R_{n-1} + R_n + R_{n+1}) \right] \psi_{n-1}(z) \\
&+ \left( N \frac{g}{2} \, R_{n-2}^{1/2} R_{n-1}^{1/2} R_n^{1/2} \right) \psi_{n-3}(z).
\end{aligned}$$

Let

$$(3.5) \qquad \vec{\Psi}_n(z) = \begin{pmatrix} \psi_n(z) \\ \psi_{n-1} \end{pmatrix} \, .$$

Then combining (3.3) with (3.4), one can obtain (cf. (3.1–7) in [FIK2]) that

$$(3.6) \qquad \begin{cases} \vec{\Psi}_{n+1}(z) = U_n(z)\vec{\Psi}_n(z), \\ \vec{\Psi}_n'(z) = N A_n(z)\vec{\Psi}_n(z), \end{cases}$$

where

$$(3.7) \qquad U_n(z) = \begin{pmatrix} R_{n+1}^{-1/2} z & -R_{n+1}^{-1/2} R_n^{1/2} \\ 1 & 0 \end{pmatrix} \, ,$$

and

$$(3.8)$$

$$A_n(z) = \begin{pmatrix} -(\frac{tz}{2} + \frac{gz^3}{2} + gzR_n) & R_n^{1/2}[t + gz^2 + g(R_n + R_{n+1})] \\ -R_n^{1/2}[t + gz^2 + g(R_{n-1} + R_n)] & \frac{tz}{2} + \frac{gz^3}{2} + gzR_n \end{pmatrix} \, .$$

Observe that

$$\operatorname{tr} A_n(z) = 0$$

and

(3.9)

$$\det A_n(z) = -\left(\frac{tz}{2} + \frac{gz^3}{2}\right)^2 + gR_n\left(t + gR_{n-1} + gR_n + gR_{n+1}\right)z^2 + R_n\theta_{n-1}\theta_n,$$

where

(3.10) $$\theta_n = t + gR_n + gR_{n+1}.$$

Due to (1.7), we can rewrite $\det A_n(z)$ as

(3.9′) $$\det A_n(z) = -\left(\frac{tz}{2} + \frac{gz^3}{2}\right)^2 + \frac{gnz^2}{N} + R_n\theta_{n-1}\theta_n.$$

The compatibility condition of equations (3.6) is

(3.11) $$U'_n(z) = NA_{n+1}(z)U_n(z) - NU_n(z)A_n(z).$$

Restricting this equation to the matrix element $U'_{n,11}(z)$ we obtain that

(3.12) $$J_{n+1} - J_n = 1,$$

where

$$J_n = NR_n[t + g(R_{n-1} + R_n + R_{n+1})].$$

Hence $J_n = n + \text{const}$. Since $J_0 = 0$, in fact, $J_n = n$. This means, that the compatibility condition (3.10), together with the initial value $J_0 = 0$ imply (1.7), and thus equations (3.6) give the Lax pair for the nonlinear difference Freud equation (1.7). In addition, equation (3.12) gives the recursive equation

(3.13) $$R_{n+1}\theta_n\theta_{n+1} = R_n\theta_{n-1}\theta_n + \frac{\theta_n}{N}.$$

Equations (3.6) have two linear independent solutions and $\vec{\Psi}_n(z)$ is one of them. We will consider another solution,

$$\vec{\Phi}_n(z) = \begin{pmatrix} \varphi_n(z) \\ \varphi_{n-1}(z) \end{pmatrix},$$

and the $2 \times 2$ matrix

(3.14) $$\Psi_n(z) = \begin{pmatrix} \psi_n(z) & \varphi_n(z) \\ \psi_{n-1}(z) & \varphi_{n-1}(z) \end{pmatrix}$$

which satisfies the same equations,

(3.15) $$\begin{cases} \Psi_{n+1}(z) = U_n(z)\Psi_n(z), \\ \Psi'_n(z) = NA_n(z)\Psi_n(z). \end{cases}$$

To define $\vec{\Phi}_n(z)$ we consider an arbitrary, linearly independent with $\vec{\Psi}_1(z)$, solution of the differential equation $\vec{\Phi}'_1(z) = NA_1(z)\vec{\Phi}_1(z)$, and then define $\vec{\Phi}_n(z)$, $n \geq 2$, with the help of the recursive equation $\vec{\Phi}_{n+1}(z) = U_n(z)\vec{\Phi}_n(z)$. Equation (3.11) then leads to the differential equation $\vec{\Phi}'_n(z) = NA_n\vec{\Phi}_n(z)$ for $n \geq 2$. Note that the equation $\vec{\Phi}_{n+1}(z) = U_n(z)\vec{\Phi}_n(z)$ means that $\varphi_n(z)$ satisfies the recursive equation (3.3); i.e.,

$$(3.16) \qquad z\varphi_n(z) = R_{n+1}^{1/2}\varphi_{n+1}(z) + R_n^{1/2}\varphi_{n-1}(z).$$

Since $\operatorname{tr} A_n(z) = 0$, the second equation in (3.15) implies that

$$(3.17) \qquad \det \Psi_n(z) = C \neq 0$$

is independent of $z$; i.e.,

$$(3.18) \qquad \psi_n(z)\varphi_{n-1}(z) - \psi_{n-1}(z)\varphi_n(z) = C, \quad C = C(n).$$

This enables us to derive a first order differential equation for $\varphi_n(z)$. Namely, from the first row in $\Psi'_n(z) = NA_n\Psi_n(z)$ we have that

$$(3.19) \qquad \begin{aligned} \varphi'_n(z) &= Na_{11}(z)\varphi_n(z) + Na_{12}(z)\varphi_{n-1}(z), \\ \psi'_n(z) &= Na_{11}(z)\psi_n(z) + Na_{12}(z)\psi_{n-1}(z), \end{aligned}$$

where $A_n(z) = \big(a_{ij}(z)\big)_{i,j=1,2}$; hence

$$\begin{aligned} \psi_n(z)\varphi'_n(z) - \varphi_n(z)\psi'_n(z) &= Na_{12}(z)[\psi_n(z)\varphi_{n-1}(z) - \psi_{n-1}(z)\varphi_n(z)] \\ &= CNa_{12}(z), \end{aligned}$$

and

$$(3.20) \qquad \left(\frac{\varphi_n(z)}{\psi_n(z)}\right)' = \frac{CNa_{12}(z)}{\psi_n^2(z)}.$$

In a similar way from the second row in (3.15), $\Psi'_n(z) = NA_n(z)\Psi_n(z)$, we obtain that

$$(3.21) \qquad \left(\frac{\varphi_{n-1}(z)}{\psi_{n-1}(z)}\right)' = -\frac{CNa_{21}(z)}{\psi_{n-1}^2(z)}, \qquad n \geq 1.$$

It is useful to note that (3.18) allows us to express $\varphi_{n-1}(z)$ in terms of $\varphi_n(z)$:

$$(3.22) \qquad \varphi_{n-1}(z) = \frac{\psi_{n-1}(z)}{\psi_n(z)}\,\varphi_n(z) + \frac{C}{\psi_n(z)}.$$

*Remark.* The system of two differential equations of the first order, $\vec{\Psi}'_n = NA_n\vec{\Psi}_n$, can be reduced to one equation of the second order (cf. [Sho]). Namely, from the first equation of the system we can express $\psi_{n-1}$ in terms of $\psi_n$,

$$(3.23) \qquad \psi_{n-1} = N^{-1}\frac{1}{a_{12}}\,\psi'_n - \frac{a_{11}}{a_{12}}\,\psi_n,$$

and then we can substitute this expression into the second equation of the system, which gives

$$(3.24) \qquad \psi_n'' - \frac{a_{12}'}{a_{12}} \psi_n' + N^2(a_{11}a_{22} - a_{12}a_{21})\psi_n - Na_{12}\left(\frac{a_{11}}{a_{12}}\right)'\psi_n = 0.$$

With the help of the substitution

$$(3.25) \qquad \psi_n = a_{12}^{1/2}\zeta_n$$

we reduce (3.24) to the Schrödinger equation

$$(3.26) \qquad -\zeta_n'' + N^2\hat{U}\zeta_n = 0,$$

where
$$(3.27)$$
$$\hat{U} = -(a_{11}a_{22} - a_{12}a_{21}) + N^{-1}\left(a_{11}' - a_{11}\frac{a_{12}'}{a_{12}}\right) - N^{-2}\left[\frac{a_{12}''}{2a_{12}} - \frac{3(a_{12}')^2}{4a_{12}^2}\right].$$

By (3.8),

$$(3.28) \qquad -a_{11} = a_{22} = \frac{tz}{2} + \frac{gz^3}{2} + gzR_n,$$
$$a_{12} = R_n^{1/2}(\theta_n + gz^2), \qquad a_{21} = -R_n^{1/2}(\theta_{n-1} + gz^2),$$

which gives

$$(3.29) \qquad \hat{U}(z) = \left[\frac{g^2z^6}{4} + \frac{tgz^4}{2} + \left(\frac{t^2}{4} - \frac{n}{N}g\right)z^2 - R_n\theta_{n-1}\theta_n\right]$$
$$- N^{-1}\left[\frac{t}{2} + \frac{3gz^2}{2} + gR_n - \frac{gz^2(t + gz^2 + 2gR_n)}{gz^2 + \theta_n}\right]$$
$$+ N^{-2}\left[\frac{g(2gz^2 - \theta_n)}{(gz^2 + \theta_n)^2}\right].$$

It is convenient to write $\hat{U}(z)$ as

$$(3.30) \qquad \hat{U}(z) = U_0(z) + U_1(z) + U_2(z),$$

where
$$(3.31)$$
$$U_0(z) = z^2\left[\left(\frac{gz^2 + t}{2}\right)^2 - \lambda'g\right], \qquad \lambda' = \frac{n + \frac{1}{2}}{N},$$
$$U_1(z) = N^{-1}\left(\frac{t}{2} + gR_n\right),$$
$$U_2(z) = -R_n\theta_{n-1}\theta_n - N^{-1}\left[\frac{\theta_n(t + gz^2 + 2gR_n)}{gz^2 + \theta_n}\right] + N^{-2}\left[\frac{g(2gz^2 - \theta_n)}{(gz^2 + \theta_n)^2}\right].$$

## 4. The Stokes phenomenon

Consider the sectors

(4.1)
$$\Omega_j = \left\{ z \colon \frac{\pi}{8} + \frac{\pi(j-1)}{2} + \varepsilon < \arg z < \frac{3\pi}{8} + \frac{\pi(j-1)}{2} - \varepsilon \right\}, \qquad j = 1, 2, 3, 4,$$

$\varepsilon > 0$, on a complex plane where the function

(4.2)
$$\psi_n(z) = \frac{1}{\sqrt{h_n}} \, P_n(z) \, e^{-N\left(\frac{t}{4}z^2 + \frac{g}{8}z^4\right)}$$

goes to infinity as $z \to \infty$. Let $b_j$ be the bisector of $\Omega_j$,

(4.3)
$$b_j = \{ z \colon z = r\omega_j, \ r > 0 \}, \qquad \omega_j = e^{i\left(\frac{\pi}{4} + \frac{\pi(j-1)}{2}\right)}, \quad j = 1, 2, 3, 4.$$

For a fixed $n \geq 1$, let $C_n > 0$ be a constant which will be chosen later, and let us consider the following special solution to equation (3.20) with $C = C_n$:

(4.4)
$$\varphi_{nj}(z) = C_n N \psi_n(z) \int_{\omega_j \infty}^{z} \frac{a_{12}(u)}{\psi_n^2(u)} \, du = C_n N h_n^{1/2} P_n(z) \, e^{-N\left(\frac{t}{4}z^2 + \frac{g}{8}z^4\right)}$$

$$\times \int_{\omega_j \infty}^{z} \frac{R_n^{1/2}[t + gu^2 + g(R_n + R_{n+1})]}{P_n^2(u)} \, e^{N\left(\frac{t}{2}u^2 + \frac{g}{4}u^4\right)} du.$$

Then $\varphi_{nj}(z) \to 0$ as $z \to \infty$ in $\Omega_j$. Evaluating the integral in (4.4) with the help of the Laplace method, we obtain the asymptotic expansion of $\varphi_{nj}(z)$ in $\Omega_j$:

(4.5)
$$\varphi_{nj}(z) \sim \left( C_n h_n^{1/2} R_n^{1/2} \right) z^{-n-1} e^{\frac{NV(z)}{2}} \left( 1 + \sum_{k=1}^{\infty} \frac{\gamma_{2k}}{z^{2k}} \right), \qquad \gamma_{2k} = \gamma_{2k}(n).$$

Let us define the function $\varphi_{n-1,j}(z)$ by the formula (3.22),

(4.6)
$$\varphi_{n-1,j}(z) = \frac{\psi_{n-1}(z)}{\psi_n(z)} \, \varphi_{nj}(z) + \frac{C_n}{\psi_n(z)}.$$

Then the vector-function

$$\vec{\Phi}_{nj}(z) = \begin{pmatrix} \varphi_{nj}(z) \\ \varphi_{n-1,j}(z) \end{pmatrix}$$

satisfies the differential equation

$$\vec{\Phi}'_{nj}(z) = N A_n(z) \vec{\Phi}_{nj}(z).$$

Hence the function $\varphi_{n-1,j}(z)$ satisfies the first-order differential equation (3.21) with $C = C_n$. In addition, (4.6) gives that $\varphi_{n-1,j}(z) \to 0$ as $z \to \infty$ in $\Omega_j$.

This implies that

$$
\begin{aligned}
(4.7) \quad \varphi_{n-1,j}(z) =& -C_n N \psi_{n-1}(z) \int_{\omega_j \infty}^{z} \frac{a_{21}(u)}{\psi_{n-1}^2(u)} \, du \\
=& \, C_n N h_{n-1}^{1/2} P_{n-1}(z) \, e^{-N\left(\frac{t}{4}z^2 + \frac{g}{8}z^4\right)} \\
& \times \int_{\omega_j \infty}^{z} \frac{R_n^{1/2}[t + gu^2 + g(R_{n-1}+R_n)]}{P_{n-1}^2(u)} \, e^{N\left(\frac{t}{2}u^2 + \frac{g}{4}u^4\right)} du \, ,
\end{aligned}
$$

which for $n \geq 2$ can be rewritten as

$$
\begin{aligned}
\varphi_{n-1,j}(z) =& \left(C_n R_n^{1/2} R_{n-1}^{-1/2}\right) N h_{n-1}^{1/2} P_{n-1}(z) \, e^{-N\left(\frac{t}{4}z^2 + \frac{g}{8}z^4\right)} \\
& \times \int_{\omega_j \infty}^{z} \frac{R_{n-1}^{1/2}[t + gu^2 + g(R_{n-1}+R_n)]}{P_{n-1}^2(u)} \, e^{N\left(\frac{t}{2}u^2 + \frac{g}{4}u^4\right)} du \, .
\end{aligned}
$$

The last expression is similar to (4.4) if we take

$$
C_{n-1} = C_n R_n^{1/2} R_{n-1}^{-1/2}, \quad n \geq 2,
$$

or

$$
C_{n-1} R_{n-1}^{1/2} = C_n R_n^{1/2}, \quad n \geq 2.
$$

We solve the last equation recursively. The initial constant $C_1$ is a free parameter. We put $C_1 = R_1^{-1/2}$. This gives $C_n = R_n^{-1/2}$, so that from (4.4) and (4.7) we derive

$$
\begin{aligned}
(4.8) \quad \varphi_{nj}(z) =& \, N h_n^{1/2} P_n(z) \, e^{-N\left(\frac{t}{4}z^2 + \frac{g}{8}z^4\right)} \\
& \times \int_{\omega_j \infty}^{z} \frac{t + gu^2 + g(R_n + R_{n+1})}{P_n^2(u)} \, e^{N\left(\frac{t}{2}u^2 + \frac{g}{4}u^4\right)} du \, , \quad n \geq 0,
\end{aligned}
$$

and

$$
(4.9) \qquad \varphi_{nj}(z) \sim h_n^{1/2} z^{-n-1} e^{\frac{NV(z)}{2}} \left(1 + \sum_{k=1}^{\infty} \frac{\gamma_{2k}}{z^{2k}}\right).
$$

Observe that $\vec{\Phi}_{nj}(z)$ satisfies the recursive equation

$$
\vec{\Phi}_{n+1,j}(z) = U_n(z) \vec{\Phi}_{nj}(z).
$$

Indeed, consider the vector-function

$$
U_n(z) \vec{\Phi}_{nj}(z) \equiv \vec{\Phi}_{n+1,j}^{0}(z) = \begin{pmatrix} \varphi_{n+1,j}^{0}(z) \\ \varphi_{nj}(z) \end{pmatrix}.
$$

Then, since $\Phi_{nj}(z)$ satisfies the differential equation

$$
\vec{\Phi}_{nj}'(z) = N A_n(z) \vec{\Phi}_{nj}(z),
$$

the equation (3.11) implies that $\vec{\Phi}^0_{n+1,j}(z)$ satisfies the differential equation

$$\left(\vec{\Phi}^0_{n+1,j}(z)\right)' = NA_{n+1}(z)\vec{\Phi}^0_{n+1,j}(z),$$

the same as for $\vec{\Phi}_{n+1,j}(z)$. In addition, $\vec{\Phi}^0_{n+1,j}(z) \to 0$ as $z \to \infty$ in $\Omega_j$. Hence $\vec{\Phi}^0_{n+1,j}(z)$ coincides with $\vec{\Phi}_{n+1,j}(z)$ up to a constant factor, but their second components are the same; hence $\vec{\Phi}^0_{n+1,j}(z)$ and $\vec{\Phi}_{n+1,j}(z)$ coincide, which was stated.

It is also worth noticing that $\vec{\Phi}_{nj}(z)$ as a solution of linear ODE with polynomial coefficients is an entire function of $z$.

As a matter of fact, the Laplace method combined with a deformation of the contour of integration gives the asymptotic expansion (4.9) in a bigger domain,

(4.10)
$$\Sigma_j = \left\{ -\frac{\pi}{8} + \frac{\pi(j-1)}{2} + \varepsilon < \arg z < \frac{5\pi}{8} + \frac{\pi(j-1)}{2} - \varepsilon \right\}, \qquad \varepsilon > 0.$$

Thus, the matrix-valued function

(4.11)
$$\Psi_{nj}(z) = \begin{pmatrix} \psi_n(z) & \varphi_{nj}(z) \\ \psi_{n-1}(z) & \varphi_{n-1,j}(z) \end{pmatrix} = \left(\vec{\Psi}_n(z), \vec{\Phi}_{nj}(z)\right)$$

is an entire function of $z$, and according to (4.2) and (4.9) it has the asymptotic expansion

(4.12)
$$\Psi_{nj}(z) \sim \left(\sum_{k=0}^{\infty} \frac{\Gamma_k}{z^k}\right) e^{-\left(\frac{NV(z)}{2} - n\ln z + \lambda_n\right)\sigma_3}, \qquad z \to \infty, \qquad z \in \Sigma_j,$$

where

(4.13)
$$\sigma_3 = \begin{pmatrix} 1 & 0 \\ 0 & -1 \end{pmatrix}$$

is the Pauli matrix,

(4.14)
$$\lambda_n = \frac{1}{2}\ln h_n,$$

and $\Gamma_k$ are some $2 \times 2$ matrices which only depend on $n$. From (4.2) and (4.9) we obtain that

(4.15)
$$\Gamma_0 = \begin{pmatrix} 1 & 0 \\ 0 & R_n^{-1/2} \end{pmatrix}, \qquad \Gamma_1 = \begin{pmatrix} 0 & 1 \\ R_n^{1/2} & 0 \end{pmatrix}.$$

The independence on $j$ of the coefficients $\Gamma_k$ in the asymptotic series (4.12) constitutes the so-called *Stokes phenomenon*, and the functions $\Psi_{nj}(z)$ form the set of canonical solutions (see, e.g., [Sib]) for the equation $\Psi'(z) = NA_n(z)\Psi(z)$. The corresponding Stokes matrices (see equation (5.3) below) are evaluated in the next section.

## 5. The Riemann-Hilbert problem

Since both $\vec{\Phi}_{nj}(z)$ for different $j$ and $\vec{\Psi}_n(z)$ satisfy the same differential equation (3.6), they are linearly dependent,

$$\vec{\Phi}_{n,j+1}(z) = q\vec{\Phi}_{nj}(z) + s\vec{\Psi}_n(z),$$

where $j$ is defined mod 4. The domains $\Sigma_j$ and $\Sigma_{j+1}$ intersect and in the intersection the functions $\vec{\Phi}_{nj}(z)$ and $\vec{\Phi}_{n,j+1}(z)$ grow to infinity and have the same asymptotic expansion (see (4.12)). On the other hand $\vec{\Psi}_n(z)$ goes to zero in this intersection, hence $q = 1$, so that

$$(5.1) \qquad \vec{\Phi}_{n,j+1}(z) = \vec{\Phi}_{nj}(z) + s\vec{\Psi}_n(z).$$

Since both $\vec{\Phi}_{nj}(z)$ and $\vec{\Psi}_n(z)$ satisfy the same recursive equation (3.6), the coefficient $s$ does not depend on $n$, but in general it depends on $j$, $s = s_j$. The equation (5.1) implies that

$$(5.2) \qquad \varphi_{n,j+1}(z) = \varphi_{nj}(z) + s_j\psi_n(z).$$

We can rewrite (5.1) in matrix form as

$$(5.3) \qquad \Psi_{n,j+1}(z) = \Psi_{nj}(z)\,S_j,$$

where

$$(5.4) \qquad S_j = \begin{pmatrix} 1 & s_j \\ 0 & 1 \end{pmatrix}.$$

To determine $s_j$ consider (5.2) at $n = 0$. From (4.8) it follows that

$$(5.5) \qquad \varphi_{0j}(z) = Nh_0^{1/2} e^{-N\left(\frac{t}{4}z^2 + \frac{g}{8}z^4\right)} \int_{\omega_j\infty}^{z} (t + gu^2 + gR_1) e^{N\left(\frac{t}{2}u^2 + \frac{g}{4}u^4\right)} du.$$

Putting $z = 0$ in (5.2) we get

$$s_j = \frac{\varphi_{0,j+1}(0) - \varphi_{0j}(0)}{\psi_0(0)}.$$

Since

$$\psi_0(0) = h_0^{-1/2}$$

(see (4.2)) and

$$\varphi_{0j}(0) = Nh_0^{1/2} \int_{\omega_j\infty}^{0} (t + gu^2 + gR_1) e^{N\left(\frac{t}{2}u^2 + \frac{g}{4}u^4\right)} du$$

we obtain that

$$(5.6) \qquad s_j = Nh_0 \int_{\omega_{j+1}\infty}^{\omega_j\infty} (t + gu^2 + gR_1) e^{N\left(\frac{t}{2}u^2 + \frac{g}{4}u^4\right)} du, \qquad j = 1, 2, 3, 4.$$

The change of variable $u \to -u$ gives

(5.7) $$s_3 = -s_1, \qquad s_4 = -s_2.$$

Another way to compute $s_j$ is to use the Cauchy type integral.

The function

(5.8) $$y_{nj}(z) = e^{-\frac{NV(z)}{2}} \varphi_{nj}(z), \qquad j = 1, 2, 3, 4,$$

is an entire function of $z$ and in $\Sigma_j$ it has the asymptotics

(5.9) $$y_{nj}(z) \sim h_n^{1/2} z^{-n-1} \left( 1 + \sum_{k=1}^{\infty} \frac{\gamma_{2k}}{z^{2k}} \right).$$

In addition,

(5.10) $$y_{n,j+1}(z) = y_{nj}(z) + s_j e^{\frac{-NV(z)}{2}} \psi_n(z).$$

Observe that the domain $\Sigma_j$ contains the $j^{\text{th}}$ quadrant,

$$\Delta_j = \left\{ z : \frac{(j-1)\pi}{2} \leq \arg z \leq \frac{j\pi}{2} \right\};$$

hence (5.9) holds in $\Delta_j$. This allows us to solve (5.10) with the help of the Cauchy type integral. Namely,

(5.11)

$$y_n(z) = \frac{s_4}{2\pi i} \int_{-\infty}^{\infty} \frac{e^{-\frac{NV(u)}{2}} \psi_n(u)}{u - z} \, du + \frac{s_1}{2\pi i} \int_{-i\infty}^{i\infty} \frac{e^{-\frac{NV(u)}{2}} \psi_n(u)}{u - z} \, du$$

$$= \frac{s_4}{2\pi i} \int_{-\infty}^{\infty} \frac{h_n^{-1/2} P_n(u) e^{-NV(u)}}{u - z} \, du + \frac{s_1}{2\pi i} \int_{-i\infty}^{i\infty} \frac{h_n^{-1/2} P_n(u) e^{-NV(u)}}{u - z} \, du$$

where $y_n(z)$ is a piecewise analytic function which coincides with $y_{nj}(z)$ in the quadrant $\Delta_j$. Expanding

$$\frac{1}{u - z} = -\frac{1}{z} \sum_{k=0}^{\infty} \frac{u^k}{z^k},$$

we obtain from (5.11) the asymptotic expansion of $y_n(z)$ as $z \to \infty$:

(5.12) $$y_n(z) \sim \frac{1}{z} \sum_{k=0}^{\infty} \frac{s_4 a_{nk} + s_1 b_{nk}}{z^k},$$

where

(5.13) $$a_{nk} = -\frac{1}{2\pi i} \int_{-\infty}^{\infty} h_n^{-1/2} P_n(u) u^k e^{-NV(u)} \, du,$$

$$b_{nk} = -\frac{1}{2\pi i} \int_{-i\infty}^{i\infty} h_n^{-1/2} P_n(u) u^k e^{-NV(u)} \, du.$$

Since by (5.9), $y_n(z) = O(z^{-n-1})$,

$$s_4 a_{nk} + s_1 b_{nk} = 0, \qquad k = 0, 1, \ldots, n-1,$$

but $a_{nk} = 0$ for $k = 0, 1, \ldots, n-1$ in virtue of the orthogonality property (1.3); hence $s_1 b_{nk} = 0$ for these $k$'s. Let us take $n = 2$ and $k = 0$. In this case

$$i \int_{-i\infty}^{i\infty} P_2(u) e^{-NV(u)} \, du = -\int_{-\infty}^{\infty} P_2(iu) e^{-NV(iu)} \, du$$

is obviously positive; hence $b_{20} \neq 0$. This implies

(5.14) $$s_1 = s_3 = 0.$$

By (5.2) this means that

$$\varphi_{n2}(z) = \varphi_{n1}(z), \qquad \varphi_{n4}(z) = \varphi_{n3}(z).$$

Let us find $s_4$. From (5.9) we know that

$$y_0(z) = h_0^{1/2}(z^{-1} + \ldots);$$

hence by (5.12), $s_4 a_{00} = h_0^{1/2}$. In addition, by (5.13),

$$a_{00} = -\frac{1}{2\pi i} h_0^{1/2}.$$

This gives

$$s_4 = -2\pi i.$$

It is interesting to notice that the values of $s_1 = s_3 = 0$ and $s_2 = s_4 = -2\pi i$ can be obtained directly from formula (5.6) as well, although the calculations are more involved. Namely, the recurrent coefficient $R_1$ is expressed in terms of the parabolic cylinder function:

$$R_1 = -\left(\frac{2}{Ng}\right)^{1/2} \frac{d}{dz} \ln\left[e^{z^2/4} D_{-1/2}(z)\right]\Bigg|_{z=t\sqrt{\frac{N}{2g}}}.$$

Then, using two different integral representations of the function $D_{-1/2}(z)$ (see, e.g., [BE]),

$$D_{-1/2}(z) = \frac{1}{\sqrt{\pi}} e^{-z^2/4} \int_0^\infty e^{-zt-t^2/2} t^{-1/2} dt$$

$$= \sqrt{\frac{2}{\pi}} e^{z^2/4} \int_0^\infty e^{-t^2/2} \cos\left(zt + \frac{\pi}{4}\right) t^{-1/2} dt,$$

one derives from (5.6) that $s_1$ and $s_4$ are expressed in terms of the Wronskians,

$$s_1 = \sqrt{2}\,\pi \begin{vmatrix} D_{-1/2}(z) & D_{-1/2}(z) \\ D'_{-1/2}(z) & D'_{-1/2}(z) \end{vmatrix} = 0,$$

$$s_4 = -\sqrt{2}\,\pi i \begin{vmatrix} D_{-1/2}(z) & D_{-1/2}(-z) \\ D'_{-1/2}(z) & D'_{-1/2}(-z) \end{vmatrix} = -2\pi i,$$

as stated above.

Now we can formulate the Riemann-Hilbert problem. Define

(5.15)
$$\Psi_{n+}(z) = \begin{pmatrix} \psi_n(z) & \varphi_{n1}(z) \\ \psi_{n-1}(z) & \varphi_{n-1,1}(z) \end{pmatrix}, \qquad \Psi_{n-}(z) = \begin{pmatrix} \psi_n(z) & \varphi_{n3}(z) \\ \psi_{n-1}(z) & \varphi_{n-1,3}(z) \end{pmatrix}.$$

Let

$$\Psi_n(z) = \begin{cases} \Psi_{n+}(z), & \text{if } \operatorname{Im} z \geq 0, \\ \Psi_{n-}(z), & \text{if } \operatorname{Im} z \leq 0. \end{cases}$$

Then $\Psi_n(z)$ has the asymptotic expansion

(5.16)
$$\Psi_n(z) \sim \left( \sum_{k=0}^{\infty} \frac{\Gamma_k}{z^k} \right) e^{-\left( \frac{NV(z)}{2} - n \ln z + \lambda_n \right)\sigma_3}, \qquad z \to \infty,$$

where $\lambda_n = \frac{1}{2} \ln h_n$ and

(5.17)
$$\Gamma_0 = \begin{pmatrix} 1 & 0 \\ 0 & R_n^{-1/2} \end{pmatrix}, \qquad \Gamma_1 = \begin{pmatrix} 0 & 1 \\ R_n^{1/2} & 0 \end{pmatrix}.$$

On the real line

(5.18)
$$\Psi_{n+}(z) = \Psi_{n-}(z)S, \qquad \operatorname{Im} z = 0,$$

where

(5.19)
$$S = \begin{pmatrix} 1 & -2\pi i \\ 0 & 1 \end{pmatrix}.$$

In addition,

(5.20)
$$\det \Psi_n(z) = R_n^{-1/2} \neq 0.$$

It must be emphasized that in the setting of the Riemann-Hilbert problem (5.16)–(5.18) the real quantities $R_n$ and $\lambda_n$ *are not the given data.* They are evaluated via the solution $\Psi_n(z)$. Indeed, suppose that $\tilde{\Psi}_n(z)$ is another function satisfying (5.16)–(5.18) with perhaps another $\tilde{R}_n$ and another $\tilde{\lambda}_n$. Consider the matrix ratio:

$$X^*(z) = [e^{\lambda_n \sigma_3} \Gamma_0^{-1} \Psi_n(z)][e^{\tilde{\lambda}_n \sigma_3} \tilde{\Gamma}_0^{-1} \tilde{\Psi}_n(z)]^{-1}.$$

Since the jump matrix $S$ for both $\Psi_n(z)$ and $\tilde{\Psi}_n(z)$ is the same, function $X^*(z)$ has no jump across the real line. Therefore, it is an entire function equal to $I$ at $z = \infty$. Hence

$$X^*(z) \equiv I,$$

or

$$[e^{\lambda_n \sigma_3} \Gamma_0^{-1} \Psi_n(z)][e^{\tilde{\lambda}_n \sigma_3} \tilde{\Gamma}_0^{-1} \tilde{\Psi}_n(z)]^{-1} \equiv I.$$

Substituting into this identity asymptotic expansion (5.16) and equating the terms of order $z^{-1}$ we come up with the matrix equation,

$$e^{\lambda_n \sigma_3} \Gamma_0^{-1} \Gamma_1 e^{-\lambda_n \sigma_3} = e^{\tilde{\lambda}_n \sigma_3} \tilde{\Gamma}_0^{-1} \tilde{\Gamma}_1 e^{-\tilde{\lambda}_n \sigma_3},$$

or

$$\begin{pmatrix} 0 & e^{2\lambda_n} \\ R_n e^{-2\lambda_n} & 0 \end{pmatrix} = \begin{pmatrix} 0 & e^{2\tilde{\lambda}_n} \\ \tilde{R}_n e^{-2\tilde{\lambda}_n} & 0 \end{pmatrix};$$

hence

$$\lambda_n = \tilde{\lambda}_n \qquad \text{and} \qquad R_n = \tilde{R}_n.$$

The Riemann-Hilbert problem (5.16)–(5.18) for the orthogonal polynomials (1.1), (1.3) first appeared in [FIK2,3] as a particular case of the more general Riemann-Hilbert problem related to the system of orthogonal polynomials on the cross $\mathbb{R} \bigcup i\mathbb{R}$ with the same weight function . Also, in [FIK2,3] an alternative approach to the derivation of the basic equations (5.15), which can be easily extended on an arbitrary system of orthogonal polynomials (cf. [FIK4]), was suggested. We present the method of [FIK] in Appendix E.

## 6. Formal asymptotic expansion for $R_n$

To guess a behavior of $R_n$ let us evaluate $R_n$ for small $n$. We have $R_0 = 0$ and

$$R_1 = \frac{h_1}{h_0} = \frac{\int_0^\infty z^2 e^{-\frac{gN}{4}(z^2 - M^2)^2} dz}{\int_0^\infty e^{-\frac{gN}{4}(z^2 - M^2)^2} dz}, \qquad M = |t|^{1/2} g^{-1/2}.$$

Using the change of variable

$$u = \left( \frac{gN}{2} \right)^{1/2} (z^2 - M^2)$$

we obtain that

$$R_1 = M^2 \frac{\int_{-1/\alpha}^\infty (1 + \alpha u)^{1/2} e^{-\frac{u^2}{2}} du}{\int_{-1/\alpha}^\infty (1 + \alpha u)^{-1/2} e^{-\frac{u^2}{2}} du}, \qquad \alpha = \frac{(2g)^{1/2}}{|t| N^{1/2}}.$$

Expanding $(1 + \alpha u)^{\pm 1/2}$ into Taylor's series and evaluating the Gaussian integrals we derive that

$$R_1 = M^2 \left( 1 - \frac{\alpha^2}{2} - \frac{3\alpha^4}{4} \right) + O(N^{-3}) = \frac{|t|}{g} - \frac{1}{|t|N} - \frac{3g}{|t|^3 N^2} + O(N^{-3}).$$

Now from the Freud equation (1.7) we obtain subsequently that

$$R_2 = \frac{2}{|t|N} + \frac{4g}{|t|^3 N^2} + O(N^{-3}),$$

$$R_3 = \frac{|t|}{g} - \frac{3}{|t|N} + O(N^{-2}),$$

$$R_4 = \frac{4}{|t|N} + O(N^{-2}).$$

These formulae suggest that for a fixed $n$ as $N \to \infty$,

$$R_n = \begin{cases} \dfrac{|t|}{g} - \dfrac{n}{|t|N} + O(N^{-2}) & \text{if} \quad n = 2k+1, \\[2ex] \dfrac{n}{|t|N} + O(N^{-2}) & \text{if} \quad n = 2k. \end{cases}$$

More generally we expect that if $\dfrac{n}{N}$ is bounded and $N \to \infty$ then

$$(6.1) \qquad R_n = \begin{cases} R\left( \dfrac{n}{N} \right) + O(N^{-2}) & \text{if} \quad n = 2k+1, \\[2ex] L\left( \dfrac{n}{N} \right) + O(N^{-2}) & \text{if} \quad n = 2k, \end{cases}$$

where $R(\lambda)$ and $L(\lambda)$ are smooth functions. Substituting this ansatz into the Freud equation (1.7) and neglecting terms of the order of $O(N^{-1})$, we obtain the equations

$$(6.2) \qquad \lambda = R[t + g(2L + R)],$$
$$\lambda = L[t + g(2R + L)].$$

Equating the expressions on the right we obtain that

$$(R - L)\left[ t + g(R + L) \right] = 0.$$

We assume that $R \neq L$; hence

$$(6.3) \qquad t + g(R + L) = 0.$$

Combining this with (6.2) we obtain

$$(6.4) \qquad \lambda = gRL,$$

so that $R, L$ are solutions of the quadratic equation

$$(6.5) \qquad u^2 + \frac{t}{g} u + \frac{\lambda}{g} = 0,$$

which are

$$(6.6) \qquad R, L = \frac{-t \pm \sqrt{t^2 - 4\lambda g}}{2g}.$$

Actually, we can find a formal asymptotic expansion of $R_n$ in powers of $N^{-2}$. Put

$$R_n = \begin{cases} R(n/N) & \text{if} \quad n = 2k+1, \\ L(n/N) & \text{if} \quad n = 2k. \end{cases}$$

Then (1.7) is equivalent to

$$(6.7) \qquad \begin{aligned} \lambda &= R(t + gR + 2gL + N^{-2}g\Delta L), \\ \lambda &= L(t + gL + 2gR + N^{-2}g\Delta R), \end{aligned}$$

where

$$\Delta f(\lambda) = \frac{f(\lambda - \frac{1}{N}) - 2f(\lambda) + f(\lambda + \frac{1}{N})}{(1/N)^2}.$$

Let us substitute the expansions

$$L(\lambda) = L_0(\lambda) + N^{-2}L_1(\lambda) + \dots; \qquad R(\lambda) = R_0(\lambda) + N^{-2}R_1(\lambda) + \dots$$

into (6.7) and equate coefficients at powers of $N^{-2}$. Then the equations on $L_0(\lambda), R_0(\lambda)$ coincide with (6.2), hence $L_0(\lambda), R_0(\lambda)$ are given by (6.6). Equating coefficients at $N^{-2}$ we obtain the system of equations

$$(6.8) \qquad \begin{cases} (R_0 + L_0)R_1 + 2R_0L_1 = -R_0\Delta L_0, \\ 2L_0R_1 + (R_0 + L_0)L_1 = -L_0\Delta R_0. \end{cases}$$

Solving this system we find $R_1, L_1$ and so on.

## 7. Proof of the main theorem:
## The asymptotic Riemann-Hilbert problem

In this section we construct explicitly, using semiclassical formulae, a piecewise analytic function $\Psi^0(z)$ which is an approximate solution to the basic Riemann-Hilbert problem (5.16)–(5.18). Then we prove that $\Psi^0(z)$ is close to $\Psi_n(z)$ (the Deift–Zhou method, see [DZ]).

Define the numbers (cf. (6.6))

$$(7.1) \qquad R_n^0 = \frac{-t - (-1)^n(t^2 - 4\lambda g)^{1/2}}{2g}, \qquad \lambda = \frac{n}{N},$$

$$R_{n\pm1}^0 = \frac{-t + (-1)^n(t^2 - 4\lambda g)^{1/2}}{2g},$$

and consider an auxiliary matrix differential equation,

$$(7.2) \qquad \Psi'(z) = NA^0(z)\Psi(z),$$

where the matrix

$$A^0 = A^0(z) = \begin{pmatrix} a_{11}^0 & a_{12}^0 \\ a_{21}^0 & a_{22}^0 \end{pmatrix}$$

is defined as

(7.3)
$$a_{11}^0 = -\left(\frac{tz}{2} + \frac{gz^3}{2} + gzR_n^0\right) = \alpha_n z - \frac{gz^3}{2}, \qquad \alpha_n = \frac{(-1)^n (t^2 - 4\lambda g)^{1/2}}{2},$$
$$a_{12}^0 = (R_n^0)^{1/2} gz^2,$$
$$a_{21}^0 = -a_{12}^0, \qquad a_{22}^0 = -a_{11}^0.$$

The matrix $A^0(z)$ is obtained from the matrix $A_n(z)$ by the substitution of $R_j^0$ for $R_j$, $j = n-1$, $n$, $n+1$. The function $\Psi^0(z)$ is constructed as a semiclassical approximation to the differential equation (7.2).

A direct calculation gives that

$$-\det A^0(z) = \frac{z^2}{4}\left[(t + gz^2)^2 - 4\lambda g\right].$$

Define the $\mu$-function as

(7.4)
$$\mu(z) = \sqrt{-\det A^0(z)} = \frac{z}{2}\sqrt{(t + gz^2)^2 - 4\lambda g} = \frac{zg}{2}\sqrt{(z^2 - z_1^2)(z^2 - z_2^2)},$$

where we take the branch of the square root which is positive on the positive half-axis. The zeros of the $\mu$-function are the turning points for the equation (7.2). There are five turning points: a double turning point at $z = 0$ and four simple turning points at $\pm z_1$ and $\pm z_2$ where

$$z_{1,2} = \left(\frac{-t \mp 2\sqrt{\lambda g}}{g}\right)^{1/2}.$$

We assume that

$$\varepsilon < \lambda < \frac{t^2}{4g} - \varepsilon$$

(cf. (1.11)), which ensures that both $z_1$ and $z_2$ are real and separated from zero.

Let $\Omega$ be a neighborhood of the interval $[z_1, z_2]$. For the sake of definiteness we define $\Omega$ as an ellipse with foci at $z_1$ and $z_2$,

$$\Omega = \left\{ z = x + iy \colon \frac{(x - z_0)^2}{a^2} + \frac{y^2}{b^2} \le 1 \right\},$$
$$z_0 = \frac{z_1 + z_2}{2}, \quad a^2 = (z_0 - z_1)^2 + b^2.$$

We will assume that $b > 0$ is independent of $N$ and it is sufficiently small so that $\Omega$ does not contain the origin. Let

$$\Omega_1 = \Omega \cap \{\operatorname{Re} z \leq z_0\}, \qquad \Omega_2 = \Omega \cap \{\operatorname{Re} z \geq z_0\},$$
$$\Omega^u = \Omega \cap \{\operatorname{Im} z \geq 0\}, \qquad \Omega^d = \Omega \cap \{\operatorname{Im} z \leq 0\},$$

and assume that the boundaries $\partial\Omega$, $\partial\Omega_1$, and $\partial\Omega^u$ are positively oriented (see Fig.2). Denote

$$C_1 = \partial\Omega \cap \{\operatorname{Re} z \leq z_0\}, \quad C_2 = \partial\Omega \cap \{\operatorname{Re} z \geq z_0\}.$$

Let $v_{1,2} = z_0 \pm ib$ be two vertices of the ellipse $\Omega$ and

$$l_0 = [v_1, v_2], \quad l_0^u = [v_1, z_0], \quad l_0^d = [z_0, v_2].$$

We shall also take the following convention: given any set $\Lambda \subset \mathbb{C}$ we denote by $\Lambda^u$ and $\Lambda^d$ the closed upper and lower parts of set $\Lambda$ with respect to the real axis, so that

$$\Omega_{1,2}^u = \Omega_{1,2} \cap \{\operatorname{Im} z \geq 0\}, \quad \Omega_{1,2}^d = \Omega_{1,2} \cap \{\operatorname{Im} z \leq 0\},$$
$$C_{1,2}^u = C_{1,2} \cap \{\operatorname{Im} z \geq 0\}, \quad C_{1,2}^d = C_{1,2} \cap \{\operatorname{Im} z \leq 0\},$$

etc. (see Fig. 2).

Our plan is the following:

(1) Outside of the domain $\Omega \cup (-\Omega)$, where $-\Omega = \{z : -z \in \Omega\}$, we define $\Psi^0(z)$ by a WKB formula.

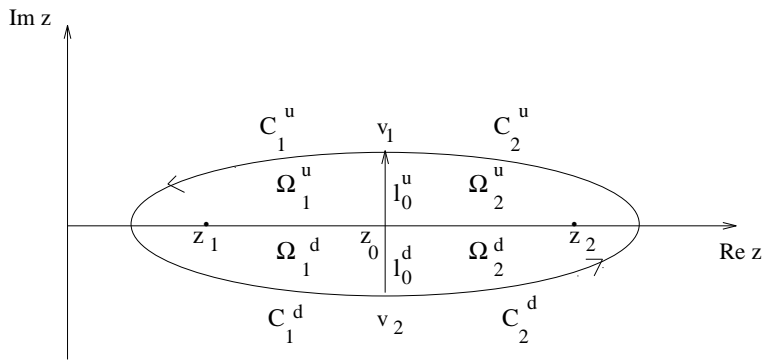

Figure 2. The set $\Omega$

(2) In the domains $\Omega_{1,2}^{u,d}$ we define $\Psi^0(z)$ by turning point formulae, with the help of the Airy function. In $-\Omega$ we define $\Psi^0(z) = (-1)^n \sigma_3 \Psi^0(-z) \sigma_3$.

(3) We check that the multiplicative jump matrix of $\Psi^0(z)$ between $\Omega^u$ and $\Omega^d$ coincides with the matrix $S$ in (5.19), and the multiplicative jump matrix on $\partial\Omega$ is close to the unit matrix, as well as the multiplicative jump matrix between $\Omega_1$ and $\Omega_2$.

This will give us an approximate explicit solution $\Psi^0(z)$ to the Riemann-Hilbert problem (5.16)–(5.18), and then we will establish the closeness of $\Psi^0(z)$ to $\Psi_n(z)$.

As the first step of our plan, we define $\Psi^0(z)$ outside of $\Omega$ by a WKB formula. To get this formula we diagonalize iteratively the equation (7.2). The diagonalization is described in detail by Kapaev [Kap]. We will use the WKB formula with the error term $O(N^{-1})$. In this approximation we get that

$$(7.5) \qquad \Psi^0(z) = \Psi_{\mathrm{WKB}}(z) \equiv C_0 T(z) e^{-N\xi(z)\sigma_3 - \tau(z) + C_1 \sigma_3},$$

where

$$(7.6) \qquad T(z) = \begin{pmatrix} 1 & \dfrac{a_{12}^0}{\mu - a_{11}^0} \\ \dfrac{(-a_{21}^0)}{\mu - a_{11}^0} & 1 \end{pmatrix}, \qquad \xi(z) = \int_{z_2}^z \mu(u)\, du,$$

$$\tau(z) = \int_{z_2}^z \operatorname{diag} T^{-1}(z) T'(z)\, dz,$$

and

$$(7.7) \qquad \mu(z) = \sqrt{-\det A^0(z)} = \left[ \left( a_{11}^0(z) \right)^2 + a_{12}^0(z) a_{21}^0(z) \right]^{1/2}.$$

In (7.6) $\operatorname{diag} M$ stands for a diagonal matrix which coincides with $M$ on the diagonal. The formula (7.5) gives an approximate solution of the differential equation (7.2), up to the terms of the order of $N^{-1}$, and this is a motivation for us to define $\Psi^0(z)$ outside of $\Omega$ by (7.5). However, it is important to underscore here that our ultimate goal is a construction of an approximate solution to the Riemann-Hilbert problem (5.16)–(5.18) and *we are not going to prove any results concerning the differential equation (7.2). It serves only as an auxiliary tool in our construction of $\Psi^0(z)$.*

Observe that

$$\mu(z) = z\sqrt{\lambda g}\sqrt{x^2 - 1}, \qquad x = \frac{t + gz^2}{2\sqrt{\lambda g}},$$

and the values $z = z_1$ and $z = z_2$ correspond to $x = -1$ and $x = 1$, respectively. We fix the branches of the multivalued functions in (7.6) by the inequalities

$$-\frac{\pi}{2} < \arg z < \frac{\pi}{2}, \quad |z| > z_2, \quad -\pi < \arg \sqrt{x^2 - 1} < \pi,$$

$$-\pi < \arg(x + \sqrt{x^2 - 1}) < \pi.$$

This defines $\mu(z)$ as an analytic function on $\mathbb{C}$, with two cuts $[-z_2, -z_1]$ and $[z_1, z_2]$. The function $\mu(z)$, in addition, satifies the symmetry equations,

$$\mu(-z) = -\mu(z), \qquad \mu(\bar{z}) = \overline{\mu(z)}.$$

The formula for $\mu(z)$ leads to the following expression for the function $\xi(z)$:
(7.7')
$$\xi(z) = \frac{\lambda}{2} \left[ x\sqrt{x^2 - 1} - \ln(x + \sqrt{x^2 - 1}) \right], \qquad x = \frac{t + gz^2}{2\sqrt{\lambda g}}, \qquad \lambda = \frac{n}{N}.$$

The function $\xi(z)$ is analytic and it has one cut $(-\infty, z_2]$.

The matrix $T(z)$ diagonalizes the matrix $A^0(z)$,

$$T^{-1}(z) A^0(z) T(z) = -\mu(z)\sigma_3,$$

so that $\mp\mu(z)$ are eigenvalues of the matrix $A^0(z)$ and $T(z)$ is the matrix of corresponding eigenvectors. The matrix $\operatorname{diag} T^{-1}(z) T'(z)$ turns out to be proportional to the unit matrix and a straightforward calculation gives that up to a nonessential constant factor,

(7.8)
$$e^{-\tau(z)} = \left( \frac{\mu - a_{11}^0}{2\mu} \right)^{1/2} I, \quad I = \begin{pmatrix} 1 & 0 \\ 0 & 1 \end{pmatrix}$$

(see Appendix A below). Thus (7.5) reduces to

(7.9)
$$\Psi^0(z) = \Psi_{\text{WKB}}(z) \equiv C_0 T_0(z) e^{-N\xi(z)\sigma_3 + C_1\sigma_3},$$

where $C_0 \neq 0$, $C_1$ are constants and

(7.10)
$$T_0(z) = \left( \frac{\mu - a_{11}^0}{2\mu} \right)^{1/2} \begin{pmatrix} 1 & \frac{a_{12}^0}{\mu - a_{11}^0} \\ \frac{a_{12}^0}{\mu - a_{11}^0} & 1 \end{pmatrix}.$$

We shall fix the branch of the square root by the condition,

$$\left( \frac{\mu - a_{11}^0}{2\mu} \right)^{1/2} > 0, \quad z > z_2.$$

It is easy to check that $T_0(z)$ has no singularities except for those at $z = \pm z_{1,2}$, and that

$$\det T_0(z) = 1.$$

The last equation in turn implies that

$$\det \Psi_{\text{WKB}}(z) = C_0^2 \neq 0.$$

Consider the region

$$\mathbb{C}^* = \mathbb{C} \setminus (\Omega \cup (-\Omega)).$$

LEMMA 7.1. *The function* $\Psi_{\mathrm{WKB}}(z)$ *defined in* (7.9)*, is a single-valued analytic function on* $\mathbb{C}^*$*, which, in addition, satisfies the symmetry equation*

$$(7.11) \qquad \Psi_{\mathrm{WKB}}(z) = (-1)^n \sigma_3 \Psi_{\mathrm{WKB}}(-z)\sigma_3.$$

Proof of Lemma 7.1 is given in Appendix B below. As $z \to \infty$,

$$(7.12) \quad T_0(z) = \begin{pmatrix} 1 & 0 \\ 0 & 1 \end{pmatrix} + z^{-1} \begin{pmatrix} 0 & (R_n^0)^{1/2} \\ (R_n^0)^{1/2} & 0 \end{pmatrix} + O\left(|z|^{-2}\right), \qquad z \to \infty.$$

In addition, from (7.6), (7.7'),

$$(7.13) \qquad \xi(z) = \frac{V(z)}{2} - \frac{n \ln z}{N} + \gamma_n + O\left(|z|^{-2}\right), \qquad z \to \infty,$$

where

$$(7.14) \qquad \gamma_n = \frac{t^2}{8g} - \frac{\lambda}{4} \ln \frac{g}{\lambda} - \frac{\lambda}{4}.$$

Thus,

$$(7.15) \qquad \Psi^0(z) \sim \left( \sum_{k=0}^{\infty} \frac{\Theta_k^0}{z^k} \right) C_0 e^{-\left( \frac{NV(z)}{2} - n \ln z + N\gamma_n - C_1 \right)\sigma_3},$$

with

$$(7.16) \qquad \Theta_0^0 = \begin{pmatrix} 1 & 0 \\ 0 & 1 \end{pmatrix}, \qquad \Theta_1^0 = \begin{pmatrix} 0 & (R_n^0)^{1/2} \\ (R_n^0)^{1/2} & 0 \end{pmatrix}.$$

*Turning Point Functions.* To construct $\Psi^0(z)$ in the sets $\Omega_{1,2}$ we use turning point functions which are expressed in terms of the Airy function. Let us consider first the set $\Omega_2$. On this set we are looking for a solution to (7.2) in the form

$$(7.17) \qquad \Psi(z) = W(z)\Phi(w(z)).$$

Following [Kap] (see also the earlier work [Ble]), we want to find an analytical matrix-valued function $W(z)$ (gauge matrix) and an analytical function $w(z)$ of the change of variable such that (7.2) reduces to the following model equation for $\Phi(w)$:

$$(7.18) \qquad \Phi'(w) = N \begin{pmatrix} 0 & 1 \\ w & 0 \end{pmatrix} \Phi(w).$$

This model equation is nothing else than the Airy equation $\psi''(w) = N^2 w \psi(w)$ written in the matrix form. The functions $W(z)$ and $w(z)$ are determined iteratively in powers of $N^{-1}$ and we consider approximation of the order of $N^{-1}$. In this approximation the functions $w(z)$ and $W(z)$ are defined as follows.

1. *Function of change of variable.* Put

$$(7.19) \qquad w(z; z_2) = \left( \frac{3}{2} \int_{z_2^N}^z \nu(z) \, dz \right)^{2/3},$$

where

$$(7.20) \qquad \nu(z) = \left[ \mu^2(z) + N^{-1} \left( \frac{t}{2} + g R_n^0 - \frac{g z^2}{2} \right) \right]^{1/2},$$

and $z_2^N$ is a root of $\nu(z)$ close to $z_2$,

$$\nu(z_2^N) = 0, \quad z_2^N - z_2 = O(N^{-1}).$$

The argument $z_2$ in $w(z; z_2)$ reminds us that this function is constructed in a neighborhood of the turning point $z_2$. Observe that the function $\nu^2(z)$ can be written in terms of the elements of the matrix $A^0(z)$ as

$$(7.20') \qquad \nu^2(z) = -\det A^0(z) + N^{-1} \left( (a_{11}^0)'(z) - a_{11}^0(z) \frac{(a_{12}^0)'(z)}{a_{12}^0(z)} \right)$$

which coincides (up to terms of the order of $N^{-2}$) with the potential of the Schrödinger equation equivalent to (7.2) (cf. (3.26,7)). In (7.19) and (7.20) we take the branches of the power functions that are positive on the positive half-axis. Since $z_2^N - z_2 = O(N^{-1})$, for large $N$ the function $w(z; z_2)$ is analytic for $z \in \Omega_2$ and

$$w'(z; z_2) \neq 0, \quad z \in \Omega_2.$$

2. *Gauge matrix.* Define

$$(7.21) \qquad W(z; z_2) = \sqrt{\frac{a_{12}^0(z)}{w'(z; z_2)}} \begin{pmatrix} 1 & 0 \\ -\frac{a_{11}^0(z)}{a_{12}^0(z)} & \frac{w'(z; z_2)}{a_{12}^0(z)} \end{pmatrix},$$

where

$$\sqrt{\frac{a_{12}^0(z)}{w'(z; z_2)}} > 0, \qquad z > z_2^N.$$

The function $W(z; z_2)$ is analytic in $z \in \Omega_2$ for large $N$.

To get a desired multiplicative jump matrix $S$ between $\Omega_2^u$ and $\Omega_2^d$ we use different solutions to the model equation (7.18) in the domains $\Omega_2^u$ and $\Omega_2^d$. They are defined as follows.

3. *Model canonical solutions.* Define

(7.22)

$$\Phi_u(z; z_2) = \begin{pmatrix} N^{1/6} & 0 \\ 0 & N^{-1/6} \end{pmatrix} \begin{pmatrix} y_0(N^{2/3}z) & y_1(N^{2/3}z) \\ y_0'(N^{2/3}z) & y_1'(N^{2/3}z) \end{pmatrix} \begin{pmatrix} (2\pi)^{-1/2} & 0 \\ 0 & (2\pi)^{1/2} \end{pmatrix},$$

$$\Phi_d(z; z_2) = \begin{pmatrix} N^{1/6} & 0 \\ 0 & N^{-1/6} \end{pmatrix} \begin{pmatrix} y_0(N^{2/3}z) & y_2(N^{2/3}z) \\ y_0'(N^{2/3}z) & y_2'(N^{2/3}z) \end{pmatrix} \begin{pmatrix} (2\pi)^{-1/2} & 0 \\ 0 & (2\pi)^{1/2} \end{pmatrix},$$

where

(7.23)
$$\begin{aligned} y_0(z) &= \operatorname{Ai}(z), \\ y_1(z) &= e^{-\pi i/6} \operatorname{Ai}\left(e^{-2\pi i/3}z\right), \\ y_2(z) &= e^{\pi i/6} \operatorname{Ai}\left(e^{2\pi i/3}z\right). \end{aligned}$$

Here $\operatorname{Ai}(z)$ is the Airy function, i.e. the solution of the model equation
(7.24)
$$\operatorname{Ai}''(z) = z \operatorname{Ai}(z); \qquad \operatorname{Ai}(z) \sim \frac{1}{2\sqrt{\pi}\, z^{1/4}} \exp\left(-\frac{2z^{3/2}}{3}\right), \quad z \to \infty.$$

It satisfies the relation

$$\operatorname{Ai}(z) + e^{2\pi i/3} \operatorname{Ai}\left(e^{2\pi i/3}z\right) + e^{-2\pi i/3} \operatorname{Ai}\left(e^{-2\pi i/3}z\right) = 0;$$

hence

$$y_1(z) - y_2(z) = -i y_0(z),$$

and this implies the equation

(7.25)
$$\Phi_u(z) = \Phi_d(z) \begin{pmatrix} 1 & -2\pi i \\ 0 & 1 \end{pmatrix}.$$

Now we define the turning point functions

(7.26)
$$\Psi_{\mathrm{TP}}^{u,d}(z; z_2) = C W(z; z_2) \Phi_{u,d}(w(z; z_2); z_2),$$

where $C$ is a constant which will be chosen later.

In $\Omega_1$ we define

(7.27)
$$\Psi_{\mathrm{TP}}^{u,d}(z; z_1) = C^{(1)} W(z; z_1) \Phi_{u,d}(w(z; z_1); z_1),$$

where $C^{(1)}$ is a constant to be chosen later, and $w(z; z_1)$, $W(z; z_1)$, and $\Phi_{u,d}(z; z_1)$ are defined as follows. The function of change of variable

(7.27′)
$$w(z; z_1) = \left(\frac{3}{2} \int_{z_1^N}^z \nu(z)\, dz\right)^{2/3},$$

where $\nu(z_1^N) = 0$, $z_1 - z_1^N = O(N^{-1})$. Observe that $\nu(z) < 0$ when $0 < z < z_1^N$, hence

$$\int_{z_1^N}^z \nu(z)\,dz > 0, \qquad 0 < z < z_1^N,$$

and we take the branch of $\zeta^{2/3}$ in (7.27') which is positive for $\zeta > 0$. The model solutions $\Phi_{u,d}(z; z_1)$ are defined as

(7.27'')

$$\Phi_u(z; z_1) = \begin{pmatrix} N^{1/6} & 0 \\ 0 & N^{-1/6} \end{pmatrix} \begin{pmatrix} y_0(N^{2/3}z) & -y_2(N^{2/3}z) \\ y_0'(N^{2/3}z) & -y_2'(N^{2/3}z) \end{pmatrix} \begin{pmatrix} (2\pi)^{-1/2} & 0 \\ 0 & (2\pi)^{1/2} \end{pmatrix},$$

$$\Phi_d(z; z_1) = \begin{pmatrix} N^{1/6} & 0 \\ 0 & N^{-1/6} \end{pmatrix} \begin{pmatrix} y_0(N^{2/3}z) & -y_1(N^{2/3}z) \\ y_0'(N^{2/3}z) & -y_1'(N^{2/3}z) \end{pmatrix} \begin{pmatrix} (2\pi)^{-1/2} & 0 \\ 0 & (2\pi)^{1/2} \end{pmatrix}.$$

Comparing this with the model solutions (7.22) in $\Omega_2$, one can notice that we change $y_1$ for $-y_2$ in the formula for $\Phi_u$ and $y_2$ for $-y_1$ in the formula for $\Phi_d$. Note that equation (7.25) still holds. The gauge matrix $W(z; z_1)$ is defined as

(7.27''') $\quad W(z; z_1) = -i\sqrt{\dfrac{a_{12}^0(z)}{w'(z; z_1)}} \begin{pmatrix} 1 & 0 \\ -\dfrac{a_{11}^0(z)}{a_{12}^0(z)} & \dfrac{w'(z; z_1)}{a_{12}^0(z)} \end{pmatrix};$

$$-i\sqrt{\frac{a_{12}^0(z)}{w'(z; z_1)}} > 0, \quad 0 < z < z_1^N,$$

and it is analytic in $z \in \Omega_1$ for large $N$. Now we define

(7.28) $$\Psi^0(z) = \begin{cases} \Psi_{\mathrm{TP}}^{u,d}(z; z_{1,2}), & \text{if } z \in \Omega_{1,2}^{u,d}, \\ (-1)^n \sigma_3 \Psi^0(-z)\sigma_3, & \text{if } z \in (-\Omega). \end{cases}$$

We reiterate that our use of equation (7.2) has been purely formal: we only refer to it as a motivation for the introduction of the function $\Psi^0(z)$ by explicit formulae (7.9) and (7.28).

*Properties of $\Psi^0(z)$.* As defined by (7.9) and (7.28), the function $\Psi^0(z)$ is a piecewise analytic function in $\mathbb{C}$ and has the following properties. By (7.11) and (7.28),

(7.29) $$\Psi^0(z) = (-1)^n \sigma_3 \Psi^0(-z)\sigma_3.$$

By (7.25),

(7.30) $\quad \Psi_+^0(z) = \Psi_-^0(z)S, \quad z \in \Omega \cap \{\mathrm{Im}\,z = 0\}; \qquad S = \begin{pmatrix} 1 & -2\pi i \\ 0 & 1 \end{pmatrix}.$

Now we want to evaluate the jump matrices on $\partial\Omega_1$ and $\partial\Omega_2$. Observe that the WKB function (7.9) has two free constants $C_0$ and $C_1$, and each of the turning point solutions (7.26), (7.27) has one free constant, $C$ and $C^{(1)}$, respectively. We show that we can adjust these constants to make the jump matrices close to $I$.

LEMMA 7.2. *If*

$$(7.31) \qquad C_0 = \frac{C}{\sqrt{2\pi}}, \qquad C_1 = -\frac{\ln 2\pi}{2}; \qquad C^{(1)} = (-1)^k C, \quad k = \left[\frac{n}{2}\right],$$

*then*

$$(7.32) \qquad \Psi^0_+(z) = \left(I + O(N^{-1})\right) \Psi^0_-(z),$$

*uniformly with respect to*

$$z \in \partial\Omega_1 \cup \partial\Omega_2 \cup (-\partial\Omega_1) \cup (-\partial\Omega_2),$$

*where*

$$-\partial\Omega_j = \{\, z\colon\, -z \in \partial\Omega_j \,\}, \quad j = 1, 2.$$

The proof of this lemma is given in Appendix C below. Consider now the asymptotics of $\Psi^0(z)$ at infinity. According to (7.15) and (7.31),

$$(7.33) \qquad \Psi^0(z) \sim \left(\sum_{k=0}^{\infty} \frac{\Theta^0_k}{z^k}\right) \frac{C}{\sqrt{2\pi}} e^{-\left(\frac{NV(z)}{2} - n\ln z + N\gamma_n + \frac{\ln 2\pi}{2}\right)\sigma_3},$$

with

$$\Theta^0_0 = \begin{pmatrix} 1 & 0 \\ 0 & 1 \end{pmatrix}, \qquad \Theta^0_1 = \begin{pmatrix} 0 & (R^0_n)^{1/2} \\ (R^0_n)^{1/2} & 0 \end{pmatrix}.$$

We want to adjust the asymptotics to the one (5.16), (5.17) in the Riemann-Hilbert problem. Define

$$(7.34) \qquad \lambda^0_n = N\gamma_n + \frac{\ln 2\pi}{2} + \frac{\ln R^0_n}{4}, \qquad C = (R^0_n)^{-1/4} (2\pi)^{1/2}.$$

Then (7.33) is rewritten as

$$(7.35) \qquad \Psi^0(z) \sim \left(\sum_{k=0}^{\infty} \frac{\Gamma^0_k}{z^k}\right) e^{-\left(\frac{NV(z)}{2} - n\ln z + \lambda^0_n\right)\sigma_3},$$

with some coefficients $\Gamma^0_k$ such that

$$(7.36) \qquad \Gamma^0_0 = \begin{pmatrix} 1 & 0 \\ 0 & (R^0_n)^{-1/2} \end{pmatrix}, \qquad \Gamma^0_1 = \begin{pmatrix} 0 & 1 \\ (R^0_n)^{1/2} & 0 \end{pmatrix}.$$

*Proof of Theorem* 1.1. We want to prove that $\Psi^0(z)$, defined in (7.9) and (7.28) with $C$, $C_0$, and $C_1$ given by (7.31), (7.34), is close to $\Psi_n(z)$. Observe that by (5.16) and (7.35), the limit

$$(7.37) \qquad \Gamma = \lim_{z \to \infty} \Psi_n(z) \left[ \Psi^0(z) \right]^{-1} = \Gamma_0 e^{(\lambda_n^0 - \lambda_n)\sigma_3} (\Gamma_0^0)^{-1}$$

exists and $\det \Gamma \neq 0$. Denote

$$(7.38) \qquad X(z) = \Gamma^{-1} \Psi_n(z) \left[ \Psi^0(z) \right]^{-1}.$$

Then

$$(7.39) \qquad \lim_{z \to \infty} X(z) = I.$$

The multiplicative jump matrices of $\Psi_n(z)$ and $\Psi^0(z)$ on $\Omega \cap \{\operatorname{Im} z = 0\}$ coincide (both are equal to $S$); hence $X(z)$ has no jump on $\Omega \cap \{\operatorname{Im} z = 0\}$.

On $\partial\Omega_j$, $j = 1, 2$, the function $\Psi_n(z)$ has no jump; hence

$$\begin{aligned}
X_+(z) &= \Gamma^{-1} \Psi_n(z) \left[ \Psi_+^0(z) \right]^{-1} = \Gamma^{-1} \Psi_n(z) \left[ \left( I + O(N^{-1}) \right) \Psi_-^0(z) \right]^{-1} \\
&= \Gamma^{-1} \Psi_n(z) \left[ \Psi_-^0(z) \right]^{-1} \left( I + O(N^{-1}) \right)^{-1} = X_-(z) \left( I + O(N^{-1}) \right)^{-1},
\end{aligned}$$

so that

$$(7.40) \qquad X_+(z) = X_-(z) \left( I + O(N^{-1}) \right), \quad \text{if} \quad z \in \partial\Omega_1 \cup \partial\Omega_2.$$

Let

$$(7.41) \qquad L = \left\{ z \colon \operatorname{Im} z = 0, \ z \notin \Omega \cup (-\Omega) \right\}.$$

On $L$, the function $\Psi^0(z)$ has no jump; hence

$$(7.42)$$
$$\begin{aligned}
X_+(z) &= \Gamma^{-1} \Psi_{n+}(z) \left[ \Psi^0(z) \right]^{-1} = \Gamma^{-1} \Psi_{n-}(z) S \left[ \Psi^0(z) \right]^{-1} \\
&= \Gamma^{-1} \Psi_{n-}(z) \left[ \Psi^0(z) \right]^{-1} \Psi^0(z) S \left[ \Psi^0(z) \right]^{-1} = X_-(z) Q(z),
\end{aligned}$$

where

$$Q(z) = \Psi^0(z) S \left[ \Psi^0(z) \right]^{-1}, \qquad z \in L.$$

When $z \in L$, $\Psi^0(z) = \Psi_{\mathrm{WKB}}(z)$; hence by (7.9),

$$\begin{aligned}
Q(z) &= T_0(z) e^{[-N\xi(z) + C_1]\sigma_3} S e^{[N\xi(z) - C_1]\sigma_3} T_0^{-1}(z) \\
&= T_0(z) \begin{pmatrix} 1 & -2\pi i \, e^{-2N\xi(z) + 2C_1} \\ 0 & 1 \end{pmatrix} T_0^{-1}(z) \\
&= I + T_0(z) \begin{pmatrix} 0 & -i \, e^{-2N\xi(z)} \\ 0 & 0 \end{pmatrix} T_0^{-1}(z).
\end{aligned}$$

Since $\det T_0(z) = 1$, we obtain from (7.10) that $|T_0(z)|$, $|T_0^{-1}(z)| \leq C$; hence

$$Q(z) = I + O\left( e^{-2N\operatorname{Re}\xi(z)} \right)$$

and

(7.43)         $X_+(z) = X_-(z)\left(I + O\left(e^{-2N\mathrm{Re}\,\xi(z)}\right)\right), \qquad z \in L.$

Thus, we have the following theorem.

THEOREM 7.3.   *The function $X(z) = \Gamma^{-1}\Psi_n(z)\left[\Psi^0(z)\right]^{-1}$ satisfies*

$$\lim_{z \to \infty} X(z) = I$$

*and*

(7.44)

$X_+(z) = X_-(z)\left(I + O(N^{-1})\right), \qquad if \quad z \in \partial\Omega_1 \cup \partial\Omega_2 \cup (-\partial\Omega_1) \cup (-\partial\Omega_2),$

$X_+(z) = X_-(z)\left(I + O\left(e^{-2N\mathrm{Re}\,\xi(z)}\right)\right), \quad if \quad z \in L;$

*i.e., there exists a constant $C > 0$ independent of $N$ and $z$ such that*

$\left|[X_-(z)]^{-1}X_+(z) - I\right| \leq CN^{-1}, \quad if \quad z \in \partial\Omega_1 \cup \partial\Omega_2 \cup (-\partial\Omega_1) \cup (-\partial\Omega_2),$

$\left|[X_-(z)]^{-1}X_+(z) - I\right| \leq Ce^{-2N\mathrm{Re}\,\xi(z)}, \quad if \quad z \in L.$

*Remark.*   Observe that there exists $c_0 > 0$ such that $\mathrm{Re}\,\xi(z) \geq c_0(1 + |z|)^4 > 0$ for all $z \in L$ (cf. (7.7′)); therefore the error term on $L$ is exponentially small as $N \to \infty$ and also as $z \to \pm\infty$.

By standard methods (see e.g. [DZ2], [DIZ] and also Appendix D below), Theorem 7.3 yields the asymptotic equation,

(7.45)         $X(z) = I + O\left(\dfrac{1}{N(1 + |z|)}\right), \qquad z \in \mathbb{C},$

so that

(7.46)         $\Psi_n(z) = \Gamma\left(I + O\left(\dfrac{1}{N(1 + |z|)}\right)\right)\Psi^0(z).$

This implies that

$$\sum_{k=0}^{\infty}\frac{\Gamma_k}{z^k} = \Gamma\left(I + O\left(\frac{1}{N(1 + |z|)}\right)\right)\left(\sum_{k=0}^{\infty}\frac{\Gamma_k^0}{z^k}\right)e^{(\lambda_n - \lambda_n^0)\sigma_3}.$$

Equating the coefficients at $z^0$ and at $z^{-1}$, we obtain

(7.47)         $\Gamma_0 = \Gamma\Gamma_0^0 e^{\theta\sigma_3}, \qquad \theta = \lambda_n - \lambda_n^0,$

$\Gamma_1 = \Gamma\left[\Gamma_1^0 + O(N^{-1})\Gamma_0^0\right]e^{\theta\sigma_3}.$

Excluding $\Gamma$ we obtain

$$e^{\theta\sigma_3}(\Gamma_0)^{-1}\Gamma_1 e^{-\theta\sigma_3} = (\Gamma_0^0)^{-1}\Gamma_1^0 + O(N^{-1}).$$

We have

$$(\Gamma_0)^{-1}\Gamma_1 = \begin{pmatrix} 0 & 1 \\ R_n & 0 \end{pmatrix}, \qquad (\Gamma_0^0)^{-1}\Gamma_1^0 = \begin{pmatrix} 0 & 1 \\ R_n^0 & 0 \end{pmatrix};$$

hence

$$e^{\theta\sigma_3}(\Gamma_0)^{-1}\Gamma_1 e^{-\theta\sigma_3} = \begin{pmatrix} 0 & e^{2\theta} \\ R_n e^{-2\theta} & 0 \end{pmatrix}.$$

Thus,

$$\begin{pmatrix} 0 & e^{2\theta} \\ R_n e^{-2\theta} & 0 \end{pmatrix} = \begin{pmatrix} 0 & 1 \\ R_n^0 & 0 \end{pmatrix} + O(N^{-1}),$$

which implies that

$$(7.47') \qquad R_n = R_n^0 + O(N^{-1}), \qquad \lambda_n = \lambda_n^0 + O(N^{-1}).$$

In view of the definition of $R_n^0$ (see (7.1)) the first equation yields (1.14). By (7.47),

$$\Gamma = \Gamma_0 e^{-\theta\sigma_3}(\Gamma_0^0)^{-1} = \begin{pmatrix} e^{-\theta} & 0 \\ 0 & e^{\theta}(R_n)^{-1/2}(R_n^0)^{1/2} \end{pmatrix} = I + O\left(N^{-1}\right);$$

hence by (7.46),

$$(7.48) \qquad \Psi_n(z) = \Gamma\left(I + O\left(\frac{1}{N(1+|z|)}\right)\right)\Psi^0(z), \qquad \Gamma = I + O\left(N^{-1}\right).$$

All the statements of our main Theorem 1.1 follow directly from equation (7.48). In what follows we are presenting the corresponding detailed derivations.

To derive the semiclassical asymptotics (1.15) of the function $\psi_n(z)$ between the turning points $z_1$ and $z_2$ we use the following lemma. Let $\delta > 0$ be an arbitrary fixed number. Define

$$\Omega_0 = \Omega \cap \{\, z\colon\ z_1 + \delta \le \operatorname{Re} z \le z_2 - \delta\,\}$$

and

$$\Omega_0^u = \Omega_0 \cap \{\, z\colon\ \operatorname{Im} z \ge 0\,\}, \quad \Omega_0^d = \Omega_0 \cap \{\, z\colon\ \operatorname{Im} z \le 0\,\}.$$

LEMMA 7.4. *The function $\Psi^0(z)$ has the following asymptotics in $\Omega_0$:*

$$(7.49) \qquad \Psi^0(z) = \frac{\left(I + O(N^{-1})\right)}{(R_n^0)^{1/4}}\left(\frac{a_{12}^0(z)}{2\mu(z)}\right)^{1/2}\begin{pmatrix} 1 & 0 \\ -\frac{a_{11}^0(z)}{a_{12}^0(z)} & \frac{\mu(z)}{a_{12}^0(z)} \end{pmatrix}Q_{u,d}^{(1)}$$

$$\times\, e^{(N\xi(z)+d(z))\sigma_3}Q_{u,d}^{(2)}P, \quad if \ \ z \in \Omega_0^{u,d},$$

*where $\xi(z)$ is defined as in (7.6),(7.7$'$),*

$$(7.49') \qquad d(z) = -\frac{1}{2}\ln\left(\frac{\mu(z) - a_{11}^0(z)}{a_{12}^0(z)}\right),$$

*and*

$$Q_u^{(1)} = \overline{Q_d^{(1)}} = \begin{pmatrix} 1 & i \\ 1 & -i \end{pmatrix}, \quad Q_u^{(2)} = \overline{Q_d^{(2)}} = \begin{pmatrix} i & 1 \\ -i & 0 \end{pmatrix},$$

$$P = \begin{pmatrix} (2\pi)^{-1/2} & 0 \\ 0 & (2\pi)^{1/2} \end{pmatrix}.$$

Proof of this lemma is given in Appendix C below. Due to (7.48), we can replace $\Psi^0(z)$ with $\Psi_n(z)$ in (7.49), which gives the semiclassical asymptotics of $\Psi_n(z)$ in $\Omega_0$. Thus, we have the following theorem.

THEOREM 7.5. *In* $\Omega_0$,

$$(7.50) \qquad \Psi_n(z) = \frac{(I + O(N^{-1}))}{(R_n^0)^{1/4}} \left( \frac{a_{12}^0(z)}{2\mu(z)} \right)^{1/2} \begin{pmatrix} 1 & 0 \\ -\frac{a_{11}^0(z)}{a_{12}^0(z)} & \frac{\mu(z)}{a_{12}^0(z)} \end{pmatrix} Q_{u,d}^{(1)}$$

$$\times \, e^{(N\xi(z)+d(z))\sigma_3} Q_{u,d}^{(2)} P, \quad if \quad z \in \Omega_0^{u,d}.$$

*Remark.* Let $\mu_0(z)$, $\xi_0(z)$, $d_0(z)$ denote the analytic continuation of the functions $\mu(z)$, $\xi(z)$, $d(z)$ from $\Omega_0^u$ to $\Omega_0$ across the interval $[z_1 + \delta, z_2 - \delta]$. (Note that $\mu_0(z) = -\mu(z)$, $\xi_0(z) = -\xi(z)$, and $d_0(z) = -d(z)$ for $z \in \Omega_0^d$.) Taking into account that by (3.14) the function $\psi_n(z)$ is the element 11 of the matrix-valued function $\Psi_n(z)$, we derive from (7.50) that

$$(7.50') \qquad \psi_n(z) = \left( I + O(N^{-1}) \right) \frac{1}{(2\pi)^{1/2}(R_n^0)^{1/4}} \left( \frac{a_{12}^0(z)}{2\mu_0(z)} \right)^{1/2}$$

$$\times \left[ i e^{N\xi_0(z)+d_0(z)} + e^{-(N\xi_0(z)+d_0(z))} \right], \quad if \quad z \in \Omega_0.$$

The function $N\xi_0(z) + d_0(z)$ is pure imaginary for real $z$ in the interval $z_1 + \delta \leq z \leq z_2 - \delta$ (cf. (7.7') and (7.49')); hence, when we restrict the asymptotics (7.50') to the interval $z_1 + \delta \leq z \leq z_2 - \delta$, the equation,

$$\psi_n(z) = \frac{1}{(2\pi)^{1/2}(R_n^0)^{1/4}} \left( \frac{a_{12}^0(z)}{2\mu(z)} \right)^{1/2}$$

$$\times \left[ i e^{N\xi_0(z)+d_0(z)} + e^{-(N\xi_0(z)+d_0(z))} + O(N^{-1}) \right],$$

follows. Since

$$a_{12}^0 = (R_n^0)^{1/2} g z^2, \quad \mu(z) = e^{\frac{i\pi}{2}} z \sqrt{\lambda g} \sqrt{1 - x^2},$$

we obtain that for $z_1 + \delta \leq z \leq z_2 - \delta$,

$$(7.51) \qquad \psi_n(z) = \frac{1}{\sqrt{\pi}} \left( \frac{g}{\lambda} \right)^{1/4} \frac{\sqrt{z}}{(1 - x^2)^{1/4}} \left[ \cos \left( \Phi(z) + \frac{\pi}{4} \right) + O(N^{-1}) \right],$$

where $\Phi(z) = i^{-1}[N\xi_0(z) + d_0(z)]$. To finish the proof of (1.15) we use the following lemma.

LEMMA 7.6. *If* $z_1 + \delta \le z \le z_2 - \delta$ *then*

$$(7.52) \qquad i^{-1}[N\xi_0(z) + d_0(z)] = \frac{(n+\frac{1}{2})}{2}\left(\frac{\sin 2\phi}{2} - \phi\right) - \frac{(-1)^n \chi}{4},$$

*where* $\phi = \arccos q$ *and* $\chi = \arccos r$ *and*

$$q = \frac{gz^2 + t}{2\sqrt{\lambda' g}}, \qquad r = \frac{2\sqrt{\lambda' g} - tq}{2\sqrt{\lambda' g}\, q - t}, \qquad \lambda' = \frac{(n+\frac{1}{2})}{N}.$$

Proof of this lemma is given in Appendix C below. Since

$$\lambda' = \frac{(n+\frac{1}{2})}{N} = \frac{n}{N} + \frac{1}{2N} = \lambda + O(N^{-1}),$$

we obtain

$$\sqrt{1-x^2} = \sin\phi\left(1 + O(N^{-1})\right);$$

hence for $z_1 + \delta \le z \le z_2 - \delta$,

$$\psi_n(z) = \frac{1}{\sqrt{\pi}}\left(\frac{g}{\lambda}\right)^{1/4}\frac{\sqrt{z}}{\sqrt{\sin\phi}}\left\{\cos\left[\frac{(n+\frac{1}{2})}{2}\left(\frac{\sin 2\phi}{2} - \phi\right)\right.\right.$$
$$\left.\left. -\frac{(-1)^n \chi}{4} + \frac{\pi}{4}\right] + O(N^{-1})\right\}.$$

Formula (1.15) is proved.

To prove formulae (1.17), (1.18) we use the following theorem.

THEOREM 7.7. *The function* $\Psi_n(z)$ *has the following asymptotics*: *in* $\mathbb{C}\setminus(\Omega\cup(-\Omega))$,

$$(7.53) \quad \Psi_n(z) = \frac{C_n C_n^0(z)}{(R_n^0)^{1/4}}\left(\frac{\mu - a_{11}^0}{2\mu}\right)^{1/2}\begin{pmatrix} 1 & \frac{a_{12}^0}{\mu - a_{11}^0} \\ \frac{a_{12}^0}{\mu - a_{11}^0} & 1 \end{pmatrix} e^{-N\xi(z)\sigma_3 - \frac{\ln 2\pi}{2}\sigma_3},$$

*where the matrix* $C_n$ *does not depend on* $z$, *and*

$$C_n = I + O\left(\frac{1}{N}\right), \qquad C_n^0(z) = I + O\left(\frac{1}{N(1+|z|)}\right).$$

*In* $\Omega_j, \ j = 1, 2,$

$$(7.54) \quad \Psi_n(z) = \frac{(I + O(N^{-1}))\sqrt{2\pi}(-1)^{\sigma_0}}{(R_n^0)^{1/4}}W(z; z_j)\,\Phi_{u,d}(w(z; z_j); z_j),$$
$$z \in \Omega_j^{u,d}, \quad j = 1, 2,$$

*where $\sigma_0 = (2 - j)\left[\frac{n}{2}\right]$,*

$$(7.55) \qquad w(z; z_j) = \left(\frac{3}{2}\int_{z_j^N}^z \nu(z)\,dz\right)^{2/3},$$

$$\nu(z) = \left[\mu^2(z) + N^{-1}\left(\frac{t}{2} + gR_n^0 - \frac{gz^2}{2}\right)\right]^{1/2},$$

*and the gauge matrix $W(z; z_j)$ and the model canonical solutions $\Phi_{u,d}(z; z_j)$ are as defined in $(7.21)$, $(7.27''')$ and $(7.22)$, $(7.27'')$ respectively.*

The proof of the theorem follows from the definition of $\Psi^0(z)$ and $(7.48)$.

Theorems 7.5 and 7.7 provide semiclassical asymptotics of $\Psi_n(z)$ in the whole complex plane. The formulae $(1.17)$ are derived from $(7.53)$ in the same way we derived $(1.15)$ from Theorem 7.5. Let us show that $(1.18)$ follows from $(7.54)$. For the sake of definiteness consider $j = 2$. Restricting $(7.54)$ to the element 11, we obtain that if $z \in \Omega_2$ then

$$\psi_n(z) = \frac{(2\pi)^{1/2}}{(R_n^0)^{1/4}}\left(\frac{a_{12}^0(z)}{w'(z; z_2)}\right)^{1/2}\Big[(1 + \varepsilon_1)N^{1/6}\,\mathrm{Ai}\,(w(z; z_2))(2\pi)^{-1/2} \\ + \varepsilon_2 N^{-1/6}\,\mathrm{Ai}\,'(w(z; z_2))\Big],$$

where $\varepsilon_{1,2} = O(N^{-1})$. This gives

$$\psi_n(z) = \frac{N^{1/6}\sqrt{g}\,z}{(w'(z; z_2))^{1/2}}\left[(1 + r_1)\,\mathrm{Ai}\,(w(z; z_2)) + r_2\,\mathrm{Ai}\,'(w(z; z_2))\right],$$

where $r_1 = O(N^{-1})$ and $r_2 = O(N^{-4/3})$. Formula $(1.18)$ is proved.

Finally, by $(4.14)$, $(7.47')$, $(7.34)$, and $(7.14)$,

$$h_n = \exp\left(2\lambda_n\right) = \exp\left(2\lambda_n^0 + O(N^{-1})\right)$$

$$= \exp\left(2N\gamma_n + \ln 2\pi + \frac{\ln R_n^0}{2} + O(N^{-1})\right)$$

$$= 2\pi\sqrt{R_n^0}\exp\left[\frac{Nt^2}{4g} - \frac{N\lambda}{2}\left(1 + \ln\frac{g}{\lambda}\right) + O(N^{-1})\right],$$

which proves $(1.23)$. Theorem 1.1 is proved.

## Appendix A. Proof of formula (7.8)

By $(7.6)$,

$$\mu^2 = a_{12}^0 a_{21}^0 + \left(a_{11}^0\right)^2,$$

and

$$\operatorname{diag} T^{-1}T' = \begin{pmatrix} \left(\dfrac{a_{21}^0}{\mu-a_{11}^0}\right)' \dfrac{a_{12}^0}{2\mu} & 0 \\[2ex] 0 & \left(\dfrac{a_{12}^0}{\mu-a_{11}^0}\right)' \dfrac{a_{21}^0}{2\mu} \end{pmatrix}.$$

In our case $a_{21}^0 = -a_{12}^0$; hence

(A.1) $$\mu^2 = \left(a_{11}^0\right)^2 - \left(a_{12}^0\right)^2$$

and

$$\operatorname{diag} T^{-1}T' = -\left(\frac{a_{12}^0}{\mu-a_{11}^0}\right)' \frac{a_{12}^0}{2\mu} \begin{pmatrix} 1 & 0 \\ 0 & 1 \end{pmatrix}.$$

Now,

$$\left(\frac{a_{12}^0}{\mu-a_{11}^0}\right)' \frac{a_{12}^0}{\mu} = \frac{\left[\left(a_{12}^0\right)'\left(\mu-a_{11}^0\right) - a_{12}^0\left(\mu' - \left(a_{11}^0\right)'\right)\right]a_{12}^0}{\left(\mu-a_{11}^0\right)^2 \mu}.$$

By (A.1),

$$(a_{12}^0)^2 = (a_{11}^0)^2 - \mu^2,$$

$$a_{12}^0(a_{12}^0)' = a_{11}^0(a_{11}^0)' - \mu\mu';$$

hence

$$\left(\frac{a_{12}^0}{\mu-a_{11}^0}\right)' \frac{a_{12}^0}{\mu} = \frac{\left[a_{11}^0(a_{11}^0)' - \mu\mu'\right]\left(\mu-a_{11}^0\right) - \left[(a_{11}^0)^2 - \mu^2\right]\left[\mu' - (a_{11}^0)'\right]}{\left(\mu-a_{11}^0\right)^2 \mu}$$

$$= \frac{a_{11}^0(a_{11}^0)'\mu + \mu\mu'a_{11}^0 - (a_{11}^0)'\mu^2 - (a_{11}^0)^2\mu'}{\left(\mu-a_{11}^0\right)^2 \mu}$$

$$= \frac{\mu(a_{11}^0)' - a_{11}^0\mu'}{\left(\mu-a_{11}^0\right)\mu} = \left(\frac{\mu}{\mu-a_{11}^0}\right)' \frac{\mu-a_{11}^0}{\mu} = \left(\ln\frac{\mu}{\mu-a_{11}^0}\right)'.$$

Thus,

$$\operatorname{diag} T^{-1}T' = \frac{1}{2}\left(\ln\frac{\mu-a_{11}^0}{\mu}\right)' I$$

and

$$\tau(z) = \int_{z_2}^z \operatorname{diag} T^{-1}(u)T'(u)\,du = \frac{1}{2}\ln\frac{\mu(u)-a_{11}^0(u)}{\mu(u)}\bigg|_{u=z_2}^z I;$$

hence

$$e^{-2\tau(z)} = C\frac{\mu(z)-a_{11}^0(z)}{\mu(z)}I.$$

Formula (7.8) is proved.

## Appendix B. Proof of Lemma 7.1

Note that

(B.1)
$$\frac{\mu - a_{11}^0}{2\mu} = \frac{x + \sqrt{x^2 - 1} + c}{2\sqrt{x^2 - 1}}, \qquad c \equiv \left(\frac{g}{\lambda}\right)^{1/2} R_n^0 = x_0 - (-1)^n \sqrt{x_0^2 - 1},$$

$$x_0 = -\frac{t}{2\sqrt{\lambda g}} > 1.$$

Let $\Omega^*$ denote the region,

$$\Omega^* = \{ z \in \mathbb{C} : \operatorname{Re} z > 0 \} \setminus [z_1, z_2],$$

and define the analytic functions on $\Omega^*$ by the equations,

(B.2)    $$\zeta_1(z) = 2\sqrt{x^2 - 1}, \quad \zeta_2(z) = x + \sqrt{x^2 - 1}; \qquad x = \frac{t + gz^2}{2\sqrt{\lambda g}},$$

and

(B.3)    $$\zeta_3(z) = x + \sqrt{x^2 - 1} + c.$$

These functions generate the conformal mappings, $z \mapsto \zeta$, of $\Omega^*$ onto the regions of the complex plane $\zeta$ depicted in Figure 3 below.

Let us also denote by $\zeta^{1/2}$ the branch of the square root of $\zeta$ defined on the complex plane $\zeta$ cut along the ray $(-\infty, 0]$ and fixed by the inequality,

$$-\pi < \arg \zeta < \pi.$$

Note that

(B.4)    $$e^{N\xi(z)} = \zeta_2^{-\frac{n}{2}}(z) e^{\frac{n}{2} x \sqrt{x^2 - 1}},$$

and

(B.5)    $$\left(\frac{\mu - a_{11}^0}{2\mu}\right)^{1/2} = \zeta_3^{1/2}(z) \zeta_1^{-1/2}(z).$$

From Figure 3 it follows that

(i) The functions $\zeta_1^{-1/2}(z)$, $\zeta_2^{-\frac{n}{2}}(z)$, and $\zeta_3^{1/2}(z)$ are single-valued on $\Omega^* \setminus [0, z_1]$.

(ii) The jumps of the functions $\zeta_1^{-1/2}(z)$, $\zeta_2^{-\frac{n}{2}}(z)$, and $\zeta_3^{1/2}(z)$ across the interval $[0, z_1]$ are described by the equations,

(B.6)    $$\zeta_1^{-1/2}(z)\Big|_+ = -\zeta_1^{-1/2}(z)\Big|_-,$$

$$\zeta_2^{-\frac{n}{2}}(z)\Big|_+ = (-1)^n \zeta_2^{-\frac{n}{2}}(z)\Big|_-,$$

$$\zeta_3^{1/2}(z)\Big|_+ = (-1)^{n+1} \zeta_3^{1/2}(z)\Big|_-.$$

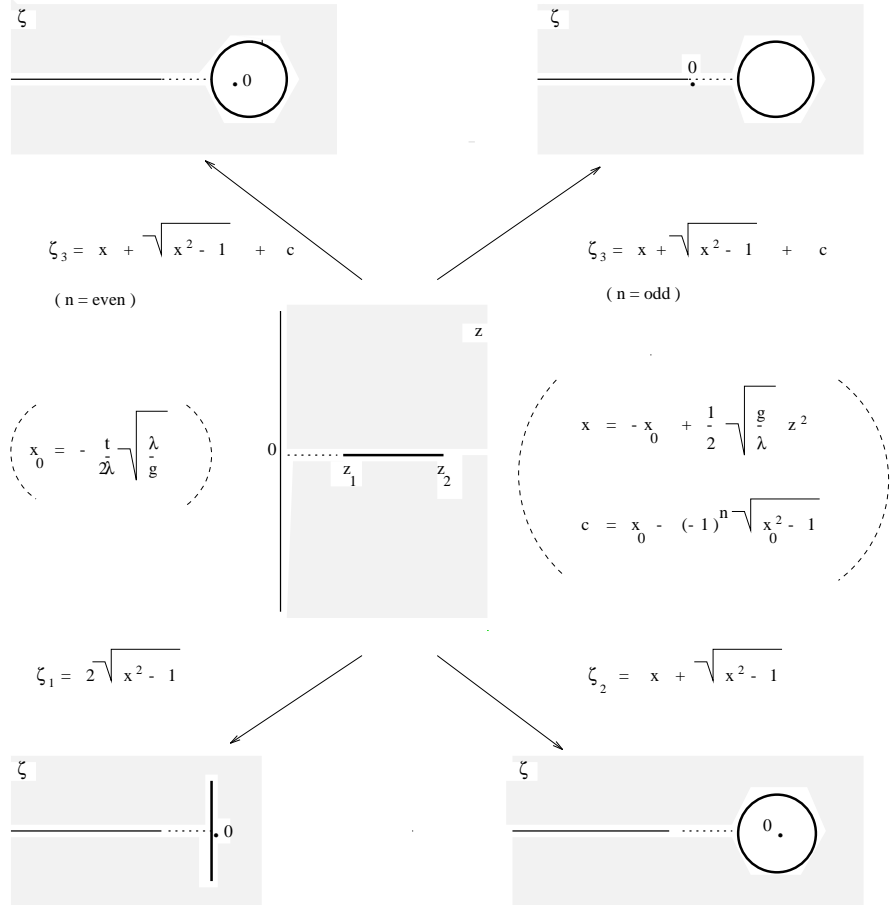

Figure 3. The conformal mappings related to $\Psi_{WKB}(z)$

(iii) The product, $\zeta_3^{1/2}(z)\zeta_1^{-1/2}(z)$, being analytically extended to $\mathbb{C} \setminus [-z_2, z_2]$ across the imaginary line becomes an even function.

(iv) The function, $\zeta_2^{-\frac{n}{2}}(z)$, being analytically extended to $\mathbb{C} \setminus [-z_2, z_2]$ across the imaginary line satisfies the symmetry equation,

$$(B.7) \qquad \zeta_2^{-\frac{n}{2}}(-z) = (-1)^n \zeta_2^{-\frac{n}{2}}(z).$$

From the properties (i)–(iv), Lemma 7.1 follows.

## Appendix C. Proofs of Lemmas 7.2, 7.4, and 7.6

*Proof of Lemma* 7.2. It is easy to see that the function $\Psi^0(z)$ defined by (7.9) and (7.28) satisfies the equation,

$$\text{(C.1)} \qquad\qquad \Psi^0(\bar{z}) = \overline{\Psi^0(z)}.$$

Taking also into account the $z \to -z$ symmetry (cf. (7.29)), we conclude that it is enough to prove the upper-right part of relation (7.32). We shall start with discussing the jump over the arc $C_2^u$ (see Fig.2).

For every $\varepsilon > 0$, in the sector $-\pi + \varepsilon < \arg z < \pi - \varepsilon$, the Airy function is written as

$$\text{Ai}\,(z) = \frac{1}{2\sqrt{\pi \eta'(z)}} \exp[-\eta(z)], \quad -\pi + \varepsilon < \arg z < \pi - \varepsilon,$$

where $\eta(z)$ is an analytic function, with the asymptotics

$$\text{(C.2)}$$
$$\eta(z) \sim \frac{2}{3} z^{3/2} \left(1 - \frac{5}{32} z^{-3} + \dots \right) = \frac{2}{3} z^{3/2} \left(1 + \sum_{j=1}^{\infty} (-1)^j \alpha_j z^{-3j} \right),$$
$$\eta(z) = \frac{2}{3} z^{3/2} \left(1 + O\left(|z|^{-3}\right)\right), \quad \eta'(z) = z^{1/2} \left(1 + O\left(|z|^{-3}\right)\right), \quad z \to \infty,$$

(cf. [Ble]). Hence

$$\text{(C.3)}$$
$$N^{1/6}\,\text{Ai}\,\left(N^{2/3} z\right) = \frac{1}{2\sqrt{\pi}\, z^{1/4}} \exp\left(-\frac{2N z^{3/2}}{3}\right) \left(1 + O\left(N^{-1}\right)\right),$$
$$N^{-1/6}\,\text{Ai}\,'\left(N^{2/3} z\right) = -\frac{z^{1/4}}{2\sqrt{\pi}} \exp\left(-\frac{2N z^{3/2}}{3}\right) \left(1 + O\left(N^{-1}\right)\right),$$
$$|z| > \varepsilon, \quad -\pi + \varepsilon < \arg z < \pi - \varepsilon, \quad N \to \infty.$$

By (7.23) this implies that

$$\text{(C.4)}$$
$$N^{1/6} y_0 \left(N^{2/3} z\right) = \frac{1}{2\sqrt{\pi}\, z^{1/4}} \exp\left(-\frac{2N z^{3/2}}{3}\right) \left(1 + O\left(N^{-1}\right)\right),$$
$$N^{-1/6} y_0' \left(N^{2/3} z\right) = -\frac{z^{1/4}}{2\sqrt{\pi}} \exp\left(-\frac{2N z^{3/2}}{3}\right) \left(1 + O\left(N^{-1}\right)\right),$$
$$|z| > \varepsilon, \quad -\pi + \varepsilon < \arg z < \pi - \varepsilon, \quad N \to \infty,$$

and

(C.5)

$$N^{1/6} y_1 \left( N^{2/3} z \right) = \frac{1}{2\sqrt{\pi}\, z^{1/4}} \exp\left( \frac{2N z^{3/2}}{3} \right) \left( 1 + O\left( N^{-1} \right) \right),$$

$$N^{-1/6} y_1' \left( N^{2/3} z \right) = \frac{z^{1/4}}{2\sqrt{\pi}} \exp\left( \frac{2N z^{3/2}}{3} \right) \left( 1 + O\left( N^{-1} \right) \right),$$

$$|z| > \varepsilon, \quad -\pi + \varepsilon < \arg z - \frac{2\pi}{3} < \pi - \varepsilon, \quad N \to \infty.$$

Hence, by (7.22),

(C.6)

$$\Phi_u(z; z_2) = \frac{1}{2\sqrt{\pi}} \begin{pmatrix} z^{-1/4} & z^{-1/4} \\ -z^{1/4} & z^{1/4} \end{pmatrix} \left( I + O\left( N^{-1} \right) \right) \exp\left( -\frac{2N z^{3/2} \sigma_3}{3} \right) P,$$

$$|z| > \varepsilon, \quad -\frac{\pi}{3} + \varepsilon < \arg z < \pi - \varepsilon, \quad N \to \infty,$$

where

$$P = e^{-\left(\frac{\ln 2\pi}{2}\right)\sigma_3} = \begin{pmatrix} (2\pi)^{-1/2} & 0 \\ 0 & (2\pi)^{1/2} \end{pmatrix}.$$

Thus,

(C.7)

$$\Phi_u(w(z; z_2); z_2) = \frac{1}{2\sqrt{\pi}} \begin{pmatrix} w^{-1/4}(z; z_2) & w^{-1/4}(z; z_2) \\ -w^{1/4}(z; z_2) & w^{1/4}(z; z_2) \end{pmatrix}$$

$$\times \left( I + O\left( N^{-1} \right) \right) \exp\left( -\frac{2N w^{3/2}(z; z_2) \sigma_3}{3} \right) P,$$

under the assumption that

(C.7′)      $$|w(z; z_2)| > \varepsilon, \quad -\frac{\pi}{3} + \varepsilon < \arg w(z; z_2) < \pi - \varepsilon, \quad N \to \infty.$$

Observe that this assumption on $w(z; z_2)$ holds for all $z \in C_2^u$ when $N$ is sufficiently large and $b > 0$ (the vertical half-axis of the ellipse $\Omega$) is sufficiently small. Indeed, from the definition of the function $w(z; z_2)$ given in Section 7 (see (7.19), (7.20)) it follows that $\pm w(z; z_2) > 0$ when $\pm(z - z_2^N) > 0$. In addition, $w'(z; z_2) \neq 0$ for $z \in \Omega_2$ and $w'(z; z_2) > 0$ for real $z \in \Omega_2$. This insures that there exist $b > 0$ and $\varepsilon > 0$, such that the assumption (C.7′) on $w(z; z_2)$ holds when $z \in C_2^u$, for all sufficiently large $N$.

By (7.19),

(C.8)
$$\frac{2}{3}\, w^{3/2}(z; z_2) = \int_{z_2^N}^{z} \sqrt{\mu^2(u) + N^{-1} U^{(1)}(u)}\, du$$

$$= \left\{ \frac{1}{2}\int_{\gamma_0} + \int_{\gamma(z)} \right\} \sqrt{\mu^2(u) + N^{-1} U^{(1)}(u)}\, du,$$

where

$$U^{(1)}(z) = (a_{11}^0(z))' - a_{11}^0(z)\, \frac{(a_{12}^0(z))'}{a_{12}^0(z)},$$

and $\gamma_0$ is a contour which starts at the point $z_0 = \frac{z_1 + z_2}{2}$ on the lower side of the cut $[z_1, z_2]$, goes around the point $z_2^N$ and ends at the point $z_0$ on the upper side of the cut $[z_1, z_2]$. The contour $\gamma(z)$ is a simple curve which goes from the point $z_0$ to the point $z$ and does not intersect any of the cuts. On $\gamma_0 \cup \gamma(z)$,

(C.9)  $\sqrt{\mu^2(u) + N^{-1} U^{(1)}(u)} = \mu(u) + \dfrac{U^{(1)}(u)}{2N\mu(u)} + O\left(N^{-2}\right), \quad u \in \gamma_0 \cup \gamma(z);$

hence

(C.10)
$$\frac{2}{3}\, w^{3/2}(z; z_2) = \left\{ \frac{1}{2}\int_{\gamma_0} + \int_{\gamma(z)} \right\} \mu(u)\, du$$

$$+ \frac{1}{N}\left\{ \frac{1}{2}\int_{\gamma_0} + \int_{\gamma(z)} \right\} \frac{U^{(1)}(u)\, du}{2\mu(u)} + O\left(N^{-2}\right)$$

$$= \int_{z_2}^{z} \mu(u)\, du + N^{-1}\int_{z_2}^{z} \frac{U^{(1)}(u)\, du}{2\mu(u)} + O\left(N^{-2}\right)$$

$$= \xi(z) + N^{-1} d(z) + O\left(N^{-2}\right)$$

where

$$\xi(z) = \int_{z_2}^{z} \mu(u)\, du = \frac{n}{2N}\left[ x\sqrt{x^2 - 1} - \ln(x + \sqrt{x^2 - 1}) \right], \qquad x = \frac{t + gz^2}{2\sqrt{\lambda g}},$$

(cf. (7.6), (7.7′)) and

(C.11)
$$d(z) = \int_{z_2}^{z} \frac{U^{(1)}(u)\, du}{2\sqrt{U(u)}} = \frac{1}{2}\int_{z_2}^{z} \frac{(a_{11}^0(u))'\, a_{12}^0(u) - a_{11}^0(u)\, (a_{12}^0(u))'}{a_{12}^0(u)\, \sqrt{(a_{11}^0(u))^2 - (a_{12}^0(u))^2}}\, du$$

$$= -\frac{1}{2}\ln\left\{ -\frac{a_{11}^0(u)}{a_{12}^0(u)} + \left[ \left(\frac{a_{11}^0(u)}{a_{12}^0(u)}\right)^2 - 1 \right]^{1/2} \right\}\Bigg|_{z_2}^{z} = -\frac{1}{2}\ln a(z),$$

where

$$a(z) = \frac{-a_{11}^0(z) + \sqrt{(a_{11}^0(z))^2 - (a_{12}^0(z))^2}}{a_{12}^0(z)} = \frac{-a_{11}^0(z) + \mu(z)}{a_{12}^0(z)}.$$

Note also that

(C.12) $$a(z) = \frac{1}{z} \left(\frac{\lambda}{g}\right)^{1/4} \frac{x + \sqrt{x^2 - 1} + c}{\sqrt{c}},$$

where

$$c \equiv \left(\frac{g}{\lambda}\right)^{1/2} R_n^0 = x_0 - (-1)^n \sqrt{x_0^2 - 1}, \quad x_0 = -\frac{t}{2\sqrt{\lambda g}}.$$

Substituting (C.10) into (C.7), we obtain the following asymptotic relation:

(C.13)
$$\Phi_u(w(z; z_2); z_2) = \frac{1}{2\sqrt{\pi}} \begin{pmatrix} w^{-1/4}(z; z_2) & w^{-1/4}(z; z_2) \\ -w^{1/4}(z; z_2) & w^{1/4}(z; z_2) \end{pmatrix}$$
$$\times \left(I + O\left(N^{-1}\right)\right) e^{-(N\xi(z) + d(z))\sigma_3} P,$$

(assuming again (C.7′)). Let us consider the function $\Psi_{\mathrm{TP}}^u(z; z_2)$ for $z$ in a small neighborhood of the arc $C_2^u$. In this case, as we discussed above, the condition (C.7′) holds (for all sufficiently large $N$); hence we can use (C.13). We want to check that for $z \in C_2^u$ the turning point solution $\Psi_{\mathrm{TP}}^u(z; z_2)$ coincides, up to terms of the order of $N^{-1}$, with the WKB solution (7.9), with some constants $C_0, C_1$. Differentiation of (C.10) gives that

(C.14) $$w'(z; z_2)w^{1/2}(z; z_2) = \xi'(z) + O(N^{-1}) = \mu(z) + O(N^{-1}).$$

Hence we obtain from (7.21) that

(C.15)
$$W(z; z_2) = \left(\frac{a_{12}^0(z)}{\mu(z)}\right)^{1/2} \begin{pmatrix} 1 & 0 \\ -\frac{a_{11}^0(z)}{a_{12}^0(z)} & \frac{\mu(z)}{a_{12}^0(z)} \end{pmatrix}$$
$$\times \begin{pmatrix} w^{1/4}(z; z_2) & 0 \\ 0 & w^{-1/4}(z; z_2) \end{pmatrix} (I + O(N^{-1})),$$

and by (C.13),

$$W(z; z_2)\Phi_u(w(z; z_2); z_2) = \frac{1}{2\sqrt{\pi}} \left(\frac{a_{12}^0(z)}{\mu(z)}\right)^{1/2} \begin{pmatrix} 1 & 0 \\ -\frac{a_{11}^0(z)}{a_{12}^0(z)} & \frac{\mu(z)}{a_{12}^0(z)} \end{pmatrix} \begin{pmatrix} 1 & 1 \\ -1 & 1 \end{pmatrix}$$
$$\times (I + O(N^{-1})) e^{-(N\xi(z) + d(z))\sigma_3} P$$
$$= \frac{1}{2\sqrt{\pi}} \left(\frac{a_{12}^0(z)}{\mu(z)}\right)^{1/2} \begin{pmatrix} 1 & 1 \\ a^{-1}(z) & a(z) \end{pmatrix}$$
$$\times (I + O(N^{-1})) e^{-(N\xi(z) + d(z))\sigma_3} P.$$

To adjust this formula to the WKB solution (7.9), we rewrite it as

$$W(z; z_2)\Phi_u(w(z; z_2); z_2) = \frac{1}{\sqrt{2\pi}} \left(\frac{\mu(z) - a_{11}^0(z)}{2\mu(z)}\right)^{1/2} \begin{pmatrix} a^{-1/2}(z) & a^{-1/2}(z) \\ a^{-3/2}(z) & a^{1/2}(z) \end{pmatrix}$$
$$\times (I + O(N^{-1}))e^{-(N\xi(z)+d(z))\sigma_3}P.$$

In virtue of (C.11), this gives that

(C.16)
$$W(z; z_2)\Phi_u(w(z; z_2); z_2)$$

$$= \frac{1}{\sqrt{2\pi}} \left(\frac{\mu(z) - a_{11}^0(z)}{2\mu(z)}\right)^{1/2} \begin{pmatrix} 1 & a^{-1}(z) \\ a^{-1}(z) & 1 \end{pmatrix} (I + O(N^{-1}))$$

$$\times e^{-N\xi(z)\sigma_3}P = \frac{1}{\sqrt{2\pi}} T_0(z) \left(I + O(N^{-1})\right) e^{-N\xi(z)\sigma_3}P$$

(cf. (7.9)). Thus, by (7.26),

(C.17)    $$\Psi_{\text{TP}}^u(z; z_2) = \frac{C}{\sqrt{2\pi}} T_0(z) \left(I + O(N^{-1})\right) e^{-N\xi(z)\sigma_3 - (1/2)(\ln 2\pi)\sigma_3},$$

uniformly for $z$ in a small neighborhood of the arc $C_2^u$. Therefore, if we take in (7.9)

(C.18)            $$C_0 = \frac{C}{\sqrt{2\pi}}, \qquad C_1 = -(1/2)(\ln 2\pi),$$

then

(C.19)        $$\Psi_{\text{TP}}^u(z; z_2) [\Psi_{\text{WKB}}(z)]^{-1} = I + O(N^{-1}), \qquad z \in C_2^u.$$

Because of symmetry relations (C.1), we obtain simultaneously that

(C.20)        $$\Psi_{\text{TP}}^d(z; z_2) [\Psi_{\text{WKB}}(z)]^{-1} = I + O(N^{-1}), \qquad z \in C_2^d.$$

Consider now the jump across the arc $C_1^u$. By (7.23),

(C.21)                    $$y_2(z) = \overline{y_1(\overline{z})};$$

hence by (C.5),

(C.22)

$$N^{1/6}y_2(N^{2/3}z) = \frac{1}{2\sqrt{\pi}\,z^{1/4}} \exp\left(\frac{2Nz^{3/2}}{3}\right)(1 + O(N^{-1})),$$

$$N^{-1/6}y_2'(N^{2/3}z) = \frac{z^{1/4}}{2\sqrt{\pi}} \exp\left(\frac{2Nz^{3/2}}{3}\right)(1 + O(N^{-1})),$$

$$|z| > \varepsilon, \quad -\pi + \varepsilon < \arg z + \frac{2\pi}{3} < \pi - \varepsilon, \quad N \to \infty.$$

Combining this formula with (C.5) and recalling the definition of the function $\Phi_u(z; z_1)$ in (7.27) we have

(C.23)

$$\Phi_u(z; z_1) = \frac{1}{2\sqrt{\pi}} \begin{pmatrix} z^{-1/4} & -z^{-1/4} \\ -z^{1/4} & -z^{1/4} \end{pmatrix} \left( I + O\left(N^{-1}\right) \right) \exp\left( -\frac{2N z^{3/2} \sigma_3}{3} \right) P,$$

$$|z| > \varepsilon, \quad -\pi + \varepsilon < \arg z < \frac{\pi}{3} - \varepsilon, \quad N \to \infty,$$

where $P$ is the same as in (C.6). Thus (cf. (C.7)),

(C.24)

$$\Phi_u(w(z; z_1); z_1) = \frac{1}{2\sqrt{\pi}} \begin{pmatrix} w^{-1/4}(z; z_1) & -w^{-1/4}(z; z_1) \\ -w^{1/4}(z; z_1) & -w^{1/4}(z; z_1) \end{pmatrix} \left( I + O\left(N^{-1}\right) \right)$$

$$\times \exp\left( -\frac{2N w^{3/2}(z; z_1) \sigma_3}{3} \right) P,$$

provided that

(C.24′) $\qquad |w(z; z_1)| > \varepsilon, \quad -\pi + \varepsilon < \arg w(z; z_1) < \frac{\pi}{3} - \varepsilon, \quad N \to \infty.$

From the definition of the function $w(z; z_1)$ given in Section 7 (see (7.27′)) it follows that $\pm w(z; z_1) > 0$ when $\pm(z - z_1^N) < 0$. In addition, $w'(z; z_1) \neq 0$ for $z \in \Omega_1$ and $w'(z; z_1) < 0$ for real $z \in \Omega_1$. This secures that there exist $b > 0$ and $\varepsilon > 0$ such that the assumption (C.24′) holds when $z \in C_1^u$, for all sufficiently large $N$.

By (7.27′),

(C.25) $$\frac{2}{3} w^{3/2}(z; z_1) = \int_{z_1^N}^{z} \sqrt{\mu^2(u) + N^{-1} U^{(1)}(u)} \, du$$

$$= \int_{z_2^N}^{z} \sqrt{\mu^2(u) + N^{-1} U^{(1)}(u)} \, du$$

$$+ \frac{1}{2} \oint_{\gamma_1} \sqrt{\mu^2(u) + N^{-1} U^{(1)}(u)} \, du,$$

where $\gamma_1$ is a loop which goes clockwise around the interval $[z_1^N, z_2^N]$. On the loop $\gamma_1$, the integrand of the last integral can be estimated as in (C.9). Hence

(C.26)

$$\frac{1}{2} \oint_{\gamma_1} \sqrt{\mu^2(u) + N^{-1} U^{(1)}(u)} \, du = \int_{z_1}^{z_2} \mu(u) \, du + N^{-1} \int_{z_1}^{z_2} \frac{U^{(1)}(u) \, du}{2\mu(u)}$$

$$+ O\left(N^{-2}\right)$$

$$= -\xi(z_1) - N^{-1} d(z_1) + O\left(N^{-2}\right),$$

where the functions $\xi(z)$ and $d(z)$ are the same as in (C.10), (C.12) and their values at $z_1$ are taken from the upper half-plane. Using the notation introduced in Appendix B, we can rewrite the functions $\xi(z)$ and $d(z)$ in the form

$$\xi(z) = \frac{n}{2N}\left[x\sqrt{x^2-1} - \ln\zeta_2(z)\right] \quad \text{and} \quad d(z) = -\frac{1}{2}\ln\frac{\zeta_3(z)}{|\zeta_3(z_1)|} - \frac{1}{2}\ln\frac{z_1}{z}.$$

This yields (cf. Fig. 3) that

$$N\xi(z_1) = -\frac{i\pi n}{2}, \quad \text{and} \quad d(z_1) = -\frac{i\pi\epsilon_n}{2},$$

where $\epsilon_n = 1$ if $n$ is even, and $\epsilon_n = 0$ if $n$ is odd. The last equations together with (C.25) and (C.26) lead to the following representation of the function $w(z;z_1)$ in the neighborhood of the arc $C_1^u$ (cf. (C.10)),

$$(\text{C.27}) \qquad \frac{2}{3}w^{3/2}(z;z_1) = \xi(z) + N^{-1}d(z) + \frac{i\pi}{2N} + \frac{i\pi}{N}\left[\frac{n}{2}\right] + O(N^{-2}).$$

Substituting (C.27) into (C.24) and assuming (C.24′), we arrive at the following asymptotic relation (cf. (C.13)):

(C.28)

$$\Phi_u(w(z;z_1);z_1) = \frac{(-1)^k}{2\sqrt{\pi}}\begin{pmatrix} w^{-1/4}(z;z_1) & -w^{-1/4}(z;z_1) \\ -w^{1/4}(z;z_1) & -w^{1/4}(z;z_1) \end{pmatrix}\left(I + O\left(N^{-1}\right)\right)$$

$$\times e^{-\frac{i\pi}{2}\sigma_3}e^{-(N\xi(z)+d(z))\sigma_3}P, \qquad k = \left[\frac{n}{2}\right].$$

As discussed above, for $z$ in a small neighborhood of the arc $C_1^u$ and sufficiently large $N$, condition (C.24′) holds; hence we can use the asymptotic formula (C.28) in equation (7.27) for $\Psi_{\mathrm{TP}}^u(z;z_1)$. To compare $\Psi_{\mathrm{TP}}^u(z;z_1)$ with $\Psi_{\mathrm{WKB}}(z)$, we also need (as in the previous case of the arc $C_2^u$) the asymptotics of the gauge factor $W(z;z_1)$ in (7.27). Similar to the previous case (cf. (C.14)), differentiation of (C.27) gives that

$$w'(z;z_1)w^{1/2}(z;z_1) = \xi'(z) + O(N^{-1}) = \mu(z) + O(N^{-1}),$$

which in turn implies the relation,

$$(\text{C.29}) \qquad (w'(z;z_1))^{1/2}w^{1/4}(z;z_1) = -\mu^{1/2}(z) + O(N^{-1}).$$

(The appearance of the sign "−" on the left hand is due to the condition $i(w'(z;z_1))^{1/2} > 0, \quad 0 < z < z_1^N$; see (7.27‴).) From (C.29) and (7.27‴) it follows (cf. (C.15)) that

(C.30)

$$W(z;z_1) = i\left(\frac{a_{12}^0(z)}{\mu(z)}\right)^{1/2}\begin{pmatrix} 1 & 0 \\ -\frac{a_{11}^0(z)}{a_{12}^0(z)} & \frac{\mu(z)}{a_{12}^0(z)} \end{pmatrix}$$

$$\times \begin{pmatrix} w^{1/4}(z;z_1) & 0 \\ 0 & w^{-1/4}(z;z_1) \end{pmatrix}(I + O(N^{-1})),$$

and by (C.28),

$$W(z; z_1)\Phi_u(w(z; z_1); z_1)$$

$$= \frac{i}{2\sqrt{\pi}}(-1)^k \left(\frac{a_{12}^0(z)}{\mu(z)}\right)^{1/2} \begin{pmatrix} 1 & 0 \\ -\frac{a_{11}^0(z)}{a_{12}^0(z)} & \frac{\mu(z)}{a_{12}^0(z)} \end{pmatrix} \begin{pmatrix} 1 & -1 \\ -1 & -1 \end{pmatrix}$$

$$\times (I + O(N^{-1}))e^{-\frac{i\pi}{2}\sigma_3}e^{-(N\xi(z)+d(z))\sigma_3}P.$$

Since $e^{-\frac{i\pi}{2}\sigma_3} = -i\sigma_3$, we obtain

$$W(z; z_1)\Phi_u(w(z; z_1); z_1) = \frac{(-1)^k}{2\sqrt{\pi}} \left(\frac{a_{12}^0(z)}{\mu(z)}\right)^{1/2} \begin{pmatrix} 1 & 0 \\ -\frac{a_{11}^0(z)}{a_{12}^0(z)} & \frac{\mu(z)}{a_{12}^0(z)} \end{pmatrix} \begin{pmatrix} 1 & -1 \\ -1 & -1 \end{pmatrix}$$

$$\times \begin{pmatrix} 1 & 0 \\ 0 & -1 \end{pmatrix} (I + O(N^{-1}))e^{-(N\xi(z)+d(z))\sigma_3}P$$

$$= \frac{(-1)^k}{2\sqrt{\pi}} \left(\frac{a_{12}^0(z)}{\mu(z)}\right)^{1/2} \begin{pmatrix} 1 & 0 \\ -\frac{a_{11}^0(z)}{a_{12}^0(z)} & \frac{\mu(z)}{a_{12}^0(z)} \end{pmatrix} \begin{pmatrix} 1 & 1 \\ -1 & 1 \end{pmatrix}$$

$$\times (I + O(N^{-1}))e^{-(N\xi(z)+d(z))\sigma_3}P$$

$$= \frac{(-1)^k}{2\sqrt{\pi}} \left(\frac{a_{12}^0(z)}{\mu(z)}\right)^{1/2} \begin{pmatrix} 1 & 1 \\ a^{-1}(z) & a(z) \end{pmatrix}$$

$$\times (I + O(N^{-1}))e^{-(N\xi(z)+d(z))\sigma_3}P.$$

The rest of the derivation is exactly the same as in the case of the arc $C_2^u$, and it yields the following formula (cf. (C.16)):

$$W(z; z_1)\Phi_u(w(z; z_1); z_1) = \frac{(-1)^k}{\sqrt{2\pi}} \left(\frac{\mu(z) - a_{11}^0(z)}{2\mu(z)}\right)^{1/2} \begin{pmatrix} 1 & a^{-1}(z) \\ a^{-1}(z) & 1 \end{pmatrix}$$

$$\times (I + O(N^{-1}))e^{-N\xi(z)\sigma_3}P$$

$$= \frac{(-1)^k}{\sqrt{2\pi}} T_0(z) (I + O(N^{-1})) e^{-N\xi(z)\sigma_3}P.$$

Thus, by (7.27),

$$(\text{C.31}) \quad \Psi_{\text{TP}}^u(z; z_1) = \frac{C^{(1)}(-1)^k}{\sqrt{2\pi}} T_0(z) (I + O(N^{-1})) e^{-N\xi(z)\sigma_3 - (1/2)(\ln 2\pi)\sigma_3},$$

uniformly for $z$ in a small neighborhood of the arc $C_1^u$ (cf. (C.17)).

To ensure the same WKB-asymptotics both for $\Psi_{\text{TP}}^u(z; z_2)$ and $\Psi_{\text{TP}}^u(z; z_1)$ we take $C^{(1)} = (-1)^k C$. Then

$$(\text{C.32}) \quad \Psi_{\text{TP}}^{u,d}(z; z_1) [\Psi_{\text{WKB}}(z)]^{-1} = I + O(N^{-1}), \qquad z \in C_{1,2}^{u,d}.$$

Our last step is to prove that the turning point solutions $\Psi_{\mathrm{TP}}^{u,d}(z; z_1)$ and $\Psi_{\mathrm{TP}}^{u,d}(z; z_2)$ coincide, up to the terms of the order of $N^{-1}$, on $l_0 = \partial\Omega_1 \cap \partial\Omega_2$ (see Fig. 2). By (C.19), (C.20) and (C.32), $\Psi_{\mathrm{TP}}^{u}(z; z_1)$ and $\Psi_{\mathrm{TP}}^{u}(z; z_2)$ coincide, up to the terms of the order of $N^{-1}$, at the end point of $l_0$, $z = z_0 + ib$. Since $b > 0$ can be chosen as small as we want, this is true for all $z = z_0 + ic$, $0 < c \le b$. The problem is to get a uniform estimate of the error term up to $c = 0$. We will consider a more general problem.

Let $\delta > 0$ be an arbitrary fixed number such that $\delta < (z_2 - z_1)/2$ and let

(C.33) $$\Omega_0 = \Omega \cap \{\, z_1 + \delta \le \mathrm{Re}\, z \le z_2 - \delta \,\}.$$

Then $l_0 \subset \Omega_0$. We will prove that

(C.34) $$\Psi_{\mathrm{TP}}^{u}(z; z_1)\, [\Psi_{\mathrm{TP}}^{u}(z; z_2)]^{-1} = I + O(N^{-1}), \qquad z \in \Omega_0,$$

and

(C.35) $$\Psi_{\mathrm{TP}}^{d}(z; z_1)\left[\Psi_{\mathrm{TP}}^{d}(z; z_2)\right]^{-1} = I + O(N^{-1}), \qquad z \in \Omega_0.$$

Observe that $w(z; z_2)$ is analytic in $z \in (\Omega \cap \{\mathrm{Re}\, z \ge z_1 + \delta\})$ if $N$ is sufficiently large, and $w'(z; z_2) \ne 0$ in this region. Hence the $\Psi_{\mathrm{TP}}^{u,d}(z; z_2)$ are analytic there as well. Similarly, the $\Psi_{\mathrm{TP}}^{u,d}(z; z_1)$ are analytic in the region $\Omega \cap \{\mathrm{Re}\, z \le z_2 - \delta\}$. Thus, all the functions $\Psi_{\mathrm{TP}}^{u,d}(z; z_{1,2})$ are analytic in $\Omega_0$. Because of the symmetry equation (C.1), it is enough to prove (C.34) only. The idea of the proof of (C.34) is to find a WKB function $\Psi_{\mathrm{WKB}}^{u}(z)$ such that for $k = 1, 2$,

(C.36) $$\Psi_{\mathrm{TP}}^{u}(z; z_k)\, [\Psi_{\mathrm{WKB}}^{u}(z)]^{-1} = I + O(N^{-1}), \qquad z \in \Omega_0.$$

To construct the function $\Psi_{\mathrm{WKB}}^{u}(z)$ let us consider an auxiliary turning point solution

(C.37) $$\Psi_{\mathrm{TP}}^{0}(z; z_2) = CW(z; z_2)\Phi_0(w(z; z_2)),$$

where
(C.38)
$$\Phi_0(z) = \begin{pmatrix} N^{1/6} & 0 \\ 0 & N^{-1/6} \end{pmatrix} \begin{pmatrix} y_1(N^{2/3}z) & y_2(N^{2/3}z) \\ y_1'(N^{2/3}z) & y_2'(N^{2/3}z) \end{pmatrix} \begin{pmatrix} (2\pi)^{-1/2} & 0 \\ 0 & (2\pi)^{1/2} \end{pmatrix}.$$

The key point here is that in contrast to $\Phi_{u,d}(z; z_2)$ in (7.22), the matrix elements of $\Phi_0(z)$ are defined in terms of $y_1(z)$, $y_2(z)$ only (and not $y_0(z)$; cf. (7.23)), and this allows us to find the asymptotics of $\Psi_{\mathrm{TP}}^{0}(z; z_2)$ as $N \to \infty$ for all $z \in \Omega_0$. The function $y_2(z)$ is an entire function. Changing the branch of the root functions on the right-hand side of (C.22) we obtain

(C.39)

$$N^{1/6} y_2(N^{2/3} z) = \frac{i}{2\sqrt{\pi}\, z^{1/4}} \exp\left(-\frac{2N z^{3/2}}{3}\right)(1 + O(N^{-1})),$$

$$N^{-1/6} y_2'(N^{2/3} z) = \frac{z^{1/4}}{2\sqrt{\pi}\, i} \exp\left(-\frac{2N z^{3/2}}{3}\right)(1 + O(N^{-1})),$$

$$|z| > \varepsilon, \quad -\pi + \varepsilon < \arg z - \frac{4\pi}{3} < \pi - \varepsilon, \quad N \to \infty.$$

Combining this formula with (C.5) we obtain

(C.40) $\quad \Phi_0(z) = \dfrac{1}{2\sqrt{\pi}} \begin{pmatrix} z^{-1/4} & i\, z^{-1/4} \\ z^{1/4} & -i\, z^{1/4} \end{pmatrix}(1 + O(N^{-1})) \exp\left(\dfrac{2N z^{3/2}\sigma_3}{3}\right) P,$

$$|z| > \varepsilon, \quad \frac{\pi}{3} + \varepsilon < \arg z < \frac{5\pi}{3} - \varepsilon, \quad N \to \infty.$$

Let $\xi_0(z)$, $\mu_0(z)$, and $d_0(z)$ be the analytic continuation of the functions $\xi(z)$, $\mu(z)$, and $d(z)$, respectively, from $\Omega_0^u$ to $\Omega_0$, across the interval $[z_1 + \delta, z_2 - \delta]$ (see equations (7.4), (7.6), and (C.11), (C.12)). Then equations (C.10), (C.14), and (C.15), with $\xi(z)$, $\mu(z)$, and $d(z)$ replaced by $\xi_0(z)$, $\mu_0(z)$, and $d_0(z)$, hold for all $z \in \Omega_0$. In addition, there exist $b > 0$ (the vertical half-axis of $\Omega$) and $\varepsilon > 0$ such that for all $z$ in $\Omega_0$ and large $N$ the inequalities

(C.41) $\qquad |w(z; z_2)| > \varepsilon, \quad \dfrac{\pi}{3} + \varepsilon < \arg w(z; z_2) < \dfrac{5\pi}{3} - \varepsilon,$

hold and hence one can use the asymptotics (C.40) for the function $\Phi_0(w(z; z_2))$. All this together leads to the asymptotic equation,

(C.42)

$$W(z; z_2)\Phi_0(w(z; z_2)) = \frac{1}{2\sqrt{\pi}} \left(\frac{a_{12}^0(z)}{\mu_0(z)}\right)^{1/2} \begin{pmatrix} 1 & 0 \\ -\frac{a_{11}^0(z)}{a_{12}^0(z)} & \frac{\mu_0(z)}{a_{12}^0(z)} \end{pmatrix} \begin{pmatrix} 1 & i \\ 1 & -i \end{pmatrix}$$

$$\times\, (I + O(N^{-1}))\, e^{(N\xi_0(z)+d_0(z))\sigma_3} P, \qquad z \in \Omega_0,$$

(cf. (C.16)). Since $iy_1(z) - iy_2(z) = y_0(z)$, we obtain that

(C.43) $\quad \Phi_u(z; z_2) = \Phi_0(z) P^{-1} \begin{pmatrix} i & 1 \\ -i & 0 \end{pmatrix} P, \qquad P = \begin{pmatrix} (2\pi)^{-1/2} & 0 \\ 0 & (2\pi)^{1/2} \end{pmatrix};$

hence

(C.44)
$$\Psi_{\text{TP}}^u(z; z_2) = CW(z; z_2)\Phi_u(w(z; z_2); z_2)$$
$$= \frac{C}{2\sqrt{\pi}} \left(\frac{a_{12}^0(z)}{\mu_0(z)}\right)^{1/2} \begin{pmatrix} 1 & 0 \\ -\frac{a_{11}^0(z)}{a_{12}^0(z)} & \frac{\mu_0(z)}{a_{12}^0(z)} \end{pmatrix} \begin{pmatrix} 1 & i \\ 1 & -i \end{pmatrix}$$
$$\times \left(I + O(N^{-1})\right) e^{(N\xi_0(z)+d_0(z))\sigma_3} \begin{pmatrix} i & 1 \\ -i & 0 \end{pmatrix} P, \qquad z \in \Omega_0.$$

Thus, if we define

(C.45)
$$\Psi_{\text{WKB}}^u(z) = \frac{C}{2\sqrt{\pi}} \left(\frac{a_{12}^0(z)}{\mu_0(z)}\right)^{1/2} \begin{pmatrix} 1 & 0 \\ -\frac{a_{11}^0(z)}{a_{12}^0(z)} & \frac{\mu_0(z)}{a_{12}^0(z)} \end{pmatrix}$$
$$\times \begin{pmatrix} 1 & i \\ 1 & -i \end{pmatrix} e^{(N\xi_0(z)+d_0(z))\sigma_3} \begin{pmatrix} i & 1 \\ -i & 0 \end{pmatrix} P,$$

then

(C.46)
$$\Psi_{\text{TP}}^u(z; z_2) = \left(I + O(N^{-1})\right) \Psi_{\text{WKB}}^u(z).$$

Consider now in $\Omega_0$ the turning point solution $\Psi_{\text{TP}}(z; z_1)$. For $z$ in $\Omega_0$ the same equations (C.27), (C.29), and (C.30), with the functions $\xi(z)$, $\mu(z)$, $d(z)$ replaced by the functions $\xi_0(z)$, $\mu_0(z)$, $d_0(z)$, hold. Also, there exist $b > 0$ and $\varepsilon > 0$ such that for all $z \in \Omega_0$ and sufficiently large $N$, $w(z; z_1)$ lies in the sector

(C.47)
$$-\frac{5\pi}{3} + \varepsilon < \arg w(z; z_1) < -\frac{\pi}{3} - \varepsilon$$

and $|w(z; z_1)| > \varepsilon$. This implies that

$$W(z; z_1)\Phi_0(w(z; z_1)) = \frac{i}{2\sqrt{\pi}}(-1)^k \left(\frac{a_{12}^0(z)}{\mu_0(z)}\right)^{1/2} \begin{pmatrix} 1 & 0 \\ -\frac{a_{11}^0(z)}{a_{12}^0(z)} & \frac{\mu_0(z)}{a_{12}^0(z)} \end{pmatrix} \begin{pmatrix} -i & 1 \\ i & 1 \end{pmatrix}$$
$$\times (I + O(N^{-1}))e^{-\frac{i\pi}{2}\sigma_3} e^{-(N\xi_0(z)+d_0(z))\sigma_3} P.$$

Since $e^{-\frac{i\pi}{2}\sigma_3} = -i\sigma_3$, we obtain that

(C.48)
$$W(z; z_1)\Phi_0(w(z; z_1))$$
$$= \frac{(-1)^k}{2\sqrt{\pi}} \left(\frac{a^0_{12}(z)}{\mu_0(z)}\right)^{1/2} \begin{pmatrix} 1 & 0 \\ -\frac{a^0_{11}(z)}{a^0_{12}(z)} & \frac{\mu_0(z)}{a^0_{12}(z)} \end{pmatrix} \begin{pmatrix} -i & 1 \\ i & 1 \end{pmatrix}$$
$$\times \begin{pmatrix} 1 & 0 \\ 0 & -1 \end{pmatrix} (I + O(N^{-1}))e^{-(N\xi_0(z)+d_0(z))\sigma_3} P$$
$$= \frac{(-1)^k}{2\sqrt{\pi}} \left(\frac{a^0_{12}(z)}{\mu_0(z)}\right)^{1/2} \begin{pmatrix} 1 & 0 \\ -\frac{a^0_{11}(z)}{a^0_{12}(z)} & \frac{\mu_0(z)}{a^0_{12}(z)} \end{pmatrix} \begin{pmatrix} -i & -1 \\ i & -1 \end{pmatrix}$$
$$\times (I + O(N^{-1}))e^{-(N\xi_0(z)+d_0(z))\sigma_3} P$$
$$= \frac{(-1)^k}{2\sqrt{\pi}} \left(\frac{a^0_{12}(z)}{\mu_0(z)}\right)^{1/2} \begin{pmatrix} 1 & 0 \\ -\frac{a^0_{11}(z)}{a^0_{12}(z)} & \frac{\mu_0(z)}{a^0_{12}(z)} \end{pmatrix} \begin{pmatrix} 1 & i \\ 1 & -i \end{pmatrix}$$
$$\times (I + O(N^{-1}))e^{(N\xi_0(z)+d_0(z))\sigma_3} \begin{pmatrix} 0 & -1 \\ -1 & 0 \end{pmatrix} P, \quad z \in \Omega_0.$$

Taking into account that (cf. (C.43))

$$\Phi_u(z; z_1) = \Phi_0(z)P^{-1} \begin{pmatrix} i & 0 \\ -i & -1 \end{pmatrix} P,$$

we derive from (C.48) the asymptotic equation

$$\Psi^u_{\mathrm{TP}}(z; z_1) = (-1)^k C W(z; z_1)\Phi_0(w(z; z_1))$$
$$= \frac{C}{2\sqrt{\pi}} \left(\frac{a^0_{12}(z)}{\mu_0(z)}\right)^{1/2} \begin{pmatrix} 1 & 0 \\ -\frac{a^0_{11}(z)}{a^0_{12}(z)} & \frac{\mu_0(z)}{a^0_{12}(z)} \end{pmatrix} \begin{pmatrix} 1 & i \\ 1 & -i \end{pmatrix}$$
$$\times (I + O(N^{-1}))e^{(N\xi_0(z)+d_0(z))\sigma_3} \begin{pmatrix} 0 & -1 \\ -1 & 0 \end{pmatrix} \begin{pmatrix} i & 0 \\ -i & -1 \end{pmatrix} P$$
$$= \frac{C}{2\sqrt{\pi}} \left(\frac{a^0_{12}(z)}{\mu_0(z)}\right)^{1/2} \begin{pmatrix} 1 & 0 \\ -\frac{a^0_{11}(z)}{a^0_{12}(z)} & \frac{\mu_0(z)}{a^0_{12}(z)} \end{pmatrix} \begin{pmatrix} 1 & i \\ 1 & -i \end{pmatrix}$$
$$\times (I + O(N^{-1}))e^{(N\xi_0(z)+d_0(z))\sigma_3} \begin{pmatrix} i & 1 \\ -i & 0 \end{pmatrix} P, \qquad z \in \Omega_0.$$

In other words,

$$\Psi^u_{\mathrm{TP}}(z; z_1) = (I + O(N^{-1})) \Psi^u_{\mathrm{WKB}}(z);$$

hence

$$\Psi^u_{\mathrm{TP}}(z; z_1) [\Psi^u_{\mathrm{TP}}(z; z_2)]^{-1} = I + O(N^{-1}), \qquad z \in \Omega_0.$$

Lemma 7.2 is proved.

*Proof of Lemma* 7.4. This lemma follows directly from the formulae (C.45), (C.46) (and symmetry (C.1)) if we substitute into these formulae the value of $C$ given in (7.34).

*Proof of Lemma* 7.6. By (C.8), (C.10),

$$
\xi_0(z) + N^{-1}d_0(z) = \left\{ \frac{1}{2} \int_{\gamma_0} + \int_{\gamma(z)} \right\} \sqrt{\mu^2(u) + N^{-1}U^{(1)}(u)} \, du + O\left(N^{-2}\right), \tag{C.49}
$$

where

$$
\mu^2(z) = \frac{z^2(gz^2 + t)^2}{4} - \frac{ngz^2}{N},
$$

$$
U^{(1)}(z) = (a_{11}^0(z))' - a_{11}^0(z) \frac{(a_{12}^0(z))'}{a_{12}^0(z)} = -\frac{gz^2}{2} + \frac{t}{2} + gR_n^0,
$$

so that

$$
\begin{aligned}
\mu^2(z) + N^{-1}U^{(1)}(z) &= z^2 \left[ \frac{(gz^2 + t)^2}{4} - \frac{\left(n + \frac{1}{2}\right)g}{N} \right] + N^{-1}\left( \frac{t}{2} + gR_n^0 \right) \\
&= z^2 \left[ \frac{(gz^2 + t)^2}{4} - \lambda' g \right] - N^{-1}\frac{(-1)^n\sqrt{t^2 - 4\lambda g}}{2} \\
&= \mu_1^2(z) - N^{-1}\alpha_n,
\end{aligned} \tag{C.50}
$$

where

$$
\mu_1^2(z) = \lambda' gz^2(q^2 - 1), \quad q = \frac{gz^2 + t}{2\sqrt{\lambda' g}}, \qquad \alpha_n = \frac{(-1)^n\sqrt{t^2 - 4\lambda g}}{2}.
$$

Hence,

$$
\begin{aligned}
\xi_0(z) + N^{-1}d_0(z) &= \left\{ \frac{1}{2} \int_{\gamma_0} + \int_{\gamma(z)} \right\} \sqrt{\mu_1^2(u) - N^{-1}\alpha_n} \, du + O\left(N^{-2}\right) \\
&= \left\{ \frac{1}{2} \int_{\gamma_0} + \int_{\gamma(z)} \right\} \left( \mu_1(u) - N^{-1}\frac{\alpha_n}{2\mu_1(u)} \right) du + O\left(N^{-2}\right) \\
&= \int_{z_{20}^N}^{z} \left( \mu_1(u) - N^{-1}\frac{\alpha_n}{2\mu_1(u)} \right) du + O\left(N^{-2}\right),
\end{aligned}
$$

where $z_{20}^N$ is a close-to-$z_2$-zero of $\mu_1(z)$. Now,

$$
\int_{z_{20}^N}^{z} \mu_1(u) \, du = \sqrt{\lambda' g} \int_{z_{20}^N}^{z} \sqrt{q^2 - 1} \, u du = i\sqrt{\lambda' g} \int_{z_{20}^N}^{z} \sqrt{1 - q^2} \, u du.
$$

The substitution $v = \frac{gu^2 + t}{2\sqrt{\lambda' g}}$ reduces this integral to

$$\int_{z_{20}^N}^z \mu_1(u)\, du = i\lambda' \int_1^q \sqrt{1 - v^2}\, dv = \frac{i\lambda'}{2}\left(q\sqrt{1 - q^2} - \arccos q\right)$$

$$= \frac{i\lambda'}{2}\left(\frac{\sin 2\phi}{2} - \phi\right), \quad \phi = \arccos q.$$

Similar computation gives

$$\int_{z_{20}^N}^z \frac{du}{\mu_1(u)} = -i\int_{z_{20}^N}^z \frac{2\,du}{u\sqrt{4\lambda' g - (gu^2 + t)^2}}$$

$$= \frac{i\arccos r}{\sqrt{t^2 - 4\lambda' g}} = (-1)^n \frac{i\chi}{2\alpha_n},$$

$$r = \frac{2\sqrt{\lambda' g} - tq}{2\sqrt{\lambda' g}\, q - t}, \quad \chi = \arccos r.$$

Thus,

$$i^{-1}\left[N\xi_0(z) + d_0(z)\right] = \frac{N\lambda'}{2}\left(\frac{\sin 2\phi}{2} - \phi\right) - \frac{(-1)^n \chi}{4}$$

and Lemma 7.6 is proved.

### Appendix D. Proof of estimate (7.45)

Denote

$$\Sigma = L \cup \partial\Omega_1 \cup \partial\Omega_2 \cup (-\partial\Omega_1) \cup (-\partial\Omega_2).$$

Theorem 7.3 implies that the function $X(z)$ solves the Riemann-Hilbert problem on the contour $\Sigma$,

(D.1) $$X(\infty) \equiv \lim_{z \to \infty} X(z) = I,$$

(D.2) $$X_+(z) = X_-(z)G(z), \qquad z \in \Sigma,$$

with the jump matrix $G(z)$ given by the equations,

(D.3)
$$G(z) = \Psi_-^0(z)\left[\Psi_+^0(z)\right]^{-1} \quad \text{if} \quad z \in \partial\Omega_1 \cup \partial\Omega_2 \cup (-\partial\Omega_1) \cup (-\partial\Omega_2),$$
$$G(z) = \Psi^0(z)S^{-1}\left[\Psi^0(z)\right]^{-1} \quad \text{if} \quad z \in L,$$

and satisfying the uniform estimates,

(D.4)
$$|I - G^{-1}(z)| \le CN^{-1}, \qquad \text{if} \quad z \in \partial\Omega_1 \cup \partial\Omega_2 \cup (-\partial\Omega_1) \cup (-\partial\Omega_2),$$
$$|I - G^{-1}(z)| \le Ce^{-2N\mathrm{Re}\,\xi(z)}, \quad \text{if} \quad z \in L.$$

The Riemann-Hilbert problem is depicted in Figure 4.

Figure 4. The Riemann-Hilbert problem for the function $X(z)$ and the contour $\Sigma$

Similar to the asymptotic problems considered in [DZ2] and [DIZ], the crucial fact, which follows from (D.4) and which is indicated in Figure 4, is that

$$(D.5) \qquad\qquad |I - G^{-1}(z)| = O(N^{-1}),$$

uniformly for $z \in \Sigma$ (compare to the similar analysis in [DZ2] and [DIZ]). Simultaneously, inequalities (D.4) yield the corresponding $L_2$-norm estimate:

$$(D.6) \qquad\qquad ||I - G^{-1}||_{L_2(\Sigma)} = O(N^{-1}).$$

*Remark* D.1. The function $\Psi^0(z)$ is defined by explicit formulae (7.9) and (7.28) which, in particular, imply that the functions $\Psi^0_\pm(z)$ and, in virtue of (D.3), the jump matrix $G(z)$ can be analytically continued in a small neighborhood of $\partial\Omega_1 \cup \partial\Omega_2 \cup (-\partial\Omega_1) \cup (-\partial\Omega_2)$. Moreover, in this neighborhood the first inequality in (D.4) holds. Indeed, the inequality follows from Lemma 7.2, whose proof is based on the estimates (C.17), (C.31), and (C.34), (C.35) taking place in the small neighborhoods of the corresponding pieces of the contour $\partial\Omega_1 \cup \partial\Omega_2$. When $z \in L$,

$$G(z) = I + T_0(z) \begin{pmatrix} 0 & -ie^{-2N\xi(z)} \\ 0 & 0 \end{pmatrix} T_0^{-1}(z),$$

(cf. (7.42), (7.43)) and hence $G(z)$ is analytic and satisfies the second inequality in (D.4) in a small neighborhood of $L$, which may include the narrow sectors (the angle less than $\pi/8$) about the intervals $(-\infty, -z_2]$ and $[z_2, \infty)$. The indicated properties of the matrix $G(z)$ mean that in the setting of the Riemann-Hilbert problem (D.1), (D.2) the choice of contour is not rigid; one can always slightly deform the contour $\Sigma$ and slightly rotate its infinite parts.

By standard techniques in Riemann-Hilbert theory (see e.g. [CG]; see also [BDT] and [DZ]) the solution $X(z)$ of the Riemann-Hilbert problem (D.1), (D.2) is given by the integral representation,

$$(D.7) \qquad X(z) = I + \frac{1}{2\pi i} \int_\Sigma \rho(\nu) \left( I - G^{-1}(\nu) \right) \frac{d\nu}{\nu - z}, \quad z \notin \Sigma,$$

where $\rho(z) \equiv X_+(z)$ solves the equation

$$(D.8) \qquad \rho(z) = I + C_+[\rho\left(I - G^{-1}\right)](z), \quad z \in \Sigma,$$

and $C_+$ is the corresponding Cauchy operator:

$$(C_+h)(z) = \lim_{z' \to z, \, z' \in (+)-\text{side}} \int_\Sigma \frac{h(\nu)}{\nu - z'} \frac{d\nu}{2\pi i}.$$

We note that by (7.38), $X(z)$ is built up from the functions which satisfy asymptotic conditions (5.16) and (7.35). Therefore, $I - \rho(z) = O\left(z^{-1}\right)$ as $z \to \infty$ and hence *a priori* it belongs to the space $L_2(\Sigma)$.

Put

$$\rho_0(z) = \rho(z) - I.$$

Then equation (D.8) can be rewritten as an equation in $L^2(\Sigma)$,

$$(D.8') \qquad\qquad\qquad (1 - B)\rho_0 = f,$$

where the function $f(z)$ and the operator $B$ are defined by the equations,

$$f = C_+(I - G^{-1}),$$
$$B\colon \phi \mapsto C_+[\phi\left(I - G^{-1}\right)], \quad \phi \in L_2(\Sigma).$$

The $L_2$-boundedness of the operator $C_+$ (see e.g. [LS]; see also [BDT] and [Zh]) and the estimates (D.5) and (D.6) imply that

$$||B||_{L_2(\Sigma) \to L_2(\Sigma)} \leq ||C_+||_{L_2(\Sigma) \to L_2(\Sigma)} ||I - G^{-1}||_{L_\infty(\Sigma)} = O(N^{-1}),$$

and

$$||f||_{L_2(\Sigma)} \leq ||C_+||_{L_2(\Sigma) \to L_2(\Sigma)} ||I - G^{-1}||_{L_2(\Sigma)} = O(N^{-1}).$$

Therefore, equation (D.8') is uniquely solvable in $L_2(\Sigma)$ for sufficiently large $N$, and its solution satisfies the estimate,

$$||\rho_0||_{L_2(\Sigma)} = O(N^{-1}),$$

or, in terms of $\rho(z)$,

$$||I - \rho||_{L_2(\Sigma)} = O(N^{-1}).$$

This equation together with (D.5) and (D.7) show that

$$(D.9) \qquad X(z) = I + O\left(\frac{1}{N(1 + |z|)}\right), \qquad z \in \mathbb{K},$$

for all closed subsets $\mathbb{K} \subset \mathbb{C} \setminus \Sigma$ satisfying the following property:

$$\frac{\text{dist}\{z; \Sigma\}}{1 + |z|} \geq c_0(\mathbb{K}) > 0, \quad \text{for all } z \in \mathbb{K}.$$

To complete the proof of estimate (7.45) we only need to notice that the domain of validity of the asymptotic equation (D.9) can be made the whole $z$-plane because of the indicated flexibility in Remark D.1 in the choice of the contour $\Sigma$.

*Remark* D.2. The kernel in integral equation (D.8) is small (estimate (D.5) again), and it is given explicitly in terms of elementary and Airy functions. Therefore, equations (D.7)–(D.9) make it possible (cf. [DZ3] and [DIZ]) to obtain the full asymptotic expansion for all the three main objects, $\lambda_n$, $R_n$, and $P_n(z)$.

## Appendix E. The Riemann-Hilbert approach to orthogonal polynomials

In this appendix , for the readers convenience, we reproduce with some more details the Riemann-Hilbert approach to the orthogonal polynomials suggested in [FIK].

Let $s_1, s_3$ be complex numbers, $|s_1| + |s_3| \neq 0$, and let $\mathbb{L}$ denote the cross,

$$\mathbb{L} = \mathbb{R} \bigcup i\mathbb{R} \,.$$

We assume that the cross $\mathbb{L}$ is oriented in a natural way, i.e. from $-\infty$ to $+\infty$, and from $-i\infty$ to $+i\infty$ (see Fig. 5).

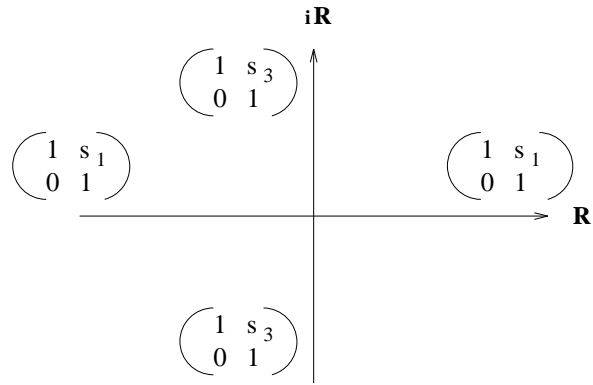

Figure 5. The Riemann-Hilbert problem for the function $Y(z) \, e^{-NV(z)\frac{\sigma_3}{2}}$

Consider the following Riemann-Hilbert problem posed on the cross $\mathbb{L}$ for $2 \times 2$ matrix function $Y(z) \equiv Y_n(z), \quad n \in \mathbb{Z}_+$:

(i) The function $Y(z)$ is analytic on $\mathbb{C} \setminus \mathbb{L}$. It has the limits, $Y_\pm(z)$, from the $\pm$ sides of $\mathbb{L}$ which satisfy the jump equation,

(E.1)
$$Y_+(z) = Y_-(z) \begin{pmatrix} 1 & se^{-NV(z)} \\ 0 & 1 \end{pmatrix} \equiv Y_-(z)e^{-\frac{NV(z)}{2}\sigma_3}Se^{\frac{NV(z)}{2}\sigma_3}, \qquad z \in \mathbb{L},$$

where

$$s = s_1 \quad \text{if} \quad z \in \mathbb{R}, \qquad s = s_3 \quad \text{if} \quad z \in i\mathbb{R},$$

and

$$S = \begin{pmatrix} 1 & s \\ 0 & 1 \end{pmatrix}.$$

(ii) The function $Y(z)$ is normalized by the asymptotic condition,

(E.2)
$$\lim_{z \to \infty} Y(z)z^{-n\sigma_3} = I.$$

The same arguments as in the case of the Riemann-Hilbert problem (5.16)–(5.18) (see Section 5) show that the solution $Y(z)$ of problem (E.1), (E.2) is unique (if it exists). In fact, we first notice that since $\det S = 1$, the scalar function $\det Y(z)$ is a bounded analytic function on $\mathbb{C}$ and such that its limit as $z \to \infty$ is 1. Hence,

(E.3)
$$\det Y(z) \equiv 1.$$

Suppose now that $\tilde{Y}(z)$ is another function satisfying (i), (ii). Since the jump matrix $S$ is the same for both $Y(z)$ and $\tilde{Y}(z)$, the matrix ratio, $\tilde{Y}(z)Y^{-1}(z)$, is a bounded analytic matrix function on $\mathbb{C}$ whose limit at $z = \infty$ equals $I$. Therefore,

$$\tilde{Y}(z)Y^{-1}(z) = I \quad \text{or} \quad \tilde{Y}(z) \equiv Y(z).$$

Consider also a system $\{P_n(z)\}$ of orthogonal polynomials on the cross $\mathbb{L}$ with the weight $\frac{i}{2\pi}se^{-NV(z)}$,

(E.4)
$$\int_{\mathbb{L}} P_n(z)P_m(z)e^{-NV(z)}dz = h_n\delta_{nm}, \qquad P_n(z) = z^n + \dots,$$

where

$$\int_{\mathbb{L}} \equiv \frac{i}{2\pi}s_1\int_{-\infty}^{+\infty} + \frac{i}{2\pi}s_3\int_{-i\infty}^{+i\infty}.$$

We shall assume that the nondegeneracy condition,

(E.5)
$$\det \left\| \left\{ \int_{\mathbb{L}} z^k z^j e^{-NV(z)}dz \right\}_{j,k=0,\dots,n} \right\| \neq 0, \quad \text{for all } n = 0, 1, 2, \dots,$$

is satisfied. Using the symmetries, $z \to -z$ and $z \to \bar{z}$, one can easily show that (E.5) is true if, for example,

$$(E.6) \qquad s_1 \in i\mathbb{R} \setminus \{0\} \quad \text{and} \quad s_3 \in i\mathbb{R}.$$

Condition (E.5) guarantees that the orthogonal set $\{P_n(z)\}$ exists. Each of the polynomials $P_n(z)$ is uniquely determined from the system of linear equations,

$$(E.7) \qquad \int_\mathbb{L} z^j P_n(z) e^{-NV(z)} dz = 0, \qquad j = 0, 1, \ldots n - 1.$$

Moreover, it also follows from (E.5) that

$$(E.8)$$
$$h_n \equiv \int_\mathbb{L} P_n^2(z) e^{-NV(z)} dz \equiv \int_\mathbb{L} z^n P_n(z) e^{-NV(z)} dz \neq 0, \quad \forall n = 0, 1, 2, \ldots .$$

Indeed, if for some $n$ condition (E.8) is violated, the coefficients of $P_n(z)$ would provide a nontrivial element of the kernel of the matrix,

$$\left\| \left\{ \int_\mathbb{L} z^k z^j e^{-NV(z)} dz \right\}_{j,k=0,\ldots,n} \right\|.$$

*Remark* E.1. A truncated condition,

$$(E.9)$$
$$\det \left\| \left\{ \int_\mathbb{L} z^k z^j e^{-NV(z)} dz \right\}_{j,k=0,\ldots,n} \right\| \neq 0, \quad \forall n = 0, 1, 2, \ldots, m-1, \quad m \in \mathbb{N},$$

yields the existence of the truncated orthogonal system $\{P_n(z)\}_{n=0}^m$, and the inequalities,

$$h_n \neq 0, \quad n = 0, \ldots, m-1.$$

The orthogonal polynomials considered in the main text correspond to the choice,

$$(E.10) \qquad s_3 = 0, \quad s_1 = -2\pi i.$$

Similar to special case (E.10), orthogonal polynomials (E.7) satisfy the linear equations (cf. (1.4)–(1.6)),

$$(E.11) \qquad z P_n(z) = P_{n+1}(z) + R_n P_{n-1}(z),$$

and

$$(E.12) \qquad P_n'(z) = N R_n [t + g(R_{n-1} + R_n + R_{n+1})] P_{n-1}(z)$$
$$+ g(N R_{n-2} R_{n-1} R_n) P_{n-3}(z), \qquad (') \equiv \frac{d}{dz},$$

where

$$(E.13) \qquad R_n = \frac{h_n}{h_{n-1}}, \quad n = 1, 2, \dots.$$

The coefficients $R_n$ satisfy the same Freud equation (1.7); i.e.,

$$(E.14) \qquad n = NR_n[t + g(R_{n-1} + R_n + R_{n+1})], \quad n = 1, 2, \dots$$

with the initial conditions,

$$R_0 = 0, \quad R_1 = \frac{h_1}{h_0} = \frac{\int_{\mathbb{L}} z^2 e^{-NV(z)} dz}{\int_{\mathbb{L}} e^{-NV(z)} dz}.$$

The relation between the Riemann-Hilbert problem (E.1), (E.2) and the orthogonal system $\{P_n(z)\}$ can be formulated as the following proposition:

PROPOSITION E.1 (cf. [FIK2,4]). *The Riemann-Hilbert problem* (E.1), (E.2) *is* (*uniquely*) *solvable for all* $n \in \mathbb{N}$ *if and only if condition* (E.5) *is satisfied, and hence the orthogonal system* $\{P_n(z)\}$ *exists. The* (*unique*) *solution* $Y(z)$ *of problem* (E.1), (E.2) *is given by the exact formula,*

$$(E.15) \qquad Y(z) = \begin{pmatrix} P_n(z) & -\int_{\mathbb{L}} \frac{P_n(\mu)e^{-NV(\mu)}}{\mu - z} d\mu \\ \frac{1}{h_{n-1}} P_{n-1}(z) & -\frac{1}{h_{n-1}} \int_{\mathbb{L}} \frac{P_{n-1}(\mu)e^{-NV(\mu)}}{\mu - z} d\mu \end{pmatrix},$$

*where the symbol* $\int_{\mathbb{L}}$ *is as introduced in* (E.4).

*Proof.* Suppose that (E.5) takes place, and define the function $Y(z)$ by the right side of equation (E.15). Then $Y(z)$ solves (E.1), (E.2). Indeed, the jump equation (E.1) follows directly from the Plemelj formula for the boundary values of Cauchy integrals, while the orthogonality conditions,

$$\int_{\mathbb{L}} z^j P_n(z) e^{-NV(z)} dz = h_n \delta_{nj}, \quad j = 0, 1, \dots, n,$$

imply that

$$Y(z) \sim \begin{pmatrix} z^n + O(z^{n-1}) & O(z^{-n-1}) \\ O(z^{n-1}) & z^{-n} + O(z^{-n-1}) \end{pmatrix}, \qquad z \to \infty,$$

and hence asymptotic equation (E.2) is valid as well.

Conversely, assume that the Riemann-Hilbert problem has a solution $Y(z)$ for all $n \in \mathbb{N}$. The triangularity of the jump matrix $S$ allows us to represent this solution in closed form (cf. [FIK4, §3.4]). In fact, asymptotic condition (E.2) implies that

$$(E.16) \qquad Y(z) \sim \begin{pmatrix} z^n + o(z^n) & o(z^{-n}) \\ o(z^n) & z^{-n} + o(z^{-n}) \end{pmatrix}, \qquad z \to \infty.$$

At the same time, the 11 and 21 components of (E.1) yield $(Y_+(z))_{11} = (Y_-(z))_{11}$ and $(Y_+(z))_{21} = (Y_-(z))_{21}$ respectively. Using these equations and (E.16) we find

$$(E.17) \qquad (Y(z))_{11} = P_n(z), \qquad (Y(z))_{21} = Q_{n-1}(z),$$

where $P_n(z)$ and $Q_{n-1}$ are the polynomials of the indicated degrees with the only restriction that

$$P_n(z) = z^n + \dots .$$

The 12 and 22 entries of equation (E.1) can be written down as the jump conditions,

$$(Y_+(z))_{12} - (Y_-(z))_{12} = se^{-NV(z)}(Y_-(z))_{11},$$
$$(Y_+(z))_{22} - (Y_-(z))_{22} = se^{-NV(z)}(Y_-(z))_{21},$$

which together with (E.17) provide us with the following representation for the function $Y(z)$:

$$(E.18) \qquad Y(z) = \begin{pmatrix} P_n(z) & -\int_{\mathbb{L}} \frac{P_n(\mu)e^{-NV(\mu)}}{\mu - z} d\mu \\ Q_{n-1}(z) & -\int_{\mathbb{L}} \frac{Q_{n-1}(\mu)e^{-NV(\mu)}}{\mu - z} d\mu \end{pmatrix}.$$

Referring again to (E.16), we conclude that the following equations,

$$(E.19) \qquad \int_{\mathbb{L}} \mu^j P_n(\mu) e^{-NV(z)} d\mu = 0, \qquad j = 0, 1, \dots n-1,$$

and

$$(E.20) \qquad \int_{\mathbb{L}} \mu^j Q_{n-1}(\mu) e^{-NV(z)} d\mu = \delta_{l,n-1}, \qquad j = 0, 1, \dots n-1,$$

hold for all $n \in \mathbb{N}$.

Our next step is to obtain the nondegeneracy conditions (E.5). Suppose that

$$(E.21)$$
$$\det \left\| \left\{ \int_{\mathbb{L}} z^k z^j e^{-NV(z)} dz \right\}_{j,k=0,\dots,m} \right\| = 0, \quad \text{for some} \quad m \in \{0, 1, 2, \dots, \}.$$

Then there is a nonzero tuple, $\{c_0, \dots, c_m\}$ such that

$$\sum_{k=0}^{m} c_k \int z^k z^j e^{-NV(z)} dz = 0, \quad \text{for all } j = 0, \dots, m.$$

This in turn implies that the polynomials,

$$\tilde{P}_{m+1}(z) \equiv P_{m+1}(z) + \sum_{k=0}^{m} c_k z^k,$$

and

$$\tilde{Q}_m(z) \equiv Q_m(z) + \sum_{k=0}^{m} c_k z^k,$$

satisfy (E.19), and (E.20) respectively. The pairs $(P_{m+1}(z), Q_m(z))$, $(\tilde{P}_{m+1}(z), \tilde{Q}_m(z))$, $(P_{m+1}(z), \tilde{Q}_m(z))$, and $(\tilde{P}_{m+1}(z), Q_m(z))$ generate, via (E.18), the different functions which solve the same Riemann-Hilbert problem (E.1), (E.2) for $n = m + 1$. This contradiction shows that assumption (E.21) is false.

Condition (E.5), which has just been established, means that equations (E.19) and (E.20) determine polynomials $P_n(z)$ and $Q_{n-1}(z)$ uniquely for all $n \in \mathbb{N}$. Simultaneously, the equation,

$$Q_{n-1}(z) = \frac{1}{h_{n-1}} P_{n-1}(z),$$

holds. This equation and equations (E.18), (E.19) complete the proof of the theorem.

*Remark* E.2. Let $m \in \mathbb{N}$. Then the solvability of the Riemann-Hilbert problem (E.1), (E.2) and representation (E.15) for $n \in \{1, \ldots, m\}$ only need truncated solvability condition (E.9).

*Remark* E.3. If $n = 0$, the Riemann-Hilbert problem (E.1), (E.2) is always solvable, and the solution is given by the equation,

$$(E.22) \qquad Y(z) = \begin{pmatrix} 1 & -\int_{\mathbb{L}} \frac{e^{-NV(\mu)}}{\mu - z} d\mu \\ 0 & 1 \end{pmatrix}.$$

From exact representation (E.15) it follows that the solution $Y(z)$ of the Riemann-Hilbert problem (E.1), (E.2) has, in fact, a complete asymptotic series at $z = \infty$,

$$(E.23) \qquad Y(z) \sim \left( I + \sum_{k=1}^{\infty} \frac{\Theta_k}{z^k} \right) z^{n\sigma_3}, \qquad z \to \infty.$$

Observe also that if $Y(z)$ is a solution of problem (E.1), (E.2), then so is the function,

$$(-1)^n \sigma_3 Y(-z) \sigma_3.$$

Because of the uniqueness of the solution of the Riemann-Hilbert problem, the symmetry equation

$$(E.24) \qquad Y(z) = (-1)^n \sigma_3 Y(-z) \sigma_3$$

holds. Equation (E.24) in turn implies that in asymptotic series (E.23) all the even coefficients, $\Theta_{2l}$, are diagonal while all the odd coefficients, $\Theta_{2l+1}$, are off-diagonal. Moreover, again from (E.15), we conclude that

(E.25)
$$(\Theta_1)_{12} = h_n, \quad (\Theta_1)_{21} = \frac{1}{h_{n-1}},$$

and hence

(E.26)
$$(\Theta_1)_{12}(\Theta_1)_{21} = R_n,$$

where the $h_n$ and $R_n$ are defined as in (E.8) and (E.13) respectively.

Define

(E.27)
$$\Psi_n(z) = e^{-\lambda_n \sigma_3}\Gamma_0 Y(z)e^{-\frac{NV(z)}{2}\sigma_3},$$

where

$$\Gamma_0 = \begin{pmatrix} 1 & 0 \\ 0 & R_n^{-1/2} \end{pmatrix}, \quad \text{and} \quad e^{2\lambda_n} = h_n.$$

The function $\Psi_n(z)$ is analytic on $\mathbb{C} \setminus \mathbb{L}$. It has the limits $\Psi_{n\pm}(z)$ from the $\pm$ sides of $\mathbb{L}$ which satisfy the jump equation (cf. (E.1)),

(E.28)
$$\Psi_{n+}(z) = \Psi_{n-}(z)S, \quad S = \begin{pmatrix} 1 & s \\ 0 & 1 \end{pmatrix}.$$

In virtue of (E.23), (E.25), and (E.26), the function $\Psi_n(z)$ has the asymptotic expansion,

(E.29)
$$\Psi_n(z) \sim \left(\sum_{k=0}^{\infty} \frac{\Gamma_k}{z^k}\right) e^{-\left(\frac{NV(z)}{2} - n\ln z + \lambda_n\right)\sigma_3}, \qquad z \to \infty,$$

where

(E.30)
$$\Gamma_0 = \begin{pmatrix} 1 & 0 \\ 0 & R_n^{-1/2} \end{pmatrix}, \qquad \Gamma_1 = \begin{pmatrix} 0 & 1 \\ R_n^{1/2} & 0 \end{pmatrix}.$$

Also, equation (E.15) yields the following representation for the function $\Psi_n(z)$,

(E.31)
$$\Psi_n(z) = \begin{pmatrix} \psi_n(z) & \varphi_n(z) \\ \psi_{n-1}(z) & \varphi_{n-1}(z) \end{pmatrix},$$

where

$$\psi_n(z) = \frac{1}{\sqrt{h_n}}P_n(z)e^{\frac{-NV(z)}{2}},$$

and

$$\varphi_n(z) = -\int_{\mathbb{L}} \frac{e^{-\frac{NV(\mu)}{2}}\psi_n(\mu)}{\mu - z}d\mu.$$

Equation (E.31) provides an orthogonal polynomial representation for the solution of the Riemann-Hilbert problem (E.28)–(E.30). In the particular case (E.10), this problem is exactly our basic Riemann-Hilbert problem (5.16)–(5.18), and formula (E.31) coincides with formula (5.15). Equation (5.15) for the solution of the Riemann-Hilbert problem (5.16)–(5.18) was derived in Section 5 via the analysis of the monodromy properties of the solutions of the Lax pair (3.6). Matrix equations (3.6) are equivalent to the pair of the scalar equations (E.11), (E.12). In the approach presented in this appendix we do not use these equations at all. It is the Riemann-Hilbert problem (E.1), (E.2), or equivalently (E.28)–(E.30), that we make now a starting point of the analysis of the orthogonal polynomial system (E.7).

The Lax pair, and hence the Freud equation, for orthogonal system (E.7) can be obtained directly from the Riemann-Hilbert problem (E.28)–(E.30) by use of the general inverse monodromy problem technique of [JMU] (see also [FIK4]). To this end, put

$$(E.32) \qquad Z(z) = e^{\lambda_n \sigma_3} \Gamma_0^{-1} \Psi_n(z) e^{W(z)\sigma_3} \equiv Y(z) z^{-n\sigma_3},$$

where we denote for the sake of brevity,

$$W(z) = \frac{NV(z)}{2} - n \ln z.$$

Then

$$(E.33) \qquad \Psi_n(z) = e^{-\lambda_n \sigma_3} \Gamma_0 Z(z) e^{-W(z)\sigma_3},$$

and by (E.23),

$$(E.34) \qquad Z(z) \sim I + \sum_{k=1}^{\infty} \frac{\Theta_k}{z^k}, \qquad z \to \infty.$$

The next lemma shows that the Riemann-Hilbert problem implies a pair of polynomial matrix differential and difference equations on $\Psi_n(z)$. We shall use the following usual notation: If

$$B(z) \sim \sum_{k=-\infty}^{m} b_k z^k, \qquad z \to \infty,$$

we denote by

$$\left\{ B(z) \right\}_+ = \sum_{k=0}^{m} b_k z^k,$$

the polynomial part of $B(z)$ at infinity.

LEMMA E.1. *Assume that $\Psi_n(z)$ is a solution of the Riemann-Hilbert problem* (E.28)–(E.30). *Then $\Psi_n(z)$ satisfies the polynomial $2\times 2$ matrix differential-difference Lax pair (cf.* (3.6)),

$$(E.35) \qquad \begin{cases} \Psi_{n+1}(z) = U_n(z)\Psi_n(z), \\ \Psi_n'(z) = NA_n(z)\Psi_n(z), \end{cases} \qquad (') \equiv \frac{d}{dz},$$

*with*

(E.36) $\qquad A_n(z) = -(1/2)e^{-\lambda_n\sigma_3}\Gamma_0\left\{Z(z)V'(z)\sigma_3 Z^{-1}(z)\right\}_+ \Gamma_0^{-1}e^{\lambda_n\sigma_3},$

*and*

(E.37)

$$U_n(z) = e^{-\lambda_{n+1}\sigma_3}\begin{pmatrix} 1 & 0 \\ 0 & R_{n+1}^{-1/2} \end{pmatrix}\left\{Z_{n+1}(z)z^{\sigma_3}Z_n^{-1}(z)\right\}_+ \begin{pmatrix} 1 & 0 \\ 0 & R_n^{1/2} \end{pmatrix}e^{\lambda_n\sigma_3},$$

*where $Z(z) \equiv Z_n(z)$ is defined as in* (E.32).

*Proof.* Observe that $\det\Psi_n(z)$ is an entire function, since

$$\det\Psi_{n+}(z) = \det\Psi_{n-}(z)\det S = \det\Psi_{n-}(z).$$

In addition,

$$\lim_{z\to\infty}\det\Psi_n(z) = R_n^{-1/2}\lim_{z\to\infty}\det Z(z) = R_n^{-1/2}\det I = R_n^{-1/2}.$$

Hence

$$\det\Psi(z) \equiv R_n^{-1/2} \neq 0.$$

Let us check that $\Psi(z)$ satisfies a matrix differential equation. Define

$$Q(z) = \Psi_n'(z)\Psi_n^{-1}(z).$$

Then by (E.28) and the fact that the jump matrix $S$ does not depend on $z$,

$$Q_+(z) = \Psi_{n+}'(z)\Psi_{n+}^{-1}(z) = \Psi_{n-}'(z)SS^{-1}\Psi_{n-}^{-1}(z) = Q_-(z),$$

so that $Q(z)$ is an entire matrix-valued function. By (E.33),

$$\begin{aligned}
Q(z) &= e^{-\lambda_n\sigma_3}\Gamma_0\left[Z'(z)e^{-W(z)\sigma_3} - Z(z)W'(z)\sigma_3 e^{-W(z)\sigma_3}\right] \\
&\quad \times e^{W(z)\sigma_3}Z^{-1}(z)\Gamma_0^{-1}e^{\lambda_n\sigma_3} \\
&= e^{-\lambda_n\sigma_3}\Gamma_0\left[Z'(z) - Z(z)W'(z)\sigma_3\right]Z^{-1}(z)\Gamma_0^{-1}e^{\lambda_n\sigma_3};
\end{aligned}$$

hence $Q(z)$ grows polynomially at infinity, and hence $Q(z)$ is a polynomial,

$$\begin{aligned}
Q(z) &= e^{-\lambda_n\sigma_3}\Gamma_0\left\{\left[Z'(z) - Z(z)W'(z)\sigma_3\right]Z^{-1}(z)\right\}_+ \Gamma_0^{-1}e^{\lambda_n\sigma_3} \\
&= -(N/2)e^{-\lambda_n\sigma_3}\Gamma_0\left\{Z(z)V'(z)\sigma_3 Z^{-1}(z)\right\}_+ \Gamma_0^{-1}e^{\lambda_n\sigma_3}.
\end{aligned}$$

Thus we get a polynomial differential equation on $\Psi_n(z)$,

$$\Psi_n'(z) = Q(z)\Psi_n(z),$$

with

$$Q(z) = -(N/2)e^{-\lambda_n \sigma_3} \Gamma_0 \left\{ Y(z)V'(z)\sigma_3 Y^{-1}(z) \right\}_+ \Gamma_0^{-1} e^{\lambda_n \sigma_3}.$$

This proves equation (E.36). Equation (E.37) is proved similarly when we consider the "discrete" logarithmic derivative,

$$\tilde{Q}(z) = \Psi_{n+1}(z)\Psi_n^{-1}(z),$$

and take this time into account the fact that the jump matrix $S$ does not depend on $n$. The proof is complete.

A straightforward calculation based on asymptotic series (E.34) and symmetry equation (cf. (E.24)),

$$Z(z) = \sigma_3 Z(-z)\sigma_3,$$

leads to representations (3.7) and (3.8) for the matrices $U_n(z)$ and $A_n(z)$ respectively. Also, substituting (E.33), (E.34) into the second equation in (E.35) and equating the terms of order $z^{-1}$, one obtains, for $R_n$, Freud equation (E.14). The latter simultaneously is a compatibility condition of the Lax pair (E.35) (see §5).

One can repeat the arguments used in Section 5 for the particular case (E.10) and show that the Riemann-Hilbert problem (E.29)–(E.31) represents the Stokes phenomenon for the matrix differential equation

(E.38) $$\Psi_n'(z) = NA_n(z)\Psi_n(z),$$

$$A_n(z) = \begin{pmatrix} -(\frac{tz}{2} + \frac{gz^3}{2} + gzR_n) & R_n^{1/2}[t + gz^2 + g(R_n + R_{n+1})] \\ -R_n^{1/2}[t + gz^2 + g(R_{n-1} + R_n)] & \frac{tz}{2} + \frac{gz^3}{2} + gzR_n \end{pmatrix},$$

with $R_n$ defined as the solution of initial problem (E.14) for the Freud equation.

For the analysis of the general monodromy problem for equation (E.38) corresponding to an arbitrary solution of the Freud equation and for its relation to the fourth Painlevé equation we refer the reader to Section 3 of [FIK2]. The monodromy problem corresponding to the polynomial $V(z)$ of an arbitrary even degree is considered in [FIK4].

*Remark* E.4. The technique used in this appendix was first suggested in [FMA] for analyzing the explicit solutions of the Painlevé equations. In [FIK4] it was applied to the case of the arbitrary even polynomial $V(z)$. In fact, using the same idea one can easily reduce the analysis of an *arbitrary* system of the orthogonal polynomials $\{P_n(z)\}$ on some contour $L$ with some weight $\omega(z)$ to the analysis of relevant $2 \times 2$ matrix Riemann-Hilbert problem. The RH problem is formulated for a $2 \times 2$ matrix function $Y(z)$ which is analytic outside the contour $L$, normalized by the asymptotic condition

$$Y(z)z^{-n\sigma_3} \to I, \qquad z \to \infty, \qquad \sigma_3 = \begin{pmatrix} 1 & 0 \\ 0 & -1 \end{pmatrix},$$

and whose boundary values $Y_\pm(z)$ satisfy equation:

$$Y_+(z) = Y_-(z) \begin{pmatrix} 1 & -2\pi i \omega(z) \\ 0 & 1 \end{pmatrix}, \qquad z \in L.$$

This Riemann-Hilbert problem can also be used to explain the appearance (see, e.g., [ASM]) of the KP-type hierarchies in the matrix models. Indeed, let us assume that the weight function $\omega(\lambda)$ is of the form

$$\omega(z) = e^{\Sigma_{k=1}^N t_k z^k}$$

and put

$$\Psi(z) = Y(z) e^{\frac{1}{2}(\Sigma_{k=1}^N t^k z^k)\sigma_3}.$$

Then, the same arguments as the ones we used for proving Lemma E.1 yield (cf. (1.12)–(1.16) in [FIK4]) the system of linear differential and difference equations for function $\Psi(z) \equiv \Psi(z; n, t_1, t_2, t_3, \ldots)$,

$$\Psi(z; n+1) = U(z)\Psi(z; n)$$
$$\partial_z \Psi(z) = A(z)\Psi(z)$$
$$\partial_{t_k} \Psi(z) = V_k(z)\Psi(z), \qquad k = 1, 2, 3, \ldots,$$

where $U(z)$, $A(z)$, and $V_k(z)$ are polynomial on $z$ (for their exact expressions in terms of the corresponding $R_n$ see [FIK4]). The first two equations constitute the Lax pair for the relevant Freud equation. The compatibility conditions of the third equation with the different $k$ generate the KP-type hierarchy of the integrable PDEs; the compatibility condition of the second and the third equations produces the Virasoro-type constraints; the compatibility condition of the first and the third equations is related to the Toda-type hierarchy and vertex operators.

*Acknowledgement.* This work was supported in part by the National Science Foundation, Grant No. DMS-9623214 (P.B.) and Grant No. DMS-9501559 (A.I.), and this support is gratefully acknowledged. The authors thank the referees for valuable remarks.

INDIANA UNIVERSITY-PURDUE UNIVERSITY INDIANAPOLIS, INDIANAPOLIS, IN
*E-mail addresses*: pbleher@math.iupui.edu
                     itsa@math.iupui.edu